 \documentclass[12pt,a4paper]{article}
 \pdfoutput=1
 \usepackage{jheppub}

\usepackage{amsmath}
\usepackage{bm}
\usepackage{wrapfig}
\usepackage{subcaption}
\usepackage{grffile}

\definecolor{rust}{rgb}{0.8,0.2,0.2}
\definecolor{green}{rgb}{0.1,0.8,0.2}
\definecolor{myblue}{rgb}{0.12,0.51,0.88}

\def\amp{{A}}
\def\per{{P}}
\def\gtt{{f}} 
\def\br{{\chi}}
\def\gxx{{\rho}}
\def\sigmaOut{{\sigma_{\text{out}}}}
\def\sigmaIn{{\sigma_{\text{in}}}}
\def\vev#1{\langle #1 \rangle}
\def\AdS#1{AdS$_{#1}$}

\def\eebdy{{\cal A}}
\def\GN{G_{\text{\tiny N}}^\text{\tiny{(4)}}}
\def\sreg{\Delta{\sf S}_{_{\cal A}} }

\title{Driven Holographic CFTs}

\author[a]{Mukund Rangamani}
\author[b]{, Moshe Rozali}
\author[b]{, Anson Wong}

\affiliation[a]{ Centre for Particle Theory \& Department of Mathematical Sciences,\\ 
Durham University, South Road, Durham DH1 3LE, UK}
\affiliation[b]{Department of Physics and Astronomy, University of British Columbia,\\
6224 Agricultural Road, Vancouver, BC V6T 1Z1, Canada}
\emailAdd{mukund.rangamani@durham.ac.uk}
\emailAdd{rozali@phas.ubc.ca}
\emailAdd{awcwong@phas.ubc.ca}

\abstract{
We study the dynamical evolution of strongly coupled field theories, initially in thermal equilibrium, under the influence of an external driving force. We model the field theory holographically using classical gravitational dynamics in an asymptotically AdS spacetime. The system is driven by a source for a (composite) scalar operator. We  focus on a scenario where the external source is periodic in time and chart out the response of several observables. We find an interesting phase structure in the response as a function of the amplitude of the source and driving frequency. Specifically the system transitions from a dissipation dominated phase, via a dynamical crossover to a highly resonant amplification phase. The diagnostics of these phases include the response of the operator in question, entropy production, energy fluctuations, and the temporal change of entanglement entropy for small subsystems. We comment on evidence for a potential phase transition in the energy fluctuations of the system.
 } 

\preprint{DCPT-15/05}

\begin{document}
\maketitle

\flushbottom

\section{Introduction and Conclusions}
\label{sec:intro}

The dynamics of quantum field theories driven far from equilibrium is a fascinating topic, owing to the complex interplay of quantum and statistical behaviours in the system. While a quantitative understanding of how field theories respond to non-linear external sources remains in general an open problem, in recent years  one has gained some insight into such phenomena.  

On the one hand progress in this direction has been driven by experimental developments which allow for a detailed study. For instance the ability to simulate many-body dynamics in cold-atom systems has led to the opening of a new frontier in dynamical simulations, cf., \cite{Polkovnikov:2010yn} for a recent review. On the other hand, theoretical horizons have been broadened with the gauge/gravity duality providing an excellent arena to explore the dynamics of strongly interacting many-body systems using 
(classical) gravitational dynamics in a suitable limit (cf., \cite{Hubeny:2010ry} for a not so recent review). Coupled with the development of excellent numerical algorithms for studying dynamical problems in AdS gravity \cite{Chesler:2008hg,Chesler:2013lia}, the confluence of ideas and techniques provides an excellent opportunity to further our understanding of 
out-of-equilibrium dynamics. 

A much studied protocol in this context is the quantum quench dynamics, wherein one takes a system initially in equilibrium, typically in the ground state, and subjects it to external sources which change the subsequent dynamics by modifying the Hamiltonian. The rate at which  sources act on the system controls the features of the subsequent relaxation, assuming that the sources are non-vanishing for a finite amount of time. The analysis of such a quench protocol has benefited both from theoretical understanding using standard quantum field theory technology in low dimensions \cite{Calabrese:2005in,Calabrese:2006rx,Calabrese:2007rg} and from a wide array of examples that have been studied holographically in the recent past 
\cite{Bhattacharyya:2009uu,Das:2010yw,Basu:2011ft,Buchel:2012gw,Bhaseen:2012gg,Basu:2012gg,Nozaki:2013wia,Buchel:2013lla,Hartman:2013qma,Basu:2013vva,Li:2013fhw,Buchel:2013gba,Auzzi:2013pca,Buchel:2014gta}. 
In most cases the interest is in the approach to equilibrium at late times and the rate at which various observables thermalize 
\cite{Danielsson:1999fa,Hubeny:2007xt,AbajoArrastia:2010yt,Albash:2010mv,Balasubramanian:2010ce,Balasubramanian:2011ur,Aparicio:2011zy,Balasubramanian:2011at,Keranen:2011xs,Galante:2012pv,Caceres:2012em,Hubeny:2013hz,Hartman:2013qma,Liu:2013iza,Balasubramanian:2013oga,Liu:2013qca,Abajo-Arrastia:2014fma}. Note that since we inject energy in the process of the quench, even an initially pure state will appear to be well approximated by a thermal ensemble asymptotically (assuming that the field theory dynamics are sufficiently ergodic).

A slightly different  but related scenario is one where we subject a system, again initially in an equilibrium configuration, to an external driving source which keeps doing work on it throughout the entire time period under study. More specifically, we will be interested in examining the behaviour when the initial state is chosen to be a thermal density matrix, so that one can simultaneously explore the response of a quantum dissipative system. For non-linear dynamical systems the response under such external driving can provide insight into the dynamics via the coherent build-up of the response. 

 Classical analogs of what we have in mind are situations where we drive a (damped) pendulum steadily or subject a viscous fluid to external forcing. The latter is particularly apposite, for the problem we study can be thought of as a hot deconfined plasma of a planar gauge theory disturbed by an external source, as studied in the hydrodynamic context in 
 \cite{Bhattacharyya:2008ji}. 
 Rather than letting the driving grow without bound, we will subject our plasma to a periodic driving by turning on the source for a relevant operator. 
 One therefore has two relevant dimensionful parameters characterizing the situation: 
(a) The amplitude of the external force whose scaling dimension is set by the conformal weight of the operator we exploit and (b) The frequency of the external driving. The third scale which is the temperature of the initial equilibrium state can be factored out, if we are interested in describing the dynamics for a conformally invariant system, which is most natural in the gauge/gravity context. This scenario was explored in \cite{Auzzi:2013pca}, who carried out a perturbative analysis  for small amplitudes of the driving source. A related analysis of periodically driving a quantum system near a critical point was undertaken in \cite{Li:2013fhw}.

\begin{figure}
\centering
\includegraphics[width=0.85\textwidth]{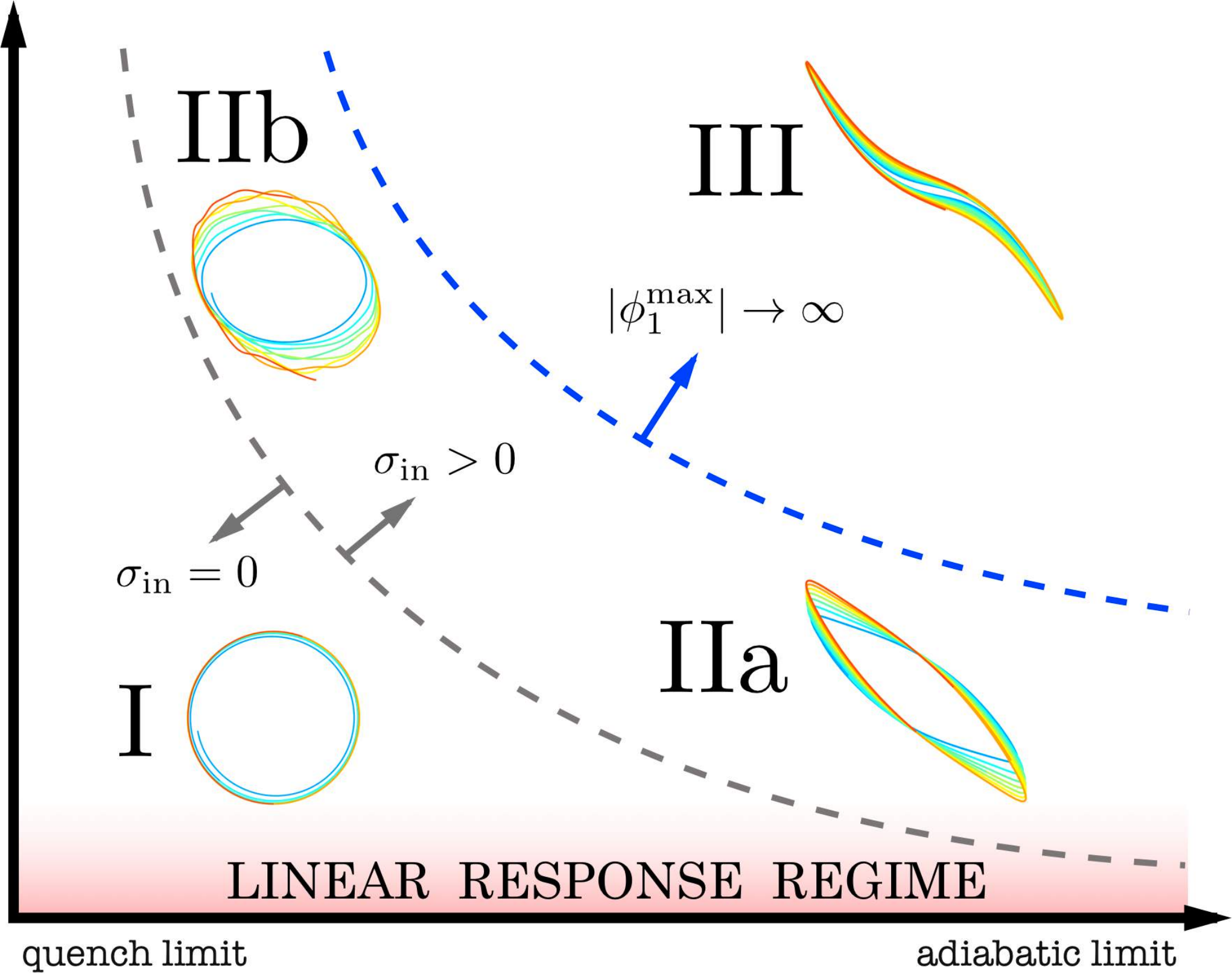}
\setlength{\unitlength}{1cm}
\begin{picture}(0.3,0.4)(0,0)
\put(0.15,0.56){\makebox(0,0){ $\per\, T_0$}}
\put(-13.04,10.5){\makebox(0,0){ $\amp/T_0$}}
 \end{picture}
\caption{The ``phase diagram'' of the driven holographic plasma characterized by the period ($\per$) and amplitude ($\amp$) of the driving force, measured in units of the initial thermal scale $T_0$. There are four distinct regimes marked on the diagram which are explained in the main text. $\sigma_\text{in}$ refers here to the imaginary (or in-phase) part of the conductivity defined in Eq.~\eqref{eq:sigdef}. As we move from southwest to northeast in the figure, the system is driven into a more non-linear regime; the crossing of the grey-dashed boundary is the turn on of the in-phase part of the conductivity $\sigma_{\text{in}}$ in regime II, and the crossing of the blue-dashed boundary signifies the entrance into the resonance phase of regime III i.e., $|\phi^{\text{max}}_{1}| \rightarrow \infty$.
The character of the different regimes is further illustrated by displaying the phase portrait of the scalar operator (expectation value against source) used to drive the plasma.}
\label{fig:PP_qualitative}
\end{figure}
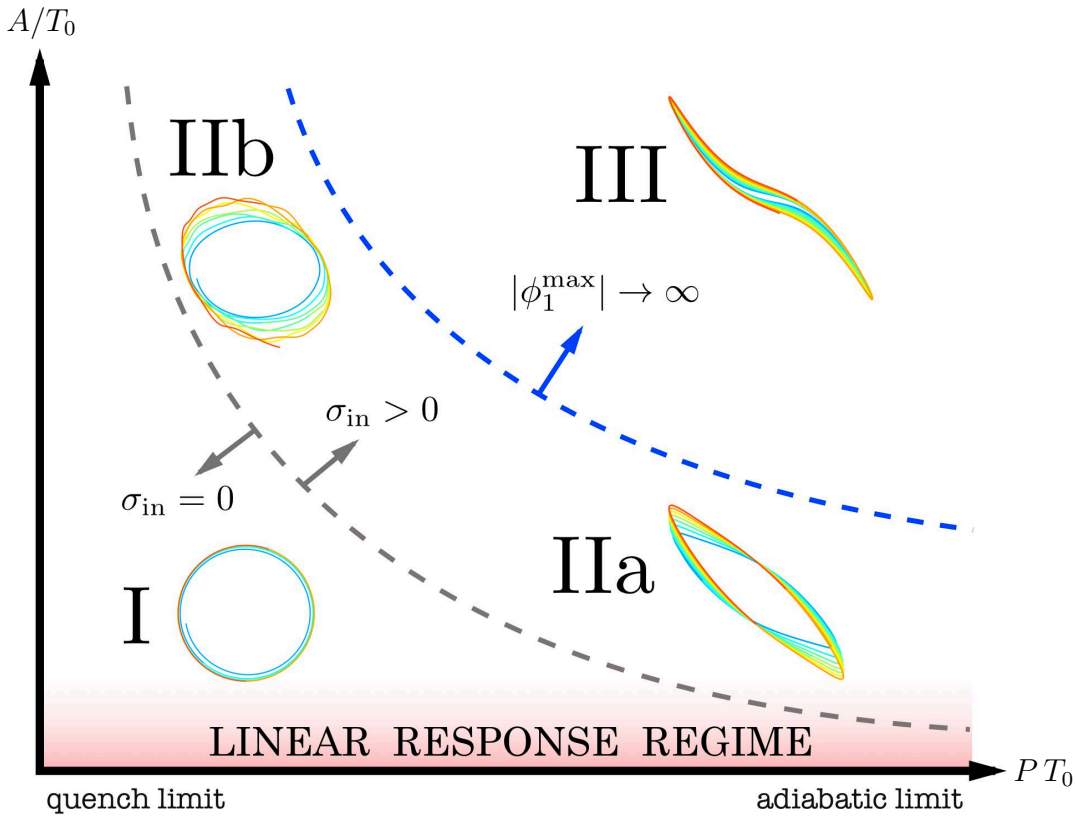

Gravitationally the problem we study is the following: we have a Schwarzschild-\AdS{4} black hole modeling our initial thermal density matrix of a three-dimensional  CFT. At some instant of time on the boundary we turn on a periodic source for a relevant scalar operator, which we specifically choose to be of dimension $2$ for simplicity.\footnote{ This choice turns out to have several advantages as the dual scalar being conformally coupled to gravity in the bulk allows a certain level of technical simplification in various holographic renormalizations we need to do to extract physical data.} The physics of the system is captured by examining the behaviour or various observables as we vary the amplitude $\amp$ and the period $\per$ of the driving (measured e.g. in units of the initial temperature).  We will in particular extend the perturbative analysis of \cite{Auzzi:2013pca} valid for  $\amp \ll 1$ to the non-perturbative regime  
$\amp \gg 1$ for a wide range of driving frequencies. We find that the system naturally exhibits at least four different phases which are depicted in phase diagram Fig.~\ref{fig:PP_qualitative}; two of these (labeled IIb and III) are non-perturbative in $\amp$.

Before we describe the different phases, let us examine for a moment the physics of the gravitational system qualitatively. Initially we have a planar black hole in \AdS{4}. When we turn on the scalar source, we are injecting energy into the bulk. This energy does work on the system and simultaneously heats it up. The latter is seen by the fact that some of the energy falls behind the horizon, which grows\footnote{ As we will be describing the dynamics of Einstein-scalar system with the scalar field satisfying the null energy condition, the areas of the event and apparent horizon (in the canonical foliation) have to grow monotonically -- a consequence of the area theorem \cite{Hayward:1993wb} (see \cite{Booth:2005qc} for an excellent overview). We will elaborate on this point in \S\ref{sec:bulksol}.} -- this is the gravitational response to the disturbance of the plasma. However, in this process we also induce an expectation value for the operator whose source we tweak. When we disturb the system `slowly enough', the operative parameter measuring this being the product of the amplitude and the period, the system has time to catch-up. This is the dissipation dominated regime indicated by I in Fig.~\ref{fig:PP_qualitative}. In this regime the injected energy  falls behind the horizon with little fanfare.

As we ramp up the disturbance, the plasma is driven more and more non-linear, with a dynamical cross-over visible as we move into phases IIa or IIb of Fig.~\ref{fig:PP_qualitative}.
Note that the entire non-linear dynamics in the system is induced by the non-linearities of gravity, for we model the system simply by a free (massive) scalar field. In this phase the response gets more in-phase with the source. It is amusing to contrast this with non-linear scalar dynamics; we find that in this phase we can model the scalar 1PI effective action induced from the gravitational interactions to be well mimicked by a polynomial potential (see  \cite{Basu:2013vva} for  previous studies of self-interacting scalars in AdS).  In this regime there is less dissipation; the entropy production by the growth of the horizon area is slowed down relative to region I. The primary distinction between the two phases IIa and IIb themselves is the lag in the response seen as the period is increased (hence the tilt in the phase portrait).  

For even larger disturbances, we enter region III, where the system response gets highly resonant and there is a steep growth in the response. As one might suspect this is the domain where the gravitational non-linearities are strongest and indeed one can check that such behaviour is not visible for a polynomially (self-) interacting scalar.  In the course of our investigation we explore not just the phase portrait, but various other physical quantities of interest, such as the growth of entropy and dissipation in the system, the  rate at which entanglement is produced, etc.. For instance, region IIb is characterized by enormous fluctuations in the energy of the system over a single period and continuous but non-differentiable behaviour in the entanglement entropy of a sub-system.

Let us contrast our results with the analysis in the perturbative regime of small amplitudes undertaken in  \cite{Auzzi:2013pca}.\footnote{ We note that \cite{Li:2013fhw} study the influence of a periodic electric fields on the phase transition between a normal and superconducting phase using holography. It is clear in this case that a  driving the system will make it exit the low temperature superconducting phase as the energy expended heats up the system past the critical point, as their analysis confirms.}
As one can see from phase diagram Fig.~\ref{fig:PP_qualitative} for small amplitudes, $A \ll 1$, one is largely in the dissipation dominated linear response regime. This is indeed consistent with the analysis of \cite{Auzzi:2013pca}, who explore the dependence of observables on both the period of the driving as well as the dimension of the perturbing operator $\Delta$. As for us the latter remains frozen and we are unable to check the detailed scaling relations they find, but in the common domain of overlap we do indeed have agreement. In particular, for perturbing operators of dimension $\Delta =2$ in CFT$_3$ we expect to see that the energy dissipation as a function of the period scales as $E_\text{diss} \sim \per^{-1}$ (for $\amp \ll 1$), independent of the initial temperature. 
Furthermore, we also expect that the work done in each cycle, measured by the entropy produced, to scale with the increased energy density. We  find that in the slow driving regime this scaling closely tracks the prediction from local thermal equilibrium, but starts to grow more steeply as we transit into more interesting non-linear regimes.

While the various response functions provide us with a useful diagnostic of the phase structure of the dynamical evolution, we also attempt to gain insight into the non-equilibrium dynamics using entanglement entropy for small sub-systems.  This non-local probe exhibits distinct characteristic features in the various regimes: for weak driving, the growth of entanglement is gradual (and appears to track the growth of thermal entropy), while for strong driving there are steep oscillations and glitches in its evolution. We should caution the reader that we have only examined entanglement entropy for relatively small sub-systems, owing to technical complications with numerical stability.  Nevertheless these results suggest a rather rich structure in the temporal growth of entanglement with driving, which deserves further detailed exploration \cite{Rangamani:2015ys}. 

The outline of this paper is as follows. We begin in \S\ref{sec:setup} by giving a quick overview of the basic set-up and the numerical solutions. Following this in \S\ref{sec:obs}, we set out the various observables we use to explore the behaviour of the system. In particular, we justify the rationale behind phase diagram Fig.~\ref{fig:PP_qualitative} and how we should physically think of the different regimes. \S\ref{sec:ee} is devoted to the study of entanglement entropy in this system where we focus on the region of an infinite strip and exploit the underlying homogeneity of the set-up. We conclude with a discussion in \S\ref{sec:discuss}. Some technical results about holographic renormalization required for computing various observables is collected in the Appendices; Appendix \ref{sec:holren} collects some useful information about holographic renormalization in our models while Appendix \ref{sec:eeapp} provides details relevant for computing entanglement entropy.

\section{Driven CFTs and their Holographic Duals }
\label{sec:setup}

We first take the opportunity to set up the basic problem of a field theory driven out of equilibrium by turning on a source for a relevant operator. We then go on to describe how to model this in the holographic set-up and present the basic methodology and results from the numerical simulations.

\subsection{Driving CFTs by Relevant Operators}
\label{sec:cftdriving}

We are interested in the dynamics of strongly coupled plasmas that are driven by an external source. The initial plasma is in equilibrium in some homogeneous thermal state at a temperature $T_0 $ for $t <0$. At $t=0$ we introduce external sources with some specified spatial-temporal profile that we control. We focus exclusively on situations where
the external sources are spatially homogeneous, but otherwise arbitrary and tunable at will. 

 To wit, the system under consideration can be modeled by an equilibrium density matrix, evolved under a time-dependent Hamiltonian, i.e., we take
\begin{equation}
S_{CFT} = S_{{\cal J}=0} + \int d^d x \sqrt{-\gamma}\;  {\cal J}(x)\, {\cal O}(x) 
\label{}
\end{equation}	
where ${\cal O}(x)$ is a single trace (gauge-invariant) relevant operator of conformal dimension 
$\Delta<d$. The source ${\cal J}(x)$ is chosen to have no spatial dependence and be temporally periodic and thus can be represented as
\begin{equation}
{\cal J}(x) = \amp \, \cos(\omega t)\, \Theta(t) \,.
\label{}
\end{equation}	
Here $\Theta(t)$ is the Heaviside step function for turning on the periodic perturbation of amplitude $\amp$ and driving frequency $\omega = 2 \pi / \per$ at $t=0$; later in actual (numerical) implementations we will choose a suitable ramp factor to smoothly turn the perturbation on. 

In the presence of the source, the Ward identities following from diffeomorphism and Weyl invariance get modified. A simple analysis shows that the boundary conservation equation now has an explicit source term
\begin{equation}
\nabla_\mu T^{\mu}_{\ \alpha} = {\cal O}\, \nabla_\alpha {\cal J} \,.
\label{eq:cward}
\end{equation}	
indicative of the work done by the driving source on the CFT. Likewise the one-point function of the trace of the energy-momentum tensor no longer vanishes but satisfies
\begin{equation}
T^\mu_\mu = \left(d-\Delta \right)\, {\cal J}(x) {\cal O}(x)
\label{eq:tward}
\end{equation}	
Since the boundary theory is conformal, it does not have any intrinsic time scale. The time scales in the problem come from only the driving force, namely its amplitude and period. 
The situation of interest is thus characterized by three scales:
\begin{itemize}
\item $T_0$: the initial thermal scale for the homogeneous plasma.
\item $\amp$: the amplitude of the source whose scaling dimension is $d-\Delta$.
\item $\omega$: the driving frequency or the time-scale set by the period 
$\per = 2\pi/\omega$.
\end{itemize}

\subsection{Holographic Driving}
\label{sec:gdual}

The gravity dual to this set-up is modeled by the dynamics of a scalar field  $\phi$ with mass  $m_\phi^2=-2$, dual to a relevant perturbation of the boundary theory.
\begin{equation}
S_{\text{bulk}} = \frac{1}{16\pi G_N}\; \int d^{d+1} x\; \sqrt{-g} \, \left( R  + d(d-1) - \frac{\alpha_g}{2} \left[ \, (\partial \phi)^2 + m^2 \phi^2 \right] \right)
\label{eq:bulkS}
\end{equation}	
In our holographic implementation of this set-up we will work in $d=3$ and consider a scalar operator with conformal dimension $\Delta =2$. While this is rather specific, we will explore the phase structure of the driven system as a function of the ratio of scales outlined above. The qualitative features we believe are independent of these actual choices.\footnote{ We have also set $\ell_\text{AdS} =1$ for simplicity.}  We have included a dimensionless gravity-scalar coupling $\alpha_g$ which we can use to tune the amount of backreaction on the geometry; for the most part we will focus on $\alpha_g = 0$ or $\alpha_g =1$, to model probe and interacting scalar fields  respectively. 

We want to study gravitational dynamics driven by a scalar field whose non-normalizable mode is turned out as dictated by the source ${\cal J}(x)$, i.e., take  $\phi_0(t) = \amp\,\cos(\omega t)$
 and study the behaviour of the theory with varying amplitude 
 $\amp$ and frequency $\omega$.  The gravitational background is an asymptotically 
 \AdS{4} spacetime, which we write in ingoing Eddington-Finkelstein coordinates (sometimes called the Bondi-Sachs form) as:
\begin{equation}
ds^2 =-2 \, \gtt (t,r)\, e^{2 \br (t,r)}\,dt^2+2 \,e^{2 \br (t,r)}\,dt \,dr + \gxx (t,r)^2\, (dx^2+dy^2)
\label{eq:bulkcy}
\end{equation}
The coordinate dependences of the metric functions $\gtt$, $\br$, $\gxx$ are explicitly indicated with homogeneity ensuring that $\partial_x$ and $\partial_y$ are Killing vector fields.

Our initial state is a planar Schwarzschild-\AdS{4} black hole with  temperature 
$T_0=3/\pi$, corresponding to horizon size $r_+ =1$. This bulk solution is given by $\gtt = r^2(1-\frac{1}{r^3})$, $\br = 0$, $\gxx = r$ with metric
\begin{equation}
ds^2_{t \le 0} = - r^2 \left(1-\frac{1}{r^3}\right)\, dt^2 +2\, dt\, dr + r^2\, \left(dx^2 + dy^2\right).
\label{}
\end{equation}	
For our choices of $m_\phi^2 = -2$ in $d=3$, the amplitude $\amp$ has mass dimension $1$.
Thus we have two interesting time scales associated with the external driving force: the period $\per$ and the inverse amplitude $\amp^{-1}$. To capture universal physics, we look at relatively late times of the non-thermalized system compared to both of these scales. 
Note also that in those late times the initial value of the temperature, $T_0$, becomes irrelevant.

There has been much interest recently in holographic {\it quenches}, in which the system is initially driven to an excited state, and then is allowed to return to equilibrium, a process which exhibits some degree of universality. In contrast, we are interested in the dynamics of the steady state system while it is being driven. Hence, in our solutions we do not turn off the driving force at late times, and seek universal features associated with the driven steady state system. We will see that such dynamical features exist, and they strongly depend on the parameter
\begin{equation}
\xi (\per,\amp) \equiv  \per\,\amp \,,
\label{eq:xidef}
\end{equation}	
the unique dimensionless parameter formed from the two time scales associated with the driving force.  Below we refer to the regime $\xi \ll 1$ as the weak driving regime, and $\xi \gg 1$ as the strong driving regime (which is further divided into two separate dynamical regimes). We also measure time in units of the period $\per$, thus we vary and discuss the dependence of observables on the two dimensionless parameters: the strength of the drive and time.

\subsection{Bulk solutions}
 \label{sec:bulksol}
We solve the equations of motion resulting from the scalar-gravity Lagrangian \eqref{eq:bulkS}
by direct numerical integration. The boundary conditions on the scalar are prescribed by the source and the metric is required to be asymptotically \AdS{4}. The AdS boundary is attained as $r\to \infty$ and the asymptotic behaviour of the fields is 
\begin{align}
\phi(t,r) &= \frac{\phi_0(t)}{r} + \frac{\phi_1(t)}{r^2} + {\cal O}(r^{-3})
\nonumber \\
\gxx (t,r) &= r + \lambda(t) - \frac{\alpha_g}{4}\, \frac{\phi_0(t)^2}{r} + {\cal O}(r^{-2})
\nonumber \\
\gtt (t,r) &= \frac{1}{2} ( r + \lambda(t))^2 - \lambda'(t) - \frac{\alpha_g}{4}\, \phi_0(t)^2
+ {\cal O}(r^{-1})
\nonumber \\
\chi(t,r) &= {\cal O}(r^{-4})
\label{}
\end{align}
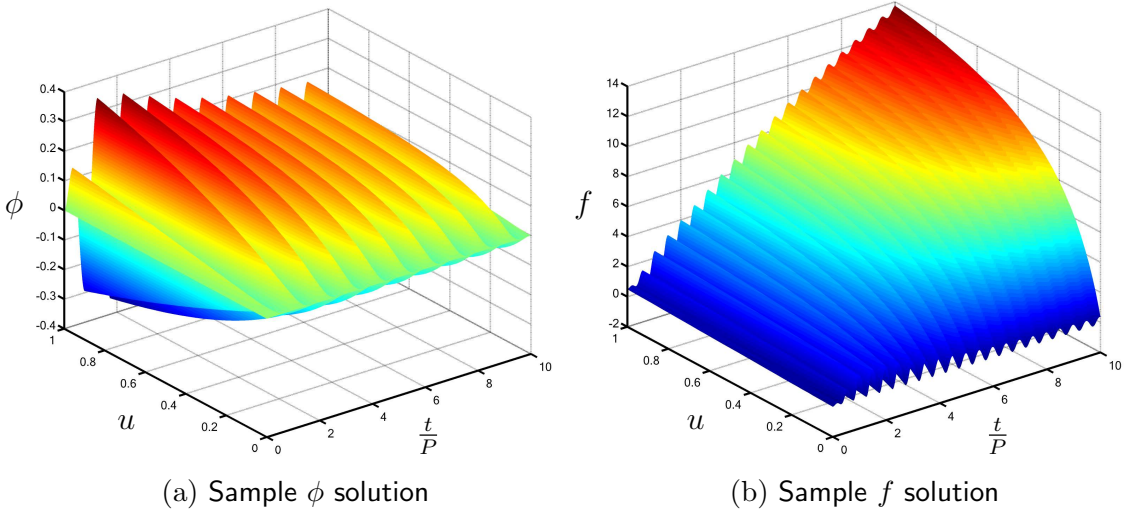
\begin{figure}
\centering
        \begin{subfigure}[b]{0.45\textwidth}
                \includegraphics[width=\textwidth]{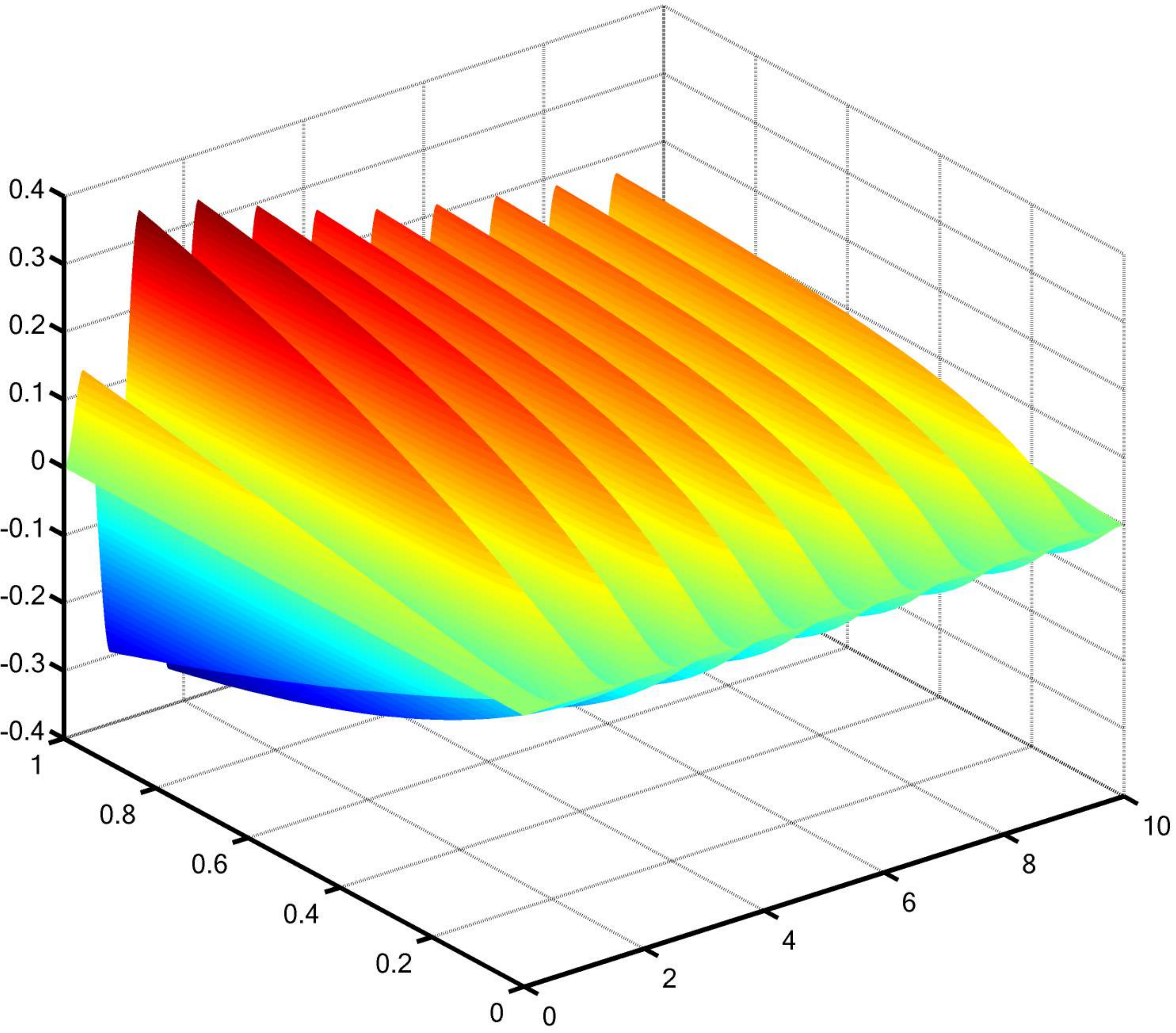}
                \put(-47,8){\makebox(0,0){\normalsize $\frac{t}{\per}$}}
                \put(-160,13){\makebox(0,0){\normalsize $u$}}
                \put(-202,95){\makebox(0,0){\normalsize $\phi$}}
        \caption{Sample $\phi$ solution}
        \label{subfig:sample_phi}
        \end{subfigure}
        ~ ~
        \begin{subfigure}[b]{0.45\textwidth}
        \includegraphics[width=\textwidth]{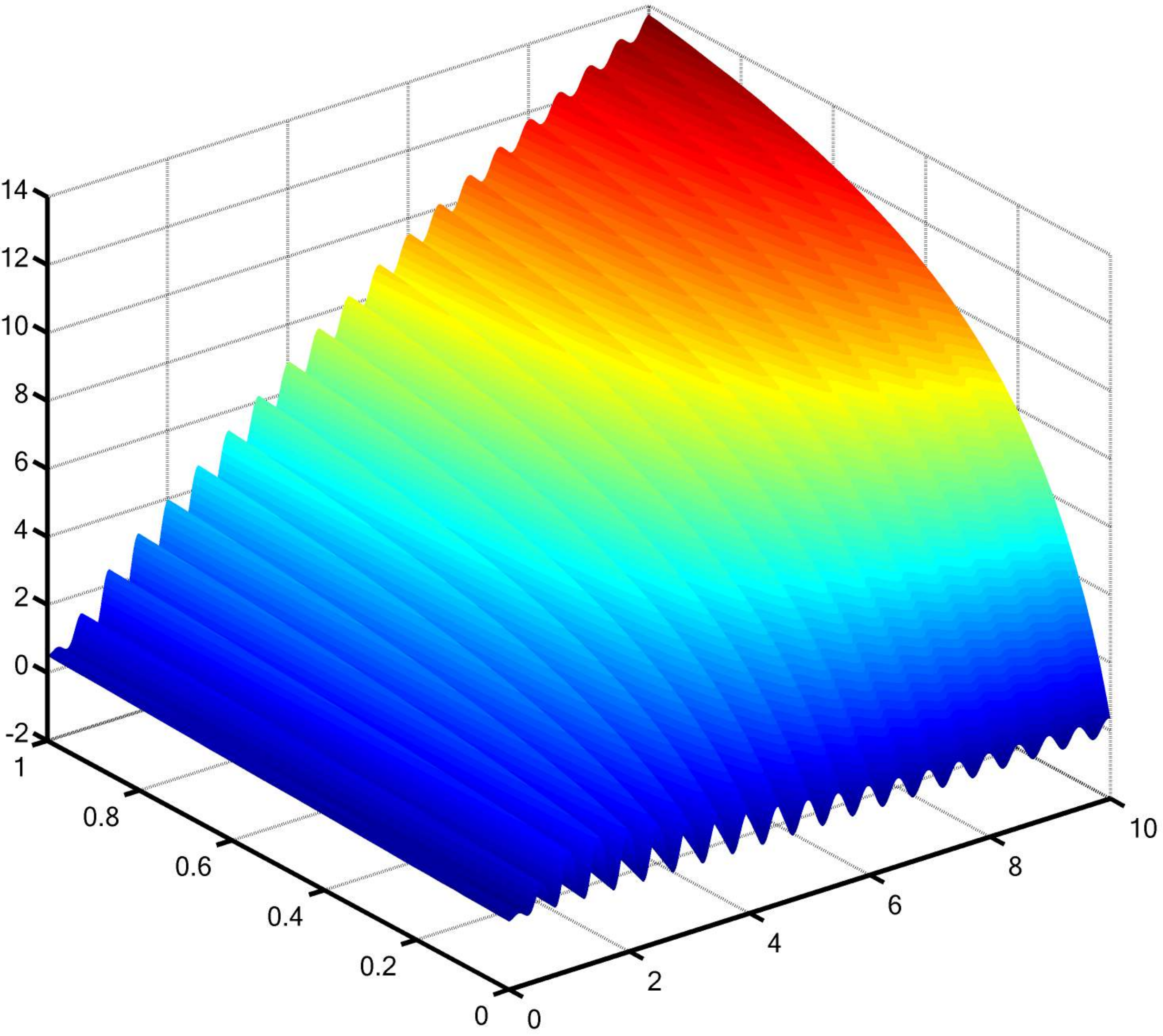}
        \put(-47,8){\makebox(0,0){\normalsize $\frac{t}{\per}$}}
        \put(-160,13){\makebox(0,0){\normalsize $u$}}
        \put(-202,95){\makebox(0,0){\normalsize $\gtt$}}
        \caption{Sample $\gtt$ solution}
        \label{subfig:sample_gtt}
        \end{subfigure}
\caption{A sample solution displaying the scalar field $\phi(t,u)$ and the temporal component of the metric function $\gtt (t,u)$ for $\xi(\per=1, \amp=1)=1$. Time is measured in units of $\per$ and the radial component is compactified as $u = 1/r$. }
\label{fig:sample_solution}
\end{figure}
More specifically, we use the characteristic formulation of the resulting partial differential equations as explained in detail in \cite{Chesler:2013lia} to numerically integrate for the solution. The advantage of the method is that it allows us to use constrained evolution: at each time step we solve a nested set of ODEs to determine the time derivatives of all dynamical quantities, and then we use one of the standard time evolution schemes to march forward in time. While we follow the general logic of \cite{Chesler:2013lia}, in our implementation we found that some of elements described in \cite{Balasubramanian:2013yqa} enabled for a more robust evolution.
 
 To solve the radial ODEs we discretize the equations using a Chebyshev basis in the radial direction, typically taking a grid of 60 points. For time evolution we use an explicit Runge-Kutta method of order 4, with an adaptive step size. We filter at each time step by throwing out the top third of the Fourier modes for each dynamical variable to avoid artificial and unphysical growth in amplitudes of short wavelength modes associated with the UV cutoff.

In the regime of strong driving, we found it necessary to turn on the perturbation gradually from zero. Therefore we include a ramp-up time of $2\,\per$, after which the amplitude reaches its intended value. Thus, the first few periods of each solution show behaviour sensitive to details of the ramp-up protocol. We look at observables only after this ramp-up time of $2\,\per$.

In Fig.~\ref{fig:sample_solution} we show one example of evolved bulk fields for a specific solution. As we perturb the system by a relevant operator, the scalar field grows towards the horizon. All fields are (at least approximately) modulated with the period of the source.

At this point it is worthwhile mentioning one important consistency check on the numerical scheme, which relies on the existence of a smooth horizon in the spacetime.  Given a metric and a Cauchy slice in the bulk spacetime, one can find the outermost trapped surface on this slice. If we have a set of Cauchy slices that foliate the spacetime, then the future outermost trapping horizon, which we simply refer to as the apparent horizon by a common abuse of terminology, is typically defined by taking the union of the outermost trapped surface on all the slices.  The apparent horizon thus defined is subject to an area law  which was originally discussed by \cite{Hayward:1993wb} -- we refer the reader to  \cite{Booth:2005qc} for a concise modern summary and proof of the statement. It is however important to note that the statement relies on the existence of a sensible foliation of the spacetime by Cauchy slices. Indeed, it is possible as discussed in \cite{Wald:1991zz} to find exotic symmetry breaking foliations (which are however incomplete) in which even the Schwarzschild black hole solutions fails to have a trapped surface.  

We mention this in passing, as \cite{Auzzi:2013pca}  quotes the result  of \cite{Wald:1991zz} to argue that apparent horizon areas need not be monotone generically. They however do not encounter such behaviour, for with the choice of 
ingoing coordinates in \eqref{eq:bulkcy}, there is a canonical choice of bulk Cauchy slices respecting the homogeneity of the disturbance. In this foliation the result quoted in \cite{Booth:2005qc} does apply and in fact simply follows from properties of null congruences using Raychaudhuri's equation.\footnote{ To be sure the statement of the area increase theorem does rely on the 
null energy condition, which we happily assume, for it is always satisfied by scalar fields with sensible kinetic terms.} Our results are indeed consistent with this expectation and we have checked that the area of the apparent horizon does grow monotonically in $t$ (which labels the leaves of the foliation chosen), as we shall extensively see in the sequel.  While initial results of \cite{Buchel:2014gta} appeared to suggest otherwise,  upon closer scrutiny, one finds that in numerical analyses so far the area of the apparent horizon does respect the second law as derived by \cite{Hayward:1993wb}.\footnote{ We thank Alex Buchel for checking this and confirming the monotone growth of the apparent horizon area.}

 \section{Driving Diagnostics}
\label{sec:obs}
 Having constructed the holographic duals we now turn to lessons that can be extracted from the geometry for the dynamics of strongly coupled field theories. A-priori there are a number of observables which are useful probes of the out-of-equilibrium situation and we will focus on those that offer most clear insight into the dynamics. Our primary goal is to quantify the behaviour of the system as a function of $\{\per,\amp\}$ and construct a phase diagram demarcating the various regimes in this phase space. Let us quickly enumerate the observables we will use and proceed to explain why they give us some insight into the dynamics:
\begin{itemize}
\item The phase portrait of response $\phi_1(t)$ as a function of the source $\phi_0(t)$. Alternatively, this relation can be codified in a conductivity $\sigma(t)$, as defined below in \eqref{eq:sigdef}.
We find 4 underlying phases regions that the system can fall into.
\item The $\phi_1$-$\phi_0$ phase portrait features for polynomial and non-polynomial potentials with the gravity-scalar coupling $\alpha_g$ switched on and off.
\item The cycle-averaged thermodynamics quantified by the energy density $\epsilon_\text{avg}(t)$ and entropy density $s_{\text{avg}}(t)$, and the scaling relation $s_{\text{avg}} \sim \epsilon^{\gamma}_\text{avg}$ between them.
\item The work done in each cycle, measured as the difference in average energy between two successive cycles, $\epsilon_\text{cycle} = \epsilon_\text{avg}^{(n+1)} - \epsilon_\text{avg}^{(n)}$.  We typically take $n$ to correspond to the penultimate cycle of our simulation.
\item Fluctuations $\epsilon_\text{fluc}(t)$ in the energy density around $\epsilon_\text{avg}(t)$ and the
 maximal response $|\phi^{\text{max}}_1(t)|$.
\item Entanglement entropy and extremal surface evolution for fixed spatial strips ${\cal A}$ on the boundary.
\end{itemize}
When the system is driven by an external source, the most basic quantity is the response, which is characterized by the scalar one-point function in the presence of the source. In linear response theory, this can be obtained from the retarded Green's function of the operator ${\cal O}(x)$ evaluated in equilibrium. We are not just interested in the linear response regime, which would correspond in our set-up to $\amp \ll T_0$, but in the full non-linear response. To visualize the response of the strongly coupled plasma, especially in the non-linear regime, where its phase relative to the source is important, we will find it instructive to exhibit the phase portrait, the trajectory traced by the system in the $\phi_0$-$\phi_1$ plane. We also codify the relation between scalar source and response by a complex conductivity, defined below.

In addition to the one-point function of the operator deforming the CFT, we are interested in the boundary energy-momentum tensor. This can be decomposed in to an energy density $\epsilon(t)$ and a pressure. In the holographic set-up one has 
\begin{equation}
\vev{{\cal O}(t)} = \phi_1(t)\,, \quad \epsilon(t) = \vev{T^t_{\ t} (t)} \,, \quad p(t) = \vev{T^i_{\ i} (t)}
\label{}
\end{equation}	
The scale Ward identity \eqref{eq:tward} implies that pressure is not an independent observable since it can be obtained from knowledge of $\epsilon(t)$ and $\phi_1(t)$, so we will not discuss the pressure separately. Additionally, to probe the local thermodynamics we will monitor the local entropy density $s(t)$, obtained by computing the area of the apparent horizon at time $t$.\footnote{ Using the area of the apparent horizon (defined as the outermost trapped surface in the foliation respecting spatial homogeneity) results a causal boundary observable. One maps points on the apparent horizon to boundary points by Lie transport along radially ingoing null geodesics, which in the ansatz \eqref{eq:bulkcy} are simply lines of constant $\{t,x,y\}$. On the other hand the teleological nature of the event horizon implies that its area would not provide a good measure for the boundary entropy density, cf., 
\cite{Chesler:2008hg, Figueras:2009iu} for a discussion of this point.}

The dynamics of the bulk gravitational fields encode the heat production resulting from supplying external energy to the system. We monitor the explicit time dependence of the energy density $\epsilon(t)$ and the entropy density $s(t)$ along with their values averaged over each driving cycle period $\per$, and find for the most part that the averaged values are increasing with time.\footnote{ Note that the averaging makes $\epsilon_\text{avg}(t)$ and $s_\text{avg}(t)$ discrete in time.} These provide a useful diagnostic of the departure from equilibrium, as one can monitor the 
 scaling relation to infer the local thermodynamic equation of state. We define the thermodynamic scaling exponent $\gamma$ when the system is in a steady state $t>t_{\text{s}}$ via
\begin{equation}
s_{\text{avg}} \sim \epsilon_{\text{avg}}^{\gamma}
\label{eq:gammadef}
\end{equation}	
Note that in thermal equilibrium, conformal invariance predicts  $\gamma_0 = \tfrac{2}{3}$.  We will encounter this and other scaling regimes in our driven system when conformal invariance is broken.

Note that one natural set of non-local observables we could use are the multi-point correlation function for gauge invariant local operators, perhaps for ${\cal O}$ itself. However, realistically this computation involves solving the wave-equation for the linearized scalar fluctuations on top of the background we have constructed, together with the imposition of suitable boundary conditions on the future horizon, to obtain sensible time-ordered correlation functions. These boundary conditions are somewhat tricky to implement  (see however \cite{CaronHuot:2011dr,Chesler:2011ds}) -- 
we will therefore postpone a discussion of correlators  to the 
future.\footnote{ We could following standard practice attempt to compute two-point correlation functions using the geodesic approximation \cite{Balasubramanian:1999zv}. However, as discussed in \cite{Louko:2000tp} and more recently in \cite{Headrick:2014cta}, this prescription doesn't generically reproduce correct time-ordered correlation functions (we really want in-in correlation functions in our set-up). As a result we will also refrain from computing geodesics in the numerical background.}

Below we describe the behaviour of the observables mentioned above in three distinct dynamical regimes, and comment on the bulk interpretation of those regimes.  Once we have gained sufficient intuition from this exercise, we will then examine the entanglement entropy for a specified boundary region. 

\subsection{Dissipation Dominated Regime}
\label{sec:dissdom}
The simplest situation occurs in the regime of weak driving $\xi \ll 1$, which is best described as the \emph{dissipation-dominated regime} (phase I). This includes the regime of small amplitudes, studied perturbatively in \cite{Auzzi:2013pca}. In this weak driving regime, the behaviour of all observables is dominated by dissipation, which we now demonstrate by looking at some specific observables.

\begin{figure}
\centering
\includegraphics[width=0.65 \textwidth]{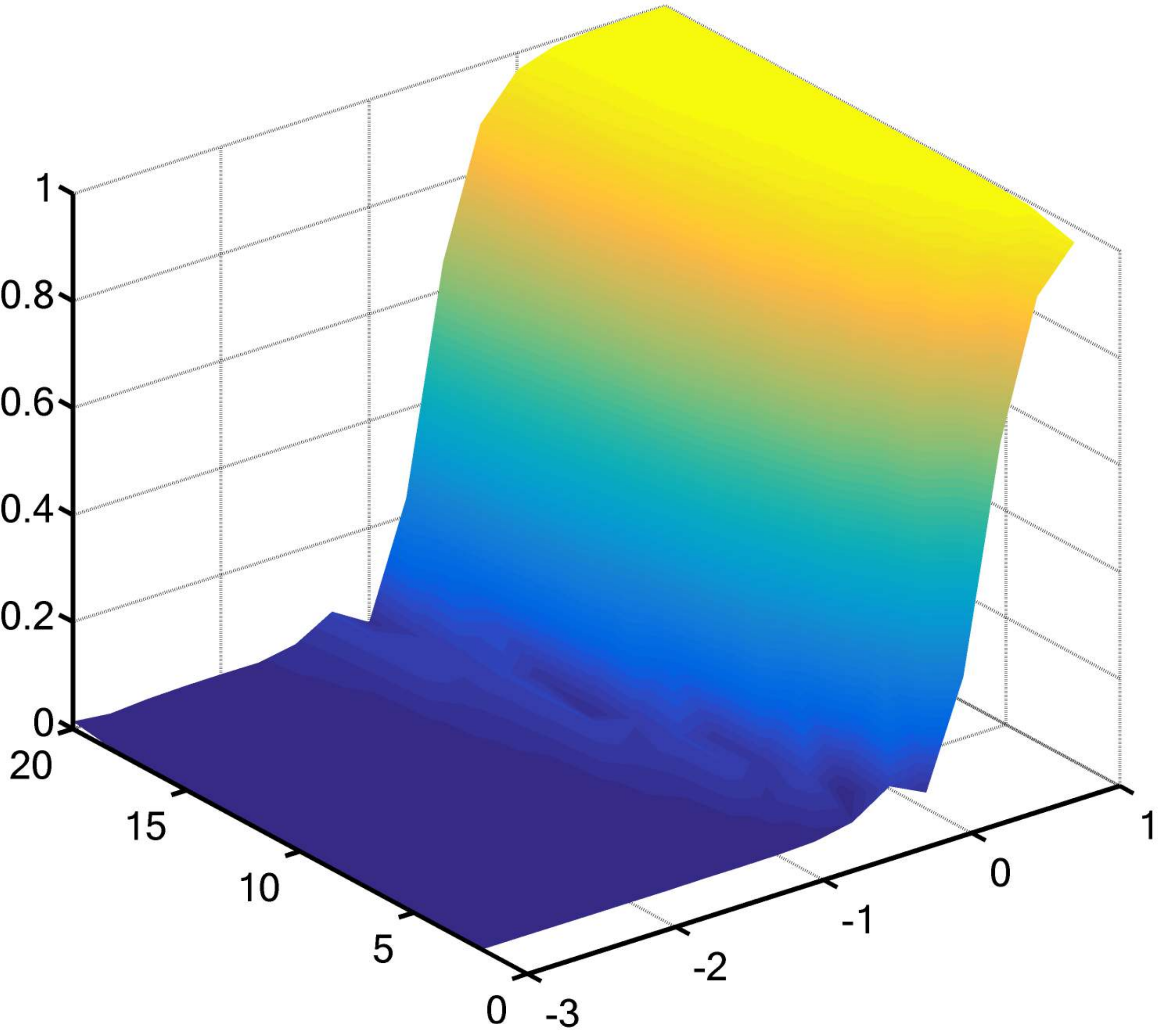}
        \put(-60,13){\makebox(0,0){\normalsize $\log_{10} \per$}}
        \put(-235,23){\makebox(0,0){\normalsize $\amp$}}
        \put(-265,225){\makebox(0,0){\Large $\left| \frac{\sigma_{\text{in}}}{\sigma} \right|$}}
        \vspace{1mm}
\caption{The fraction of the complex conductivity $\sigma_{\text{in}}$ over the entire $(\per,\amp)$ phase diagram where $|\sigma|^2 = \sigma^2_{\text{in}} + \sigma^2_{\text{out}}$.}
\label{fig:conductivity_phasediagram}
\end{figure}

As we drive the system by the scalar non-normalizable mode $\phi_0$ it is instructive to divide the scalar response $\phi_1$ to the part in-phase with the driving force, and the part completely out-of-phase with the perturbation. In analogy with an electromagnetic perturbation in linear response, we can complexify the time dependence of the scalar field\footnote{ That is, regard $\cos \omega t$ and $\sin \omega t$ as the real and imaginary parts of $e^{i \omega t}$.} and define a complex conductivity 
\begin{equation}
\sigma(t) \equiv \frac{1}{i \omega} \frac{\phi_1(t)}{\phi_0(t)} = \sigmaOut(t) + i \, \sigmaIn(t) .
\label{eq:sigdef}
\end{equation}	
With this notation the out-of-phase and in-phase parts of the response correspond to the real and imaginary parts of the complex conductivity, $\sigmaOut(t)$ and $\sigmaIn(t)$, respectively. 
This is the usual convention for the more familiar conductivity, related to electromagnetic perturbations.
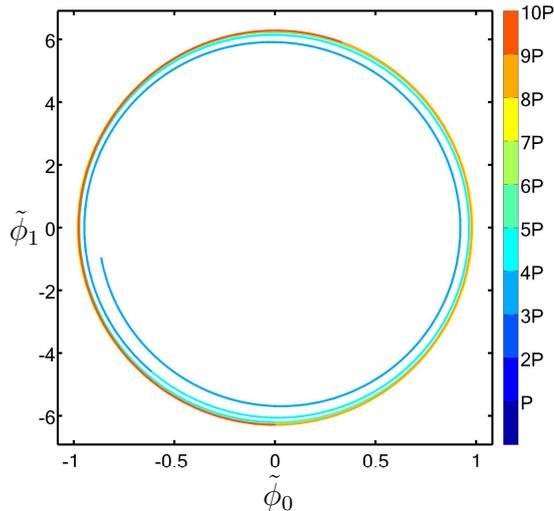
\begin{figure}
\centering
\includegraphics[width=0.45 \textwidth]{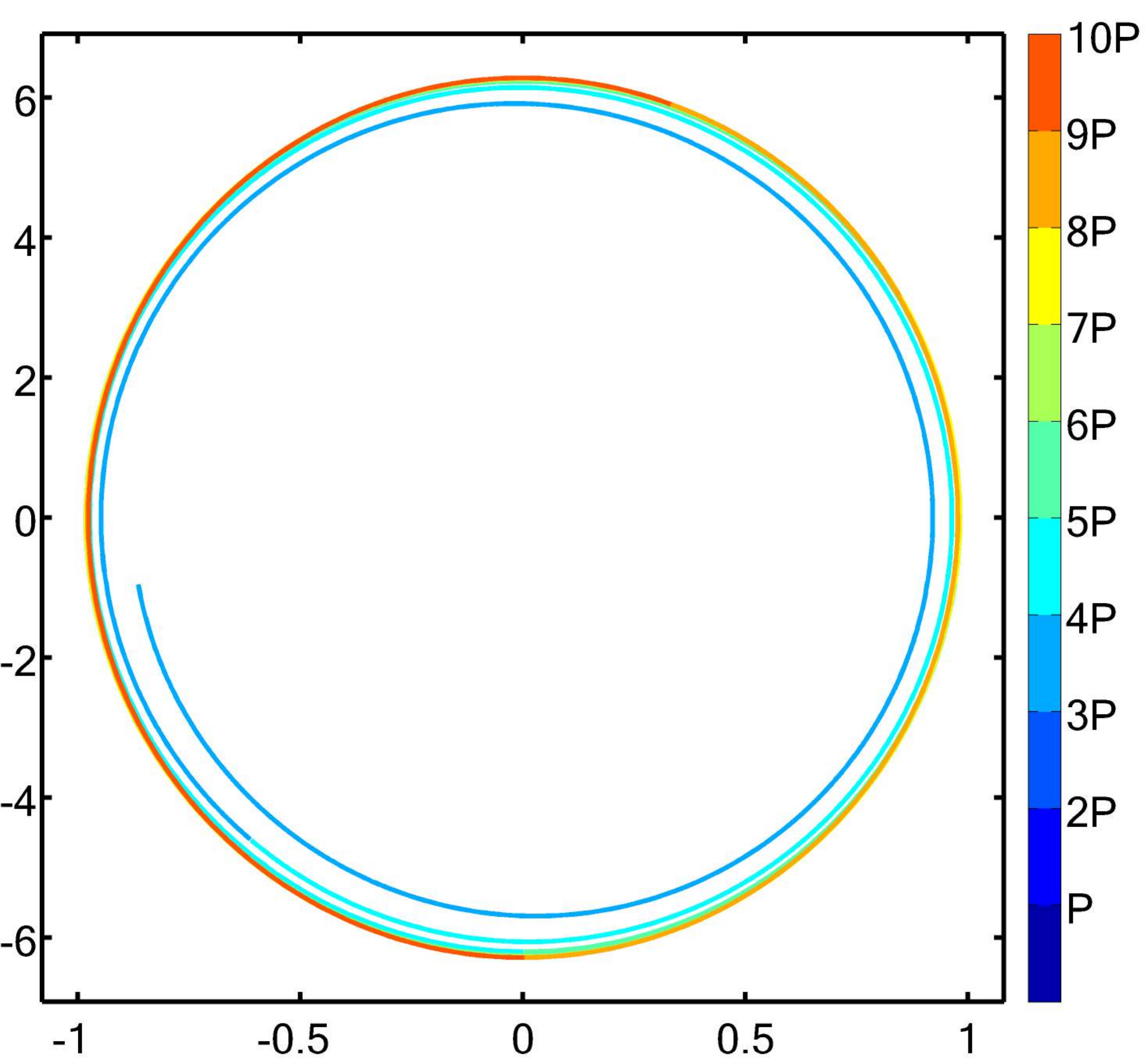}
\begin{picture}(0.3,0.4)(0,0)
\put(-108,-8.0){\makebox(0,0){\normalsize $\tilde{\phi}_0$}}
\put(-203,92){\makebox(0,0){\normalsize $\tilde{\phi}_1$}}
\end{picture}
\vspace{2 mm}
\caption{The phase portrait of the dimensionless response 
$\tilde{\phi}_1 \equiv \tfrac{\per}{\amp} \,\phi_1$ versus the dimensionless source 
$\tilde{\phi}_0 \equiv \tfrac{1}{\amp}\, \phi_0$ 
for $\xi = 0.001 \ll 1$ in the dissipation dominated regime ($\per = 0.001, \amp = 1$)
which we label as phase I.  We evolve the solution for 10 periods with each colour segment representing one period. 
The early times $t < 2\,\per$ show the effect of the perturbation ramp-up, 
and thus are numerical artefacts that we omit from the plot.}
\label{fig:PP_boring}
\end{figure}
As shown in Fig.~\ref{fig:PP_boring} in the low driving regime the scalar response is precisely out of phase with the scalar source, $\sigmaIn=0$, meaning all the energy is dissipated and none of it used to excite the internal energy associated with the scalar field i.e., no work is being done on the system.
This is the quench limit and it matches with what we expect from the behaviour of the perturbation in linear response.
The complex conductivity $\sigma=\sigmaOut$ is purely real and has constant amplitude as a function of time at high frequencies.\footnote{ This is similar to the behaviour of the conductivity for electromagnetic perturbations in asymptotically AdS space.} 
This is manifested in the final steady state being reached almost immediately and consisting of closed untilted trajectories in phase space. 
As we shall see below, tilting of the trajectories in phase space is indicative of non-trivial response and work done onto the system.
Fig.~\ref{fig:conductivity_phasediagram} shows what fraction of the complex conductivity $\sigma_{\text{out}}$ is present on each point on the $(\per,\amp)$ phase diagram, and for what we are concerned with currently, the system has the response being completely out-of-phase with the source when the period is low.

\begin{figure}
\centering
        \begin{subfigure}[b]{0.43\textwidth}
                \includegraphics[width=\textwidth]{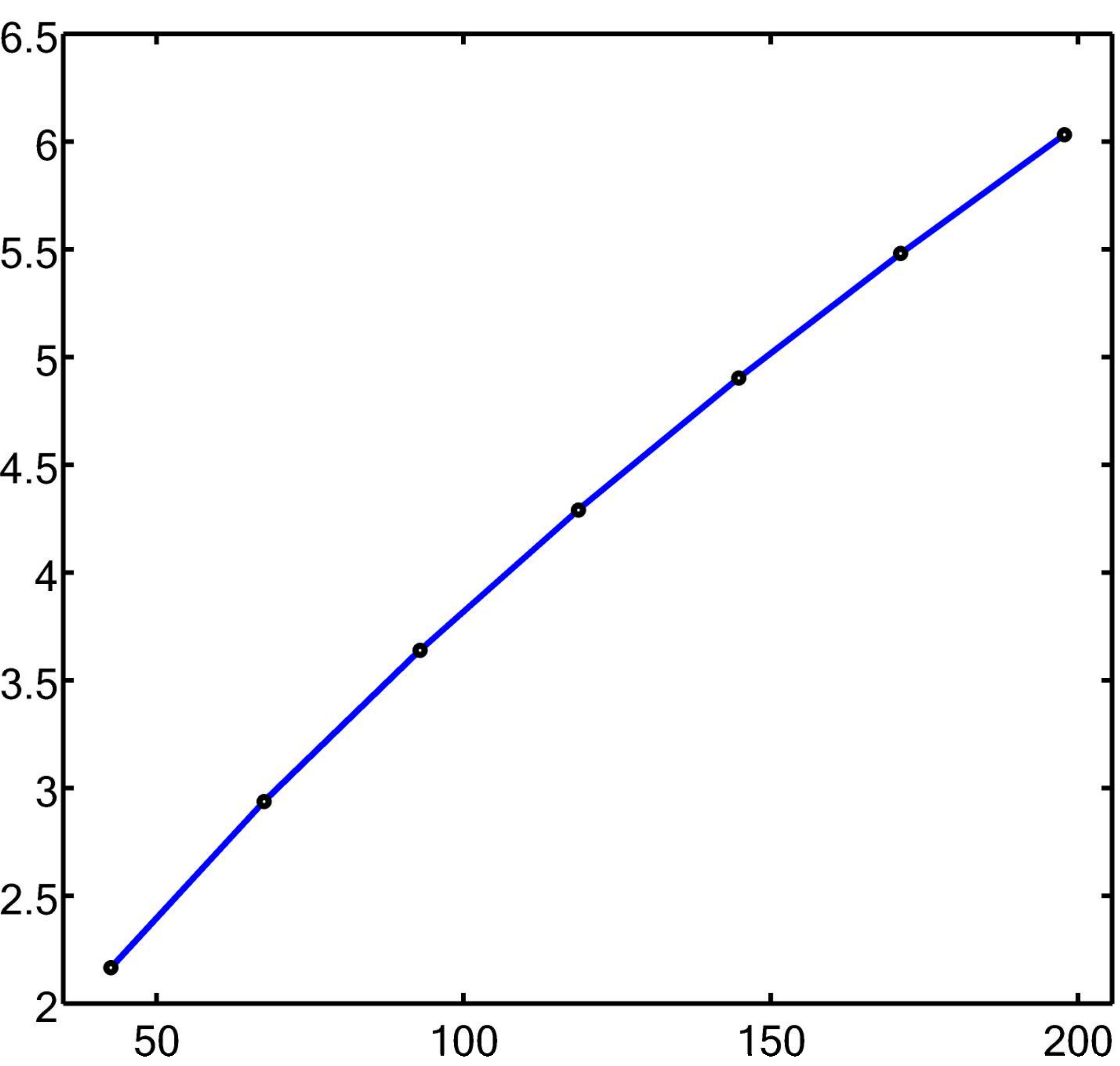}
        \put(-82,-8.0){\makebox(0,0){\normalsize $\tilde{\epsilon}_{\text{avg}}$}}
        \put(-194,90){\makebox(0,0){\normalsize $\tilde{s}_{\text{avg}}$}}
        \vspace{1mm}
        \caption{$s_{\text{avg}}(t)$ versus $\epsilon_{\text{avg}}(t)$.}
        \label{subfig:S_vs_E_plot}
        \end{subfigure}
        ~ ~ 
        \begin{subfigure}[b]{0.465\textwidth}
        \includegraphics[width=\textwidth]{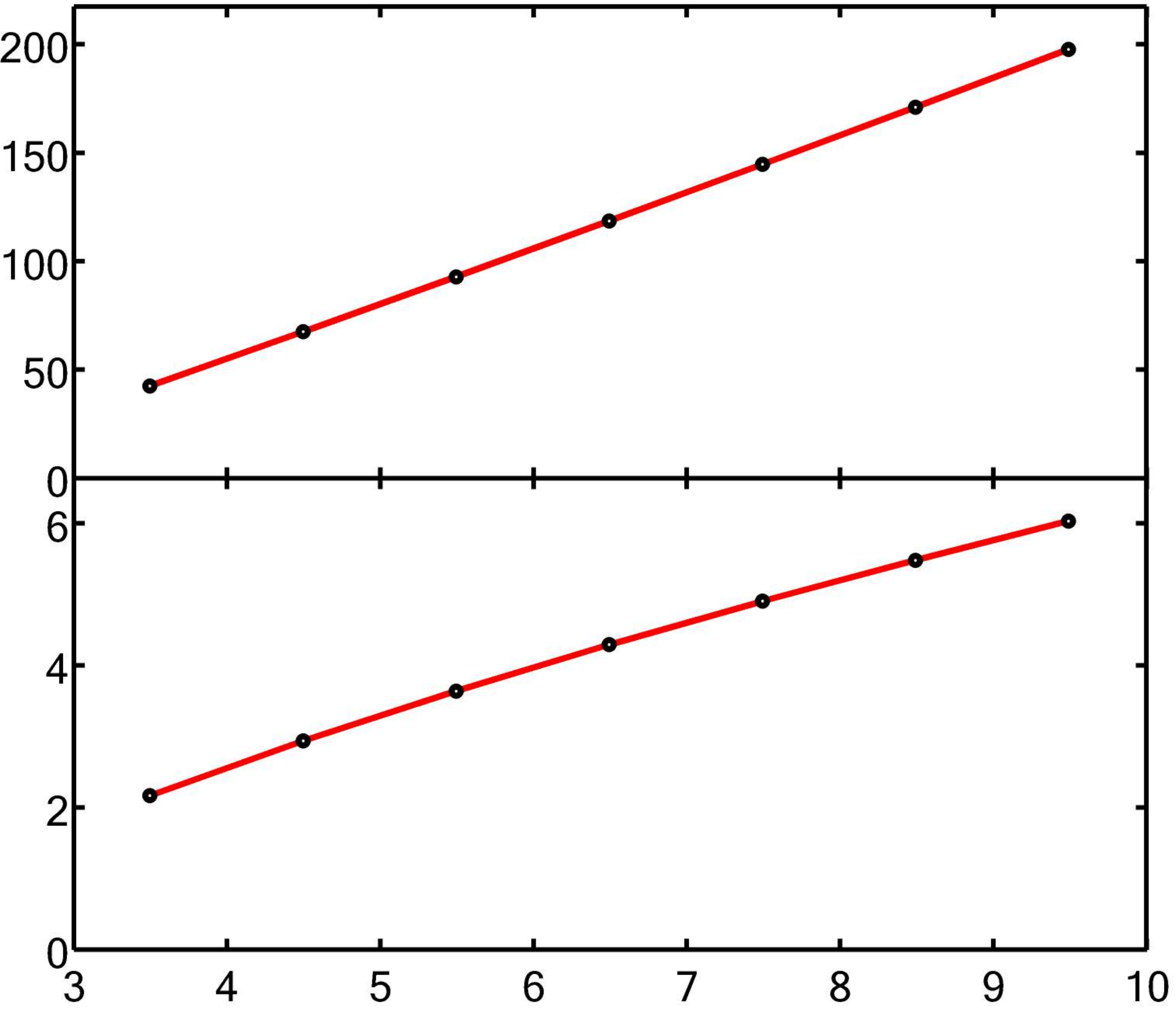}
        \put(-200,122){\makebox(0,0){\normalsize $\tilde{\epsilon}_{\text{avg}}$}}
        \put(-200,49){\makebox(0,0){\normalsize $\tilde{s}_{\text{avg}}$}}
        \put(-90,-8.0){\makebox(0,0){\normalsize $\frac{t}{\per}$}}
        \vspace{1mm}
        \caption{$s_{\text{avg}}(t)$ and $\epsilon_{\text{avg}}(t)$ versus time.}
        \label{subfig:SE_vs_t_plot}
        \end{subfigure}
\caption{The fitted average entropy $s_{\text{avg}}$ versus the average energy $\epsilon_{\text{avg}}$ (left) 
and their individual values as a function of time (right) for $\xi(\per = 0.01, \amp = 1) = 0.01$. 
Fitting for $s_{\text{avg}} \sim \epsilon_{\text{avg}}^{\gamma}$, we find a fitted value of 
$\gamma=0.6682 \pm 0.0023 \gtrsim \tfrac{2}{3}$ with 95\% confidence.}
\label{fig:S_vs_E_boring}
\end{figure}
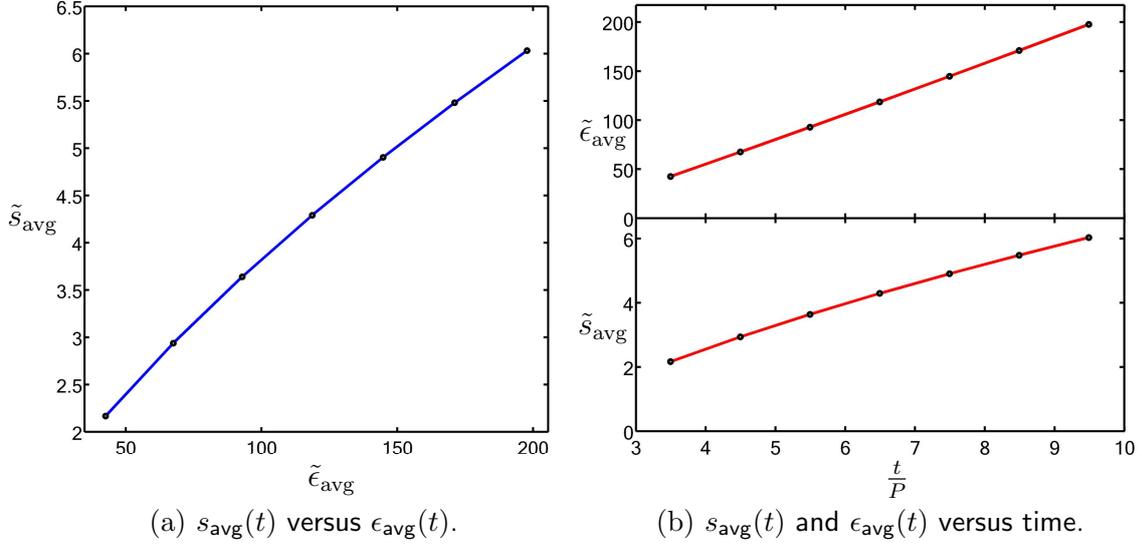

Both the energy and entropy density, averaged over each cycle, grow linearly with time in the dissipation-dominated regime .  As the black hole grows, its entropy growth tracks its energy growth at a slightly higher rate than the equilibrium relation $s_{\text{avg}} \sim \epsilon_{\text{avg}}^{2/3}$, i.e., $\gamma \gtrsim 2/3$. This entropy-energy scaling is shown in Fig.~\ref{fig:S_vs_E_boring} along with their own evolution with time. Note that the expansion of the black hole horizon is not necessarily adiabatic (as measured e.g., by the rate of entropy increase $\frac{1}{T}\frac{\dot{S}}{S}$).

\begin{figure}
\centering
\includegraphics[width=0.45 \textwidth]{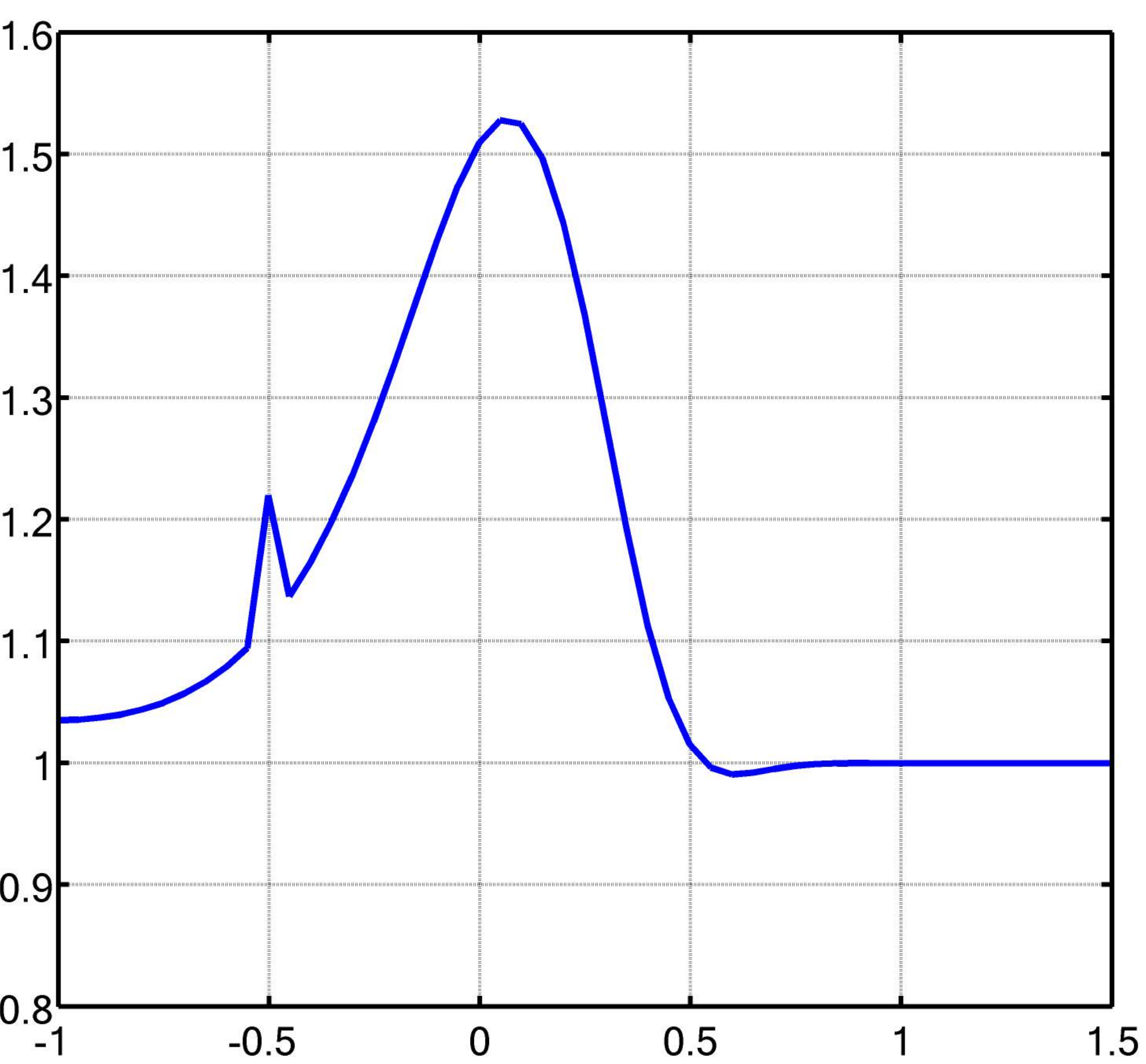}
\begin{picture}(0.3,0.4)(0,0)
\put(-91,-8.0){\makebox(0,0){\normalsize $\text{log}_{10} \ \omega$}}
\put(-207,98){\makebox(0,0){\normalsize $\alpha$}}
\end{picture}
\vspace{2 mm}
\caption{The dimensionless scaling parameter $\alpha(\omega)$ from fitting $\epsilon_{\text{cycle}} \sim \omega^{\alpha}$ for a small amplitude $\amp = 1$ in the linear response regime. It is expected for our choice of the scalar and dimension ($\Delta = 2$ and $d=3$) that $\alpha \rightarrow 1$ in both the small ($\epsilon_{\text{cycle}} \sim \omega$) and large frequency ($\epsilon_{\text{cycle}} \sim \omega^{2 \Delta - d}$) limits. }
\label{fig:alpha_omega_low_A}
\end{figure}
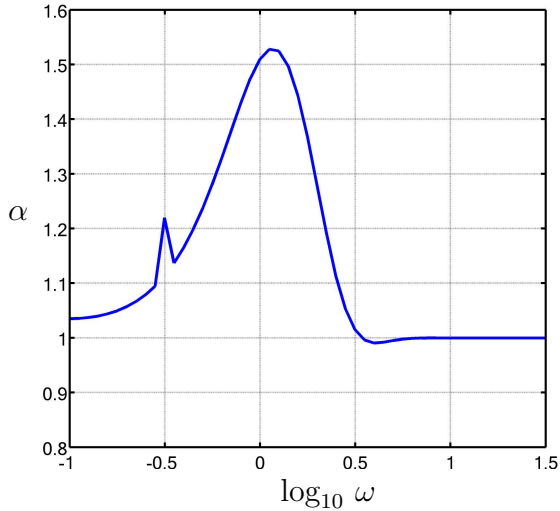

In the low amplitude regime, one can also estimate in perturbation theory the amount of energy dissipated per cycle $\epsilon_\text{cycle}$ which we define as the difference of the average energy $\epsilon_\text{avg}$ between two successive cycles; for simplicity we take the result for the last two cycles of our evolution in quoting the results below.
One expects the relation to take a scaling form $\epsilon_\text{cycle} \sim \omega^\alpha$. The scaling exponent $\alpha$ should be a non-trivial function of frequency itself; for low frequencies it is independent of the driving operator, but the high frequency limit cares about the spectral properties about the operator in question. Specifically, one finds that  \cite{Auzzi:2013pca}:
$\epsilon_\text{cycle} \sim \omega$ for small frequencies and 
$\epsilon_\text{cycle} \sim \omega^{2\,\Delta - d}$ for high frequencies. Since we are not scanning over different choices of the driving operator, we have a single shot at determining this result. As depicted in Fig.~\ref{fig:alpha_omega_low_A} we indeed find that the energy dissipated is linear both at low and high frequencies: $\alpha(\omega) \to 1$ both for $\omega \gg 1$ and for $\omega \ll 1$ 
(a  coincidence owing to our choice $\Delta =2$ and $d=3$). Interestingly there is some non-trivial intermediate frequency behaviour which appears to amplify the energy dissipated in a single cycle.

The bulk picture of the process is also very simple: as we send energy pulses, which are either weak or infrequent, they interact very rarely before falling into the black hole horizon. All injected energy from the boundary goes towards steadily increasing the black hole mass and the scalar field remains unexcited. The more diverse behaviour observed below can be attributed to gravitational interactions of those energy pulses before they fall into the black hole.

 \subsection{Dynamical Crossover Tilted Regime}
\label{sec:dc}
\begin{figure}[h]
\centering
        \begin{subfigure}[b]{0.45\textwidth}
        \includegraphics[width=\textwidth]{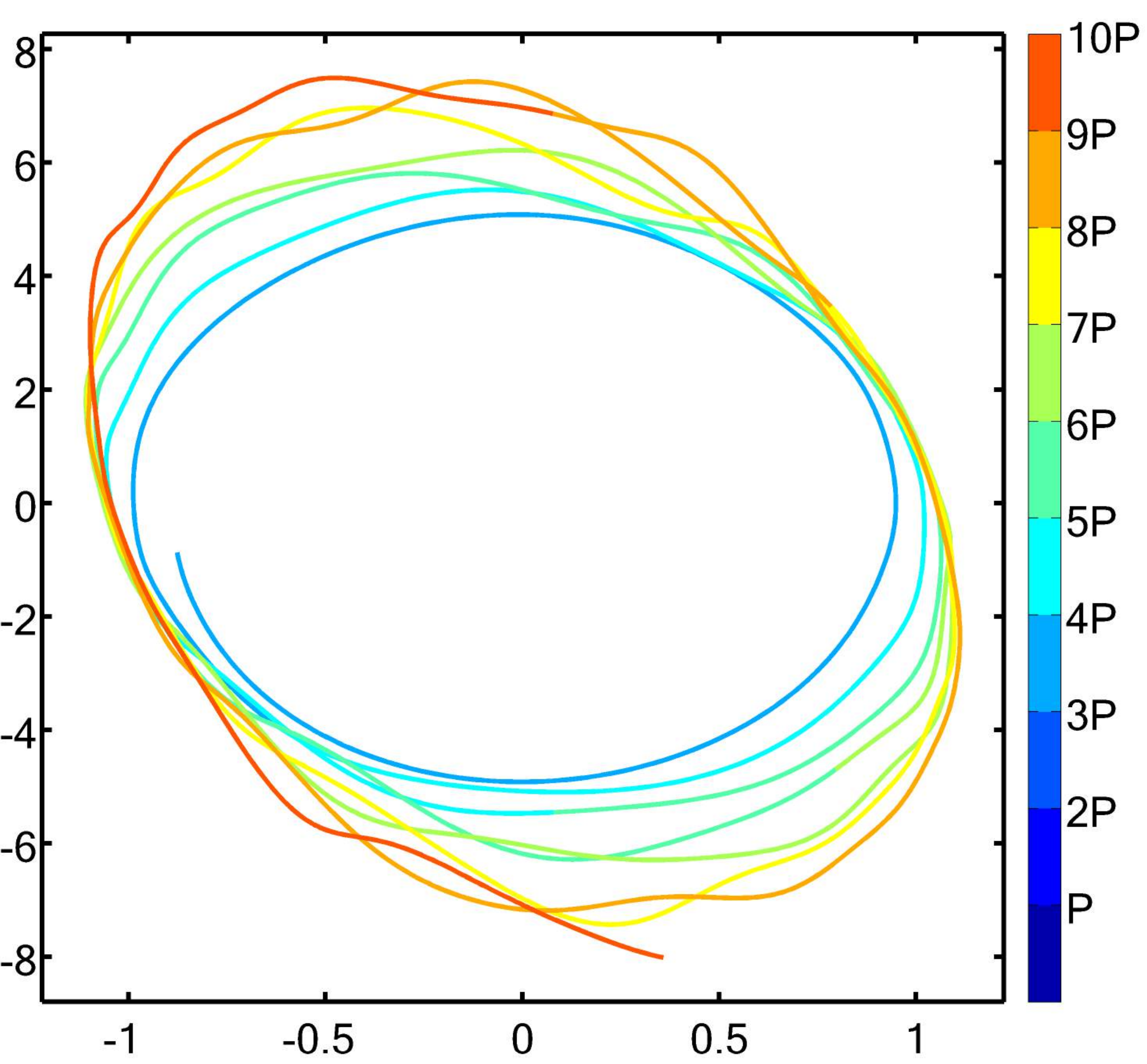}
        \put(-104,-8.0){\makebox(0,0){\normalsize $\tilde{\phi}_0$}}
        \put(-199,95){\makebox(0,0){\normalsize $\tilde{\phi}_1$}}
        \vspace{1mm}
        \caption{$\xi(\per = 0.1, \amp = 20) = 2$}
        \label{subfig:PP_boring_high_freq}
        \end{subfigure}
        ~ ~
        \begin{subfigure}[b]{0.45\textwidth}
        \includegraphics[width=\textwidth]{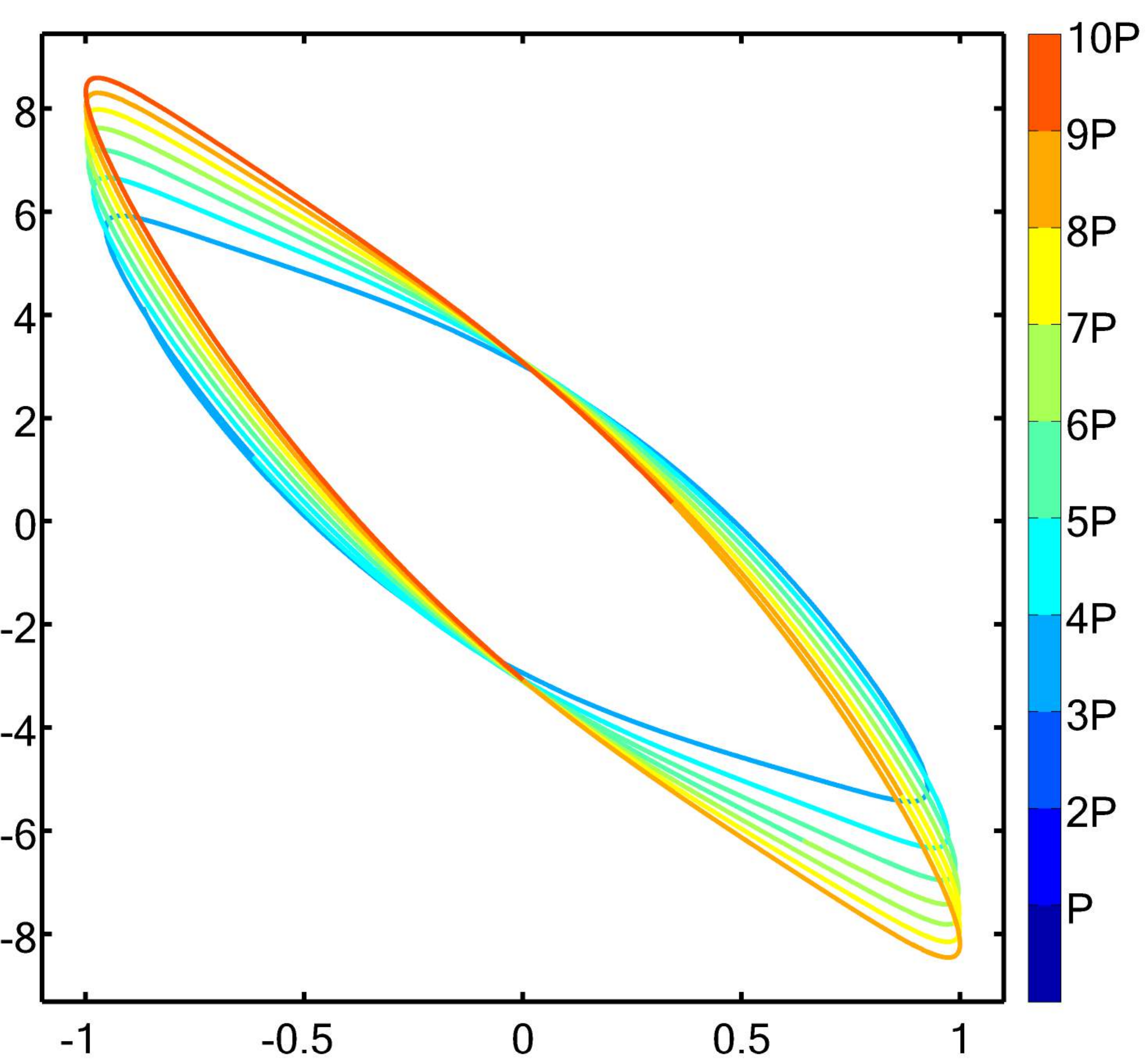}
        \put(-104,-8.0){\makebox(0,0){\normalsize $\tilde{\phi}_0$}}
        \put(-199,95){\makebox(0,0){\normalsize $\tilde{\phi}_1$}}
        \vspace{1mm}
        \caption{$\xi(\per = 10, \amp = 1)=10$}
        \label{subfig:PP_boring_low_freq}
        \end{subfigure}
\caption{The dimensionless phase portrait of the response $\tilde{\phi}_1$ versus the source $\tilde{\phi}_0$ for $\xi(\per = 0.1, \amp = 20)=2$ (left) and $\xi( \per = 10, \amp = 1) = 10$ (right). The conventions are as in Fig.~\ref{fig:PP_boring}.  The left panel shows the behaviour in phase IIb while the right panel pertains to phase IIa.}
\label{fig:PP_wobbly}
\end{figure}

We now discuss the qualitative changes in the system as we begin to move from the weak driving $\xi \ll 1$ to the strong driving regime $\xi \gg 1$ (from regime I to regime II through the grey-dashed line in phase diagram Fig.~\ref{fig:PP_qualitative}). Fig.~\ref{fig:PP_wobbly} depicts a typical phase portrait of the system as we cross into the new dynamical regime. We see that this regime is characterized by an onset of excitations of the scalar field and breaking of discrete time translation symmetry.
The left panel of Fig.~\ref{fig:PP_wobbly} shows the transition from $\xi \ll 1 \rightarrow \xi \gg 1$ at high amplitudes: the trajectories are no longer closed, rather they precess as a function of time and are slightly tilted. The breaking of discrete time-translation invariance is an interesting effect of the gravitational interactions of the scalar field.

In the right panel of Fig.~\ref{fig:PP_wobbly} we see the effect of moving into the new dynamical regime at low amplitudes: there is a clear tilt in the phase portrait from the one in Fig.~\ref{fig:PP_boring} with $\xi \gg 1$ which indicates that the response is no longer completely out of phase with the source. 
The tilting of the trajectories at lower frequencies corresponds to the emergence of a finite in-phase contribution $\sigmaIn > 0$ in the conductivity; this sets the system somewhere between one with a purely out-of-phase conductivity (closed circular trajectories) and one with a purely in-phase conductivity (straight diagonal line trajectories). In other words not all of the injected energy is dissipated as was the case in regime I, but rather, work is actually being done on the system.

\begin{figure}
\centering
\includegraphics[width=0.65 \textwidth]{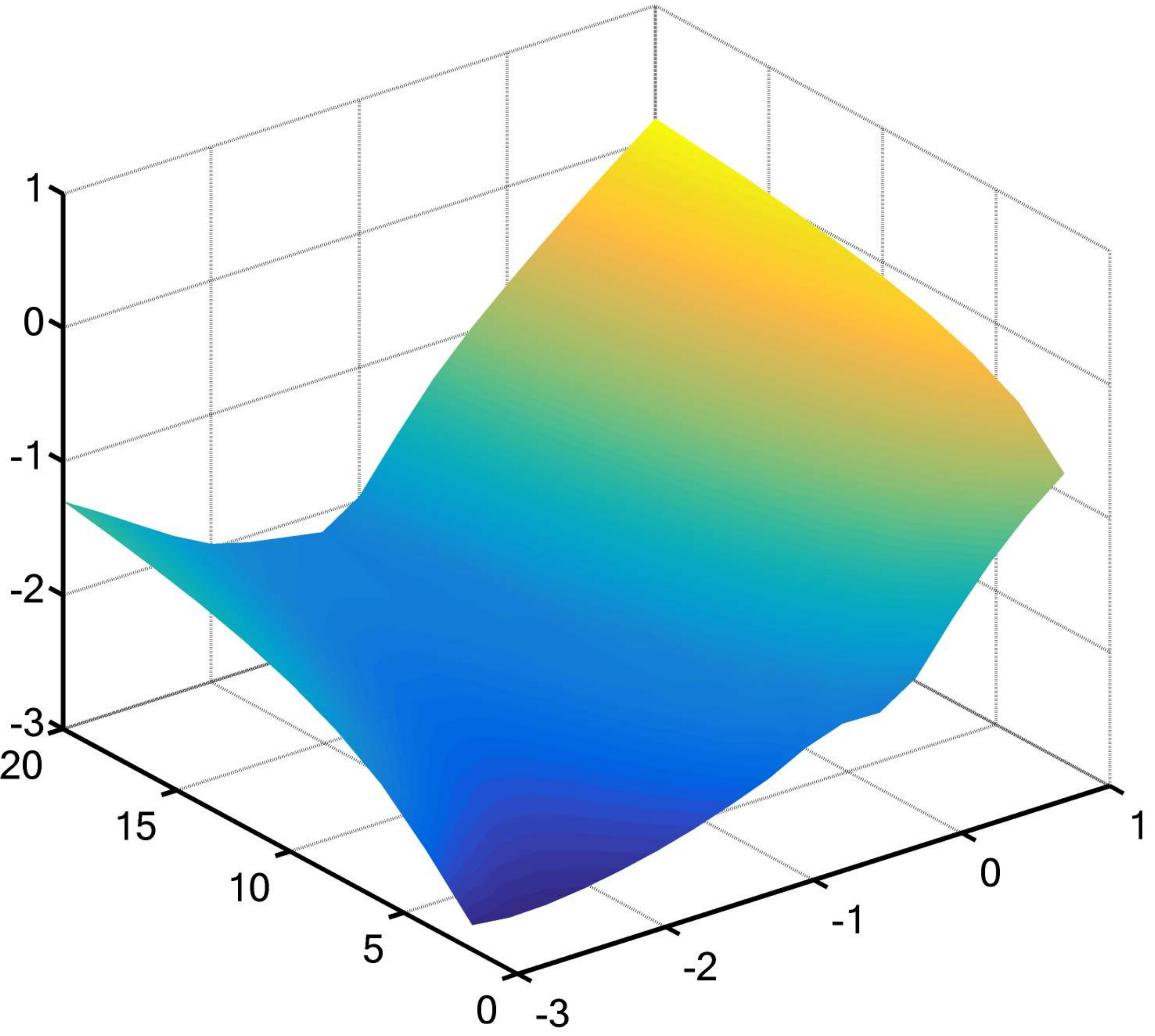}
        \put(-60,13){\makebox(0,0){\normalsize $\log_{10} \per$}}
        \put(-235,23){\makebox(0,0){\normalsize $\amp$}}
        \put(-265,222){\makebox(0,0){\normalsize $\log_{10}(\gamma-\gamma_0)$}}
        \vspace{1mm}
\caption{The increase in the scaling exponent $\gamma$ in $s_\text{avg} \sim \epsilon^{\gamma}_\text{avg}$ from the equilibrium value of $\gamma_{\text{0}} = \tfrac{2}{3}$ over the entire $(\per,\amp)$ phase diagram. We find that $\gamma > \gamma_{\text{0}}$ holds for all scanned values on the phase diagram.}
\label{fig:gamma_phasediagram}
\end{figure}

As a result of having less dissipation in this regime, the energy and entropy of the black hole grow more slowly with time. Moreover, we find the scaling behaviour between the average energy and entropy, with a thermodynamic scaling exponent $\gamma > \tfrac{2}{3}$, for all values of $(\per,\amp)$, as shown in Fig.~\ref{fig:gamma_phasediagram}. In other words, while the work done in the system slows down the energy increase of the black hole, the entropy production is affected less.

\begin{figure}[h]
\centering
        \begin{subfigure}[b]{0.45\textwidth}
        \includegraphics[width=\textwidth]{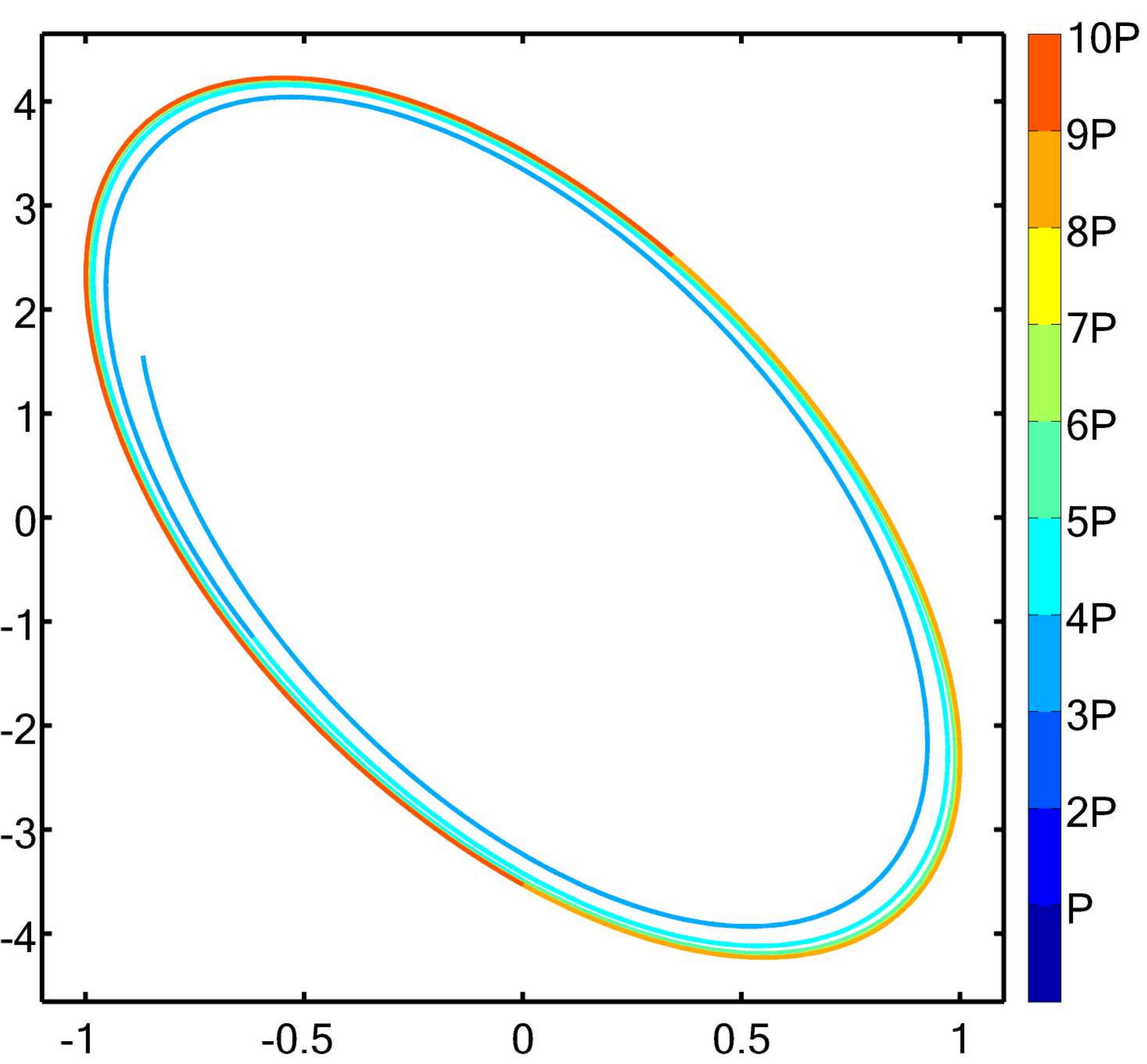}
        \put(-104,-8.0){\makebox(0,0){\normalsize $\tilde{\phi}_0$}}
        \put(-198,93){\makebox(0,0){\normalsize $\tilde{\phi}_1$}}
        \vspace{1mm}
        \caption{$V_2(\phi) = -2 \phi^2$}
        \label{subfig:PP_phi2_xi10}
        \end{subfigure}
        ~ ~
        \begin{subfigure}[b]{0.46\textwidth}
                \includegraphics[width=\textwidth]{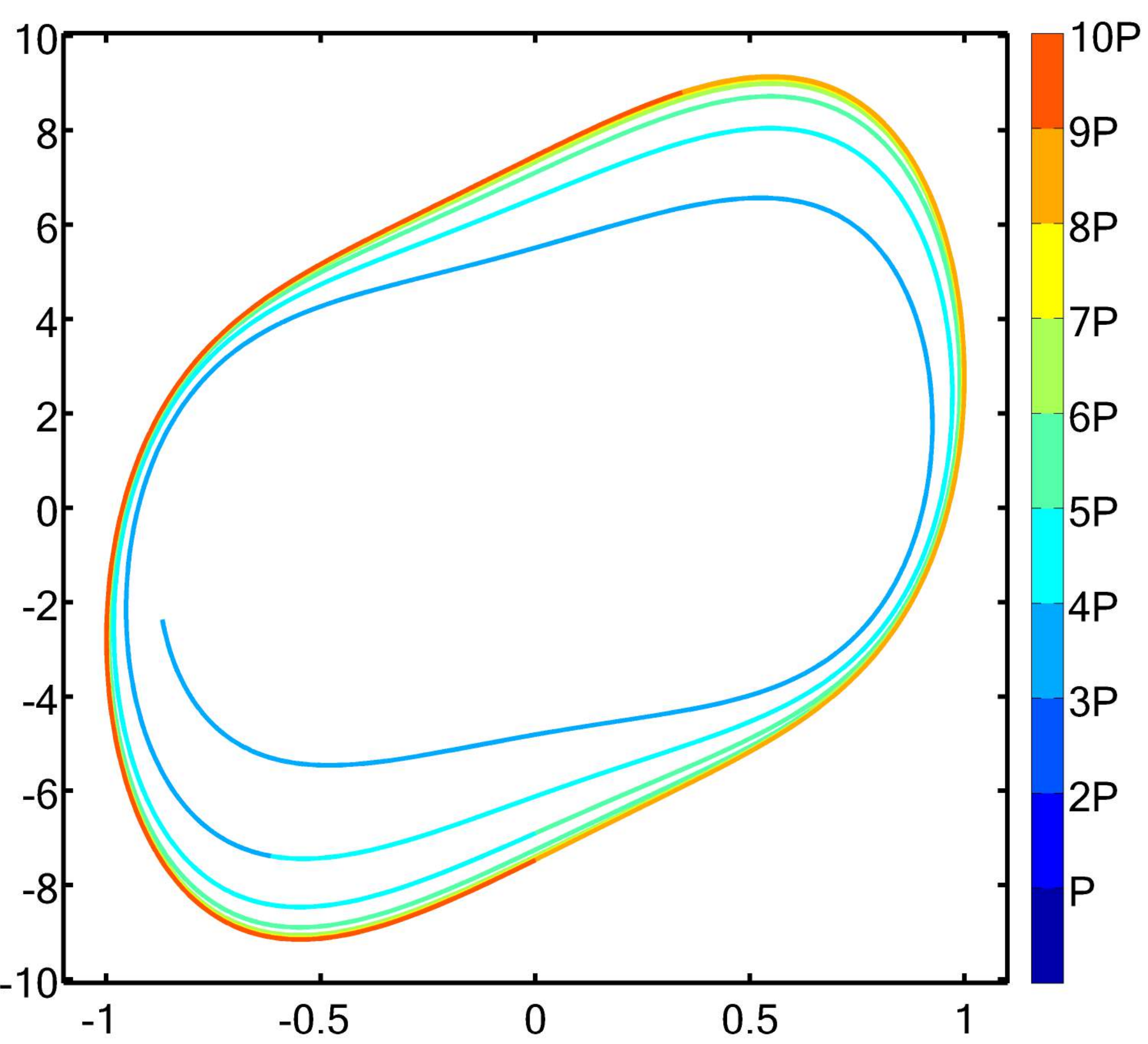}
        \put(-104,-8.0){\makebox(0,0){\normalsize $\tilde{\phi}_0$}}
        \put(-197,92){\makebox(0,0){\normalsize $\tilde{\phi}_1$}}
        \vspace{1mm}
        \caption{$V_4(\phi) = -2 \phi^2-\frac{1}{2}\phi^4$}
        \label{subfig:PP_phi4_xi10}
        \end{subfigure}
\caption{The phase portrait of the dimensionless response 
$\tilde{\phi}_1 \equiv \tfrac{\per}{\amp} \,\phi_1$ versus the dimensionless source 
$\tilde{\phi}_0 \equiv \tfrac{1}{\amp}\, \phi_0$ 
for $\xi(\per = 10, \amp = 1)=10$ with $\alpha_g=0$ and different polynomial potentials $V(\phi)$. The conventions are 
as described in Fig.~\ref{fig:PP_boring}.}
\label{fig:PP_phi2}
\end{figure}

To understand this regime further, it is instructive to reproduce this type of phase portrait for a system without gravity. To that effect, we can study the special case of scalar field evolution in a fixed black hole background, with no backreaction on the geometry (i.e., $\alpha_g =0$). To include non-linearity into the problem, we add self-coupling to the scalar field, to mimic the effect of the non-linearities due to gravitational interactions (see also \cite{Basu:2013vva}). Fig.~\ref{fig:PP_phi2} depicts the phase portrait of a self-coupled scalar field with two types of {\it polynomial} potentials, which we took to be our original form (free massive scalar) and also one with quartic self-interactions:
\begin{equation}
V_{\text{poly},4}(\phi)=-2 \phi^2-\frac{1}{2}\phi^4.
\label{eqn:V4}
\end{equation}
We can see that  without non-linearity as in Fig.~\ref{subfig:PP_phi2_xi10}, the phase portrait is tilted, but sharp features of the phase portrait are lost compared to the case with the same driving but also gravitational backreaction, depicted in Fig.~\ref{subfig:PP_boring_low_freq}. Adding a polynomial non-linearity, as done in Fig.~\ref{subfig:PP_phi4_xi10}, gives a phase portrait that starts to form slightly sharper features along with some amplification of the response. Thus, the simple system of self-interacting scalar field allows us sufficiently separate the two effects in regime II: we see that the tilt in the phase diagram is associated with decreased frequency, whereas the breaking of time-translation invariance is associated with increased amplitude. We note also that for this simple system, the third dynamical regime of unbounded amplification discussed in the next subsection seems to be absent. 

Thus, the bulk interpretation of this dynamical regime becomes clear: the pulses of energy injected at the boundary interact gravitationally before falling into the black hole. This results in additional physics to that of simple dissipation, modeled here by infalling the black hole. The gravitational interaction is due to perturbative exchange of gravitons, and can be mimicked by a polynomial self-interaction of the scalar field. In the next subsection we will see the effect of the gravitational interactions becoming strong when both $\amp$ and $\per$ are large.

\subsection{Unbounded Amplification Regime}
\label{sec:unamp}
\begin{figure}
\centering
        \begin{subfigure}[b]{0.45\textwidth}
        \includegraphics[width=\textwidth]{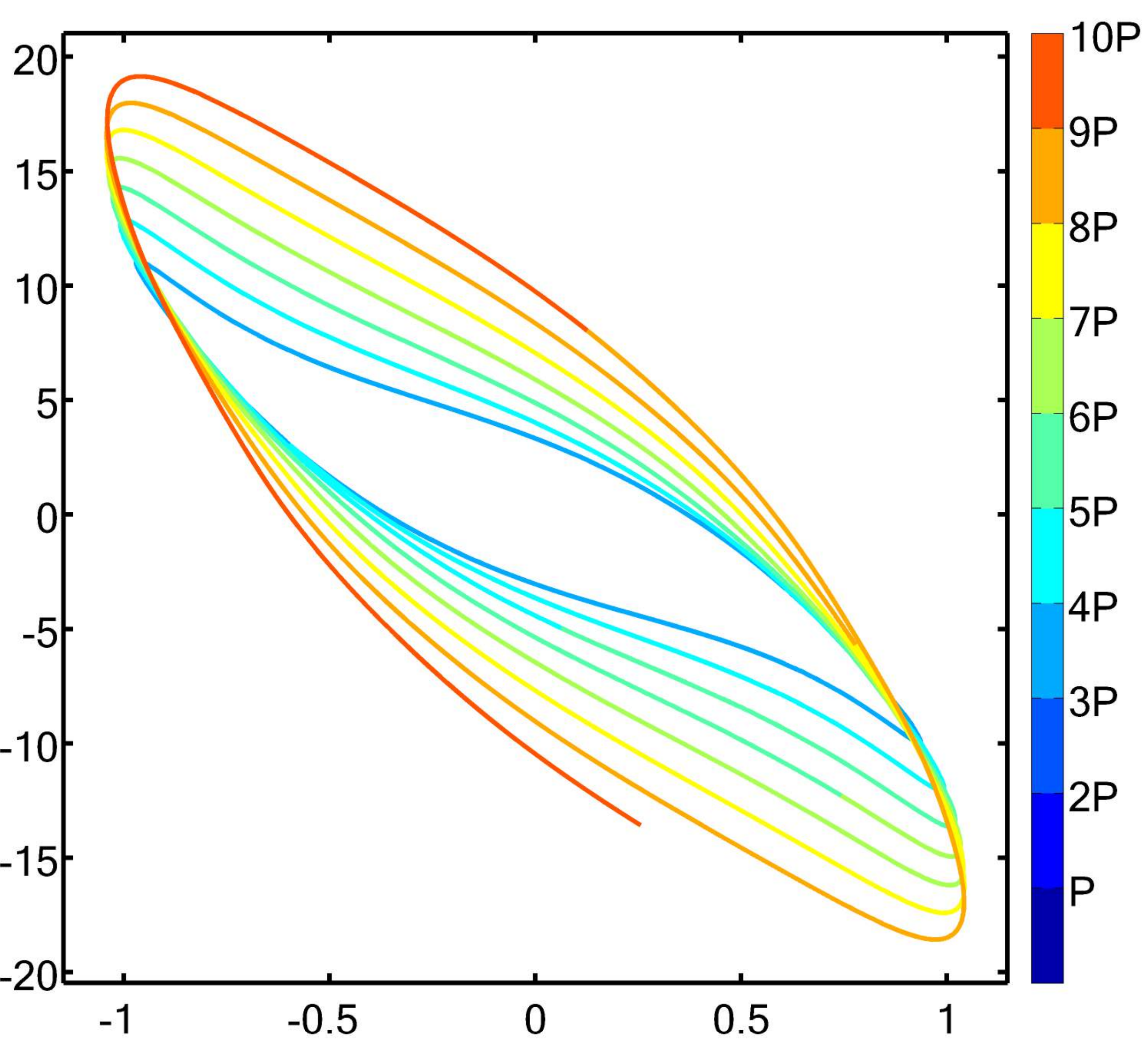}
        \put(-102,-8.0){\makebox(0,0){\normalsize $\tilde{\phi}_0$}}
        \put(-195,90){\makebox(0,0){\normalsize $\tilde{\phi}_1$}}
        \vspace{1mm}
        \caption{$\xi(\per = 1, \amp = 20)=20$}
        \label{subfig:PP_sharp}
        \end{subfigure}
        ~ ~
        \begin{subfigure}[b]{0.46\textwidth}
        \includegraphics[width=\textwidth]{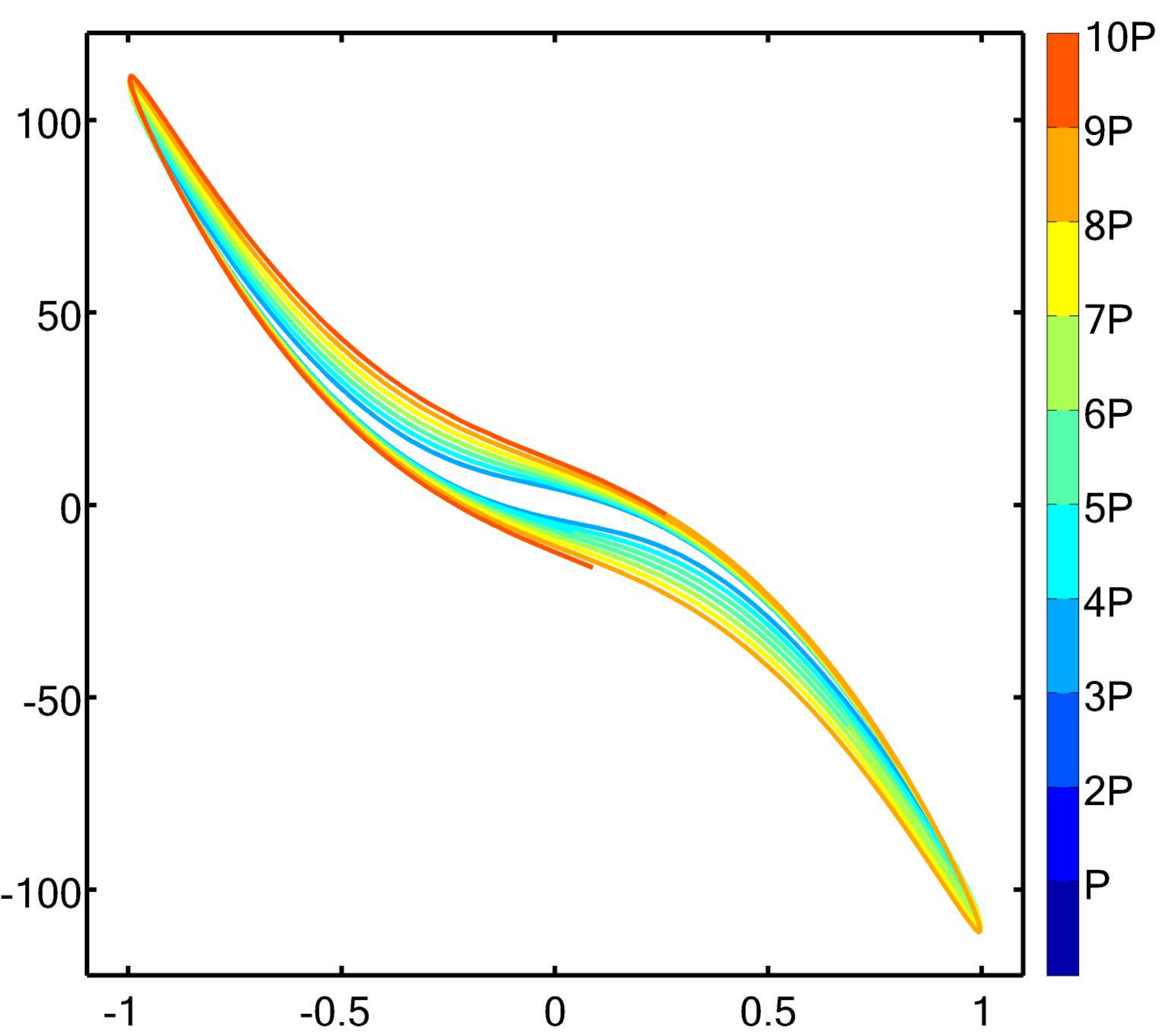}
        \put(-102,-8.0){\makebox(0,0){\normalsize $\tilde{\phi}_0$}}
        \put(-194,90){\makebox(0,0){\normalsize $\tilde{\phi}_1$}}
        \vspace{1mm}
        \caption{$\xi(\per = 10, \amp= 20) = 200$}
        \label{subfig:PP_sharper}
        \end{subfigure}
\caption{The phase portrait of the response $\tilde{\phi}_1$ versus the source $\tilde{\phi}_0$ for $\xi=20$ (left) and $\xi= 200$ (right) in the non-perturbative dynamical regime (regime III). The conventions are as in 
Fig.~\ref{fig:PP_boring}.}
\label{fig:PP_interesting}
\end{figure}

As we increase the driving strength further in both $\amp$ and $\per$ directions (from regime II to regime III through the blue-dashed line in phase diagram Fig.~\ref{fig:PP_qualitative}), we enter a dynamical regime no longer reproducible by polynomial self-interactions of the scalar field. 
We see the phase portrait of the scalar field in Fig.~\ref{fig:PP_interesting}
for two instances of parameters in this regime. Moreover, we find this dynamical regime to be characterized by unbounded response and restoration of time translation symmetry. 

As we increase the strength of the driving force $\xi$, the phase portrait becomes sharper and tilted, corresponding to an increased response and, again, less lag with the source as seen in Fig.~\ref{fig:conductivity_phasediagram}.
The `slowness' of the energy injection from the boundary allows the scalar field to heat up as if the entire process were adiabatic, consequently allowing the scalar response to respond relatively quicker to the source.
Note that although Fig.~\ref{fig:conductivity_phasediagram} shows $\left| \sigma_{\text{in}} / \sigma \right| \approx 1$ in this regime, the absolute value $\left| \sigma \right|$ is actually very large in this unbounded amplification regime so that a small $\left| \sigmaOut / \sigma \right|$ is still strong enough to keep the black hole perpetually growing in size.

\begin{figure}
\centering
\includegraphics[width=0.65 \textwidth]{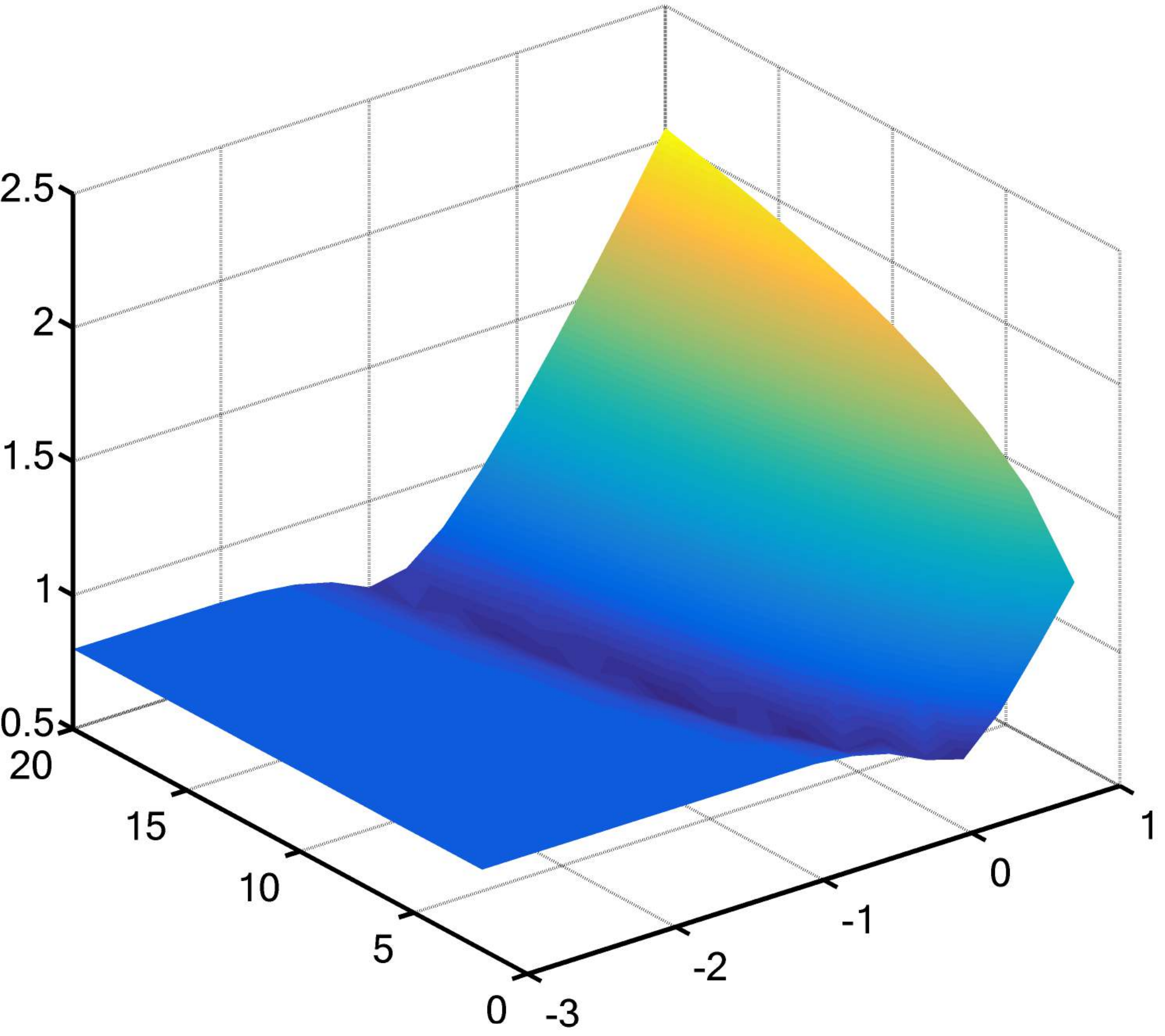}
        \put(-60,13){\makebox(0,0){\normalsize $\log_{10} \per$}}
        \put(-235,23){\makebox(0,0){\normalsize $\amp$}}
        \put(-270,224){\makebox(0,0){\normalsize $\log_{10} \left| \tilde{\phi}^{\text{max}}_1 \right| $}}
        \vspace{1mm}
\caption{The maximal response $\left| \tilde{\phi}^{\text{max}}_1 \right| = \frac{\per}{\amp} \left| \phi^{\text{max}}_1 \right|$ over the entire $(\per,\amp)$ phase diagram.}
\label{fig:maxresponse_phasediagram}
\end{figure}

The maximal response $|\phi^{\text{max}}_1|$ over our ten cycles of driving is plotted in Fig.~\ref{fig:maxresponse_phasediagram} 
throughout the phase diagram.  It is seen to increase rapidly with $\xi$ past the dissipation-dominated regime. This seems to indicate the presence of a non-linear resonance, which allows the scalar response to grow without bound. An interesting feature of 
Fig.~\ref{fig:maxresponse_phasediagram} is that the maximal response does not grow in the high frequency regime regardless of how large $\xi$ is by increasing $\amp$. 
It seems unlikely that unbounded behaviour is attainable even for amplitudes drastically higher than the bounds of numerical explorations reported in Fig.~\ref{fig:maxresponse_phasediagram}. 
Physically, this means that a rapid pulsing of small packets of energies can barely amplify the response of the system; the frequency of driving has to be below a certain bound for resonance to be possible -- or in other words, a certain slowness in the sourcing is required.
We conjecture that one should would see unbounded amplification only in the combined  large $\per$,  large $\amp$ regime which is slightly different from the traditional definition of resonance that depends only on frequency. An interesting curiousity is a slight dip in the response  for moderate values of $\xi$  preceding the rapid growth. This trough appears to demarcate the domains of bounded (regime II) and unbounded responses (regime III) empirically.  It would be interesting to come up with a explanation for this phenomenon.

\begin{figure}
\centering
        \begin{subfigure}[b]{0.45\textwidth}
        \includegraphics[width=\textwidth]{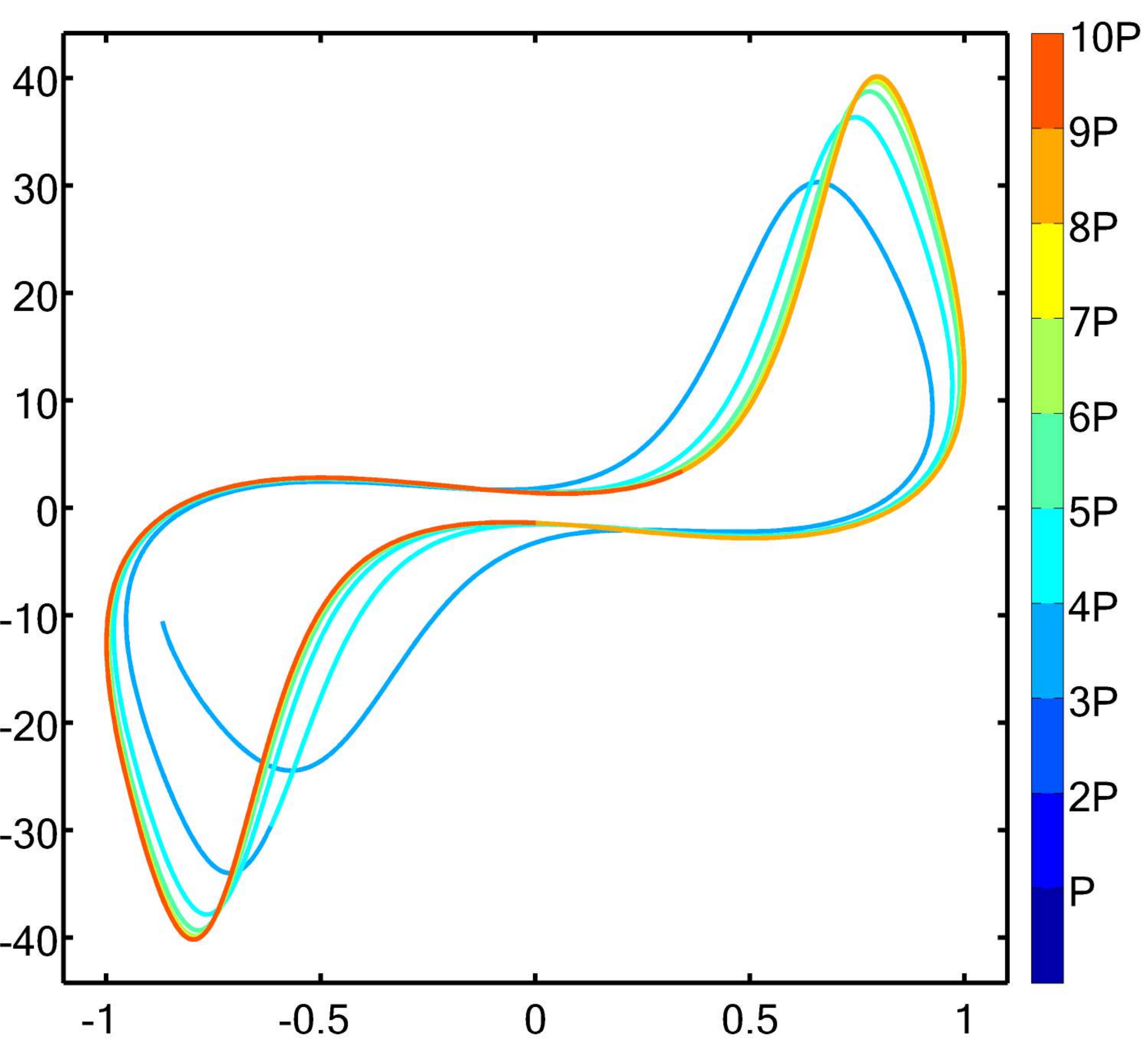}
        \put(-102,-8.0){\makebox(0,0){\normalsize $\tilde{\phi}_0$}}
        \put(-195,91){\makebox(0,0){\normalsize $\tilde{\phi}_1$}}
        \vspace{1mm}
        \caption{$\xi(\per = 10, \amp = 1.5)=15$}
        \label{subfig:V3_A1p5_P10}
        \end{subfigure}
        ~ ~
        \begin{subfigure}[b]{0.45\textwidth}
        \includegraphics[width=\textwidth]{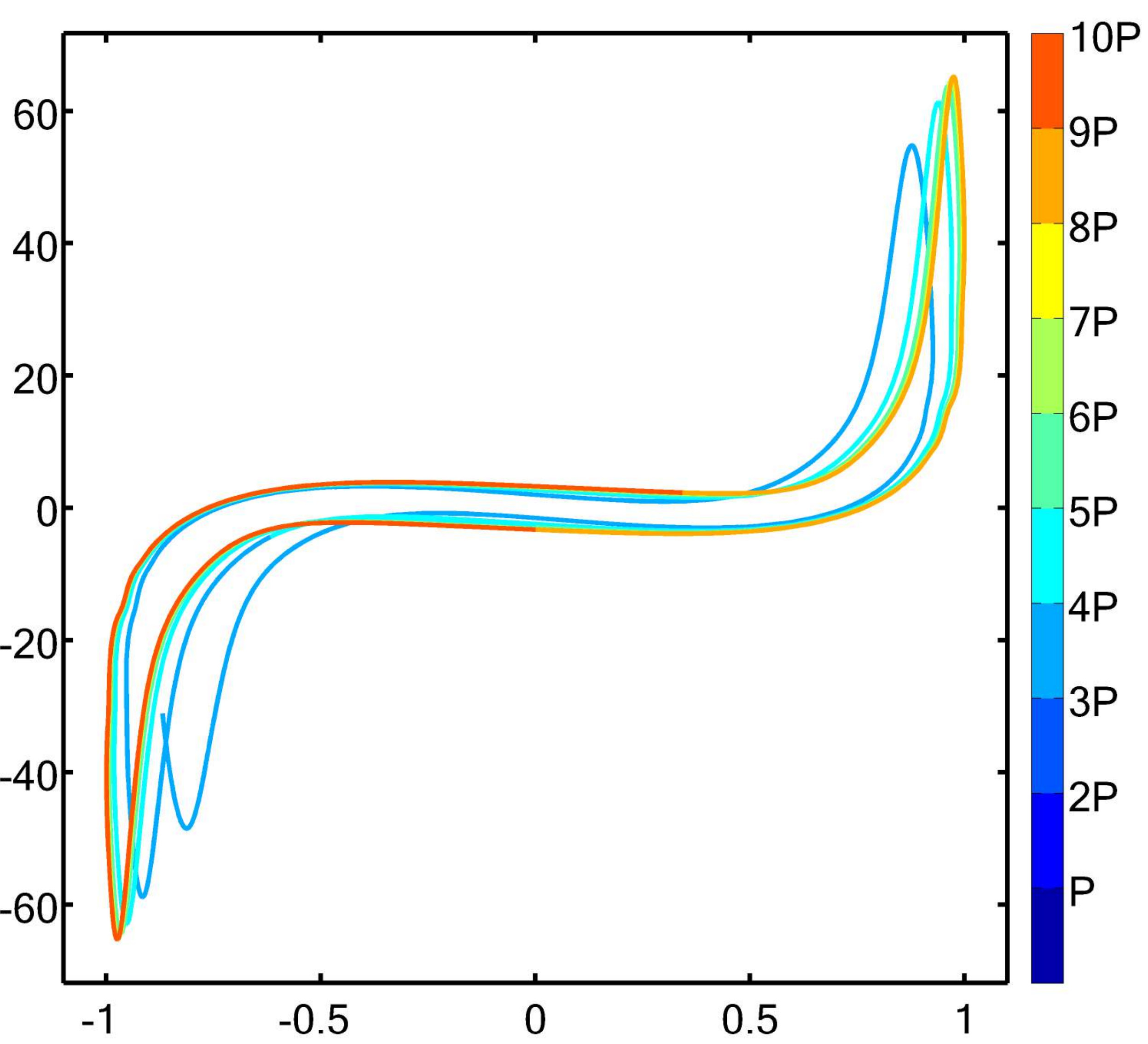}
        \put(-102,-8.0){\makebox(0,0){\normalsize $\tilde{\phi}_0$}}
        \put(-195,91){\makebox(0,0){\normalsize $\tilde{\phi}_1$}}
        \vspace{1mm}
        \caption{$\xi(\per = 10, \amp = 2) = 20$}
        \label{subfig:V3_A2_P10}
        \end{subfigure}
\caption{The phase portrait of the response $\tilde{\phi}_1$ versus the source $\tilde{\phi}_0$ for $\xi= 15$ (left) and $\xi= 20$ (right) for the non-polynomial potential Eq.~\eqref{eqn:V_nonpoly}, in the conventions of Fig.~\ref{fig:PP_boring}. }
\label{fig:PP_nonpoly}
\end{figure}

Finally, it is amusing to model the non-linear effects of gravity in terms of an effective scalar potential to see what is necessary to attain regime III.  We find that while a scalar field with polynomial self-interaction does not seem to posses this regime, one can reproduce similar features by {\it non-polynomial} potentials. For example, we can discuss a self-interacting scalar field probe, with 
\begin{equation}
V_{\text{non-poly}}(\phi)=-2 \sinh^2\phi+\frac{1}{6}\sinh^4\phi\,.
\label{eqn:V_nonpoly}
\end{equation}
This choice of scalar self-interaction is chosen to agree with our previous example \eqref{eqn:V4} in the small field regime, but of course behaves differently for large field values. In Fig.~\ref{fig:PP_nonpoly} we see that indeed similar features of the phase diagram are reproduced: narrow closed trajectories and resonant response. We conclude therefore that the features of this dynamical regime are due to strong, non-perturbative gravitational effects occurring outside the black hole horizon. The fact that the non-linearities induced by gravity can be extremely strong, should perhaps be borne in mind while attempting to come up with simplified models of gravitational dynamics in AdS spacetime.

\subsection{Energy Fluctuations}

Another observable we monitor is the behaviour of energy fluctuations. More precisely, we consider the deviations from the average energy in a each cycle, $\epsilon_\text{fluc}(t) = |\epsilon(t) - \epsilon_\text{avg}(t)|$. 
 These cycle fluctuations are a crude proxy for genuine fluctuation information that can be extracted, for instance, by considering symmetrized two-point functions of the boundary energy momentum tensor. Such ensemble-averaged fluctuations are known to exhibit phase transitions in periodically driven systems \cite{nature}. Some indication those transitions are possible in holographic systems is given in \cite{Auzzi:2013pca}.
 
\begin{figure}
\centering
\includegraphics[width=0.65 \textwidth]{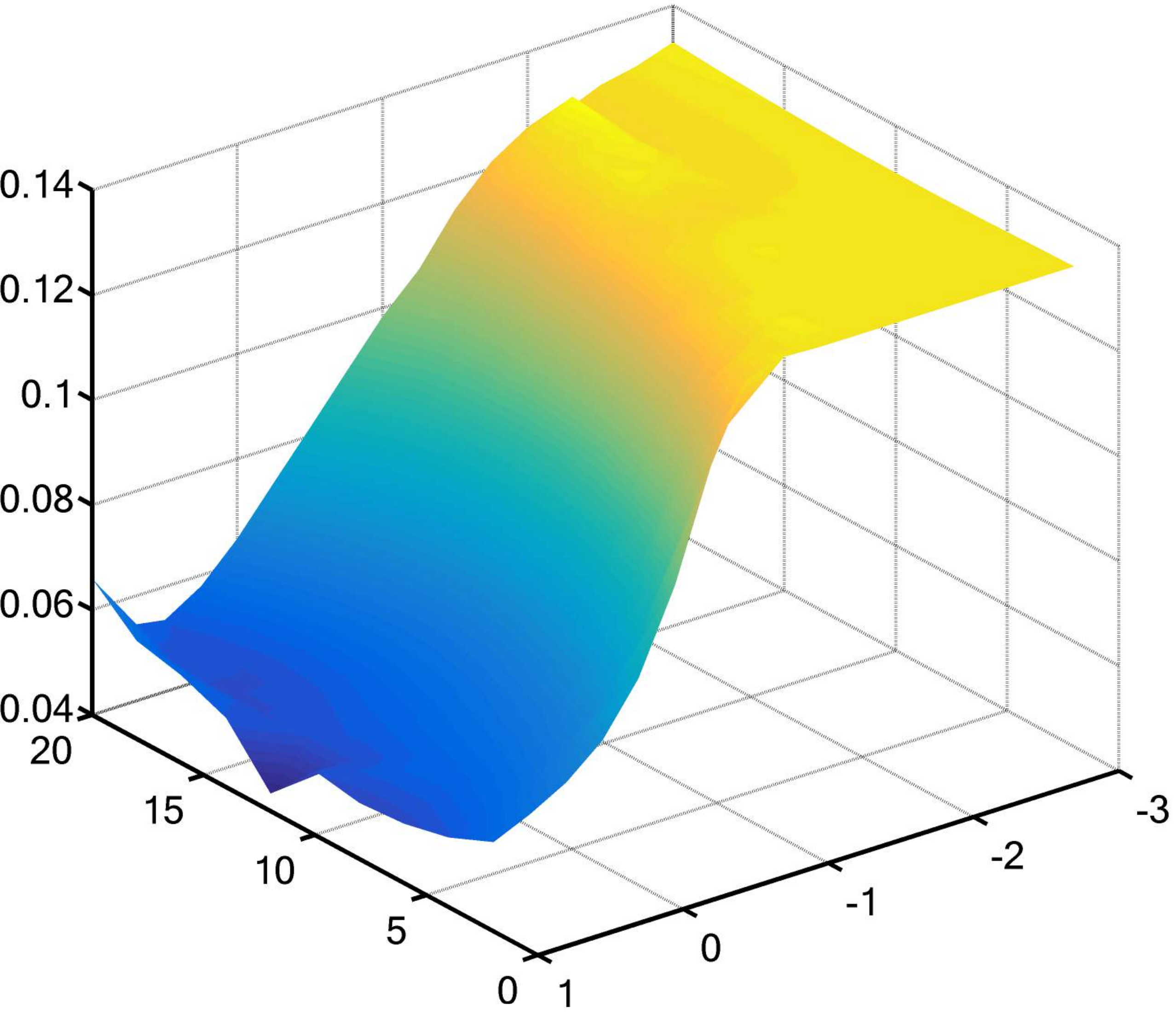}
        \put(-60,13){\makebox(0,0){\normalsize $\log_{10} \per$}}
        \put(-235,23){\makebox(0,0){\normalsize $\amp$}}
        \put(-260,217){\makebox(0,0){\large $\tilde{\epsilon}_{\text{fluc}}$}}
        \vspace{1mm}
\caption{Energy density fluctuations (last cycle) $\tilde{\epsilon}_{\text{fluc}}$ in units of $\amp^2 / \per$ over the entire $(\per,\amp)$ phase diagram.}
\label{fig:energyfluc_phasediagram}
\end{figure}

 The results for our simulations in various regimes are plotted in Fig.~\ref{fig:energyfluc_phasediagram}. We observe a qualitative change in these cycle fluctuations between different regimes. While in the dissipation-dominated phase we do not  see a lot of deviation from the mean, there is a steep growth in fluctuations as we enter the non-linear phases. The fluctuations are maximal in the unbounded amplification regime (regime III). We note that in contrast to the maximal scalar response, which also grows dramatically in that phase, the fluctuations do track the driving frequency, with there being more deviations in the large period limit.
 
It would be useful to confirm this behaviour directly with the computation of correlation functions, a task we leave for future investigation. 

%

\section{Entanglement entropy}
\label{sec:ee}
Thus far we have discussed various local observables (response functions and thermodynamic data) which have served to help us chart the phase diagram of the driven system in Fig.~\ref{fig:PP_qualitative}. We now turn to other non-local field theory observables that are sensitive to the non-equilibrium dynamics. Since we are not going to examine the behaviour of higher point correlation functions, we will dive right into the dynamical behaviour of entanglement entropy.

In the boundary we have a density matrix $\rho(t)$ which is time-evolving with respect to the perturbed Hamiltonian. At any given instant of (boundary) time, we pick a spatial region ${\cal A}$ and construct the matrix elements of the reduced density matrix 
$\rho_{\cal A} (t) = {\rm Tr}_{{\cal A}^c}\left(\rho(t)\right)$ by tracing out the degrees of freedom in the complement (on the chosen Cauchy slice). The entanglement entropy is given by the von Neumann entropy of $\rho_{\cal A}$, i.e., $S_{\cal A}(t) = -{\rm Tr}_{\cal A} \left(\rho_{\cal A}\, \log \rho_{\cal A}\right)$ which we can monitor as a function of time.

Holographically computing the entanglement entropy for boundary regions in time dependent situations involves finding bulk codimension-2 extremal surfaces $\cal E_{\cal A}$ anchored on the said boundary region ${\cal A}$ \cite{Hubeny:2007xt}. 
We study the evolution of entanglement entropy focusing in particular on  translationally invariant strip regions:
\begin{equation}
{\cal A} = \{ t = t_{\cal A},  -a \leq x \leq a , y \in {\mathbb R} \}\,.
\label{eqn:spatial_strip_defn}
\end{equation}	
The bulk codimension-2 surface ends at $x=\pm a$ at some chosen instance of boundary time $t_{\cal A}$ and is obtained by solving effectively a set of geodesic-like equations with our interpolated metric functions $\Sigma$, $f$, and $\chi$ (see Appendix \ref{sec:extrdet} for details). The covariant holographic entanglement entropy prescription \cite{Hubeny:2007xt}  generalizing \cite{Ryu:2006bv,Ryu:2006ef} states that 
\begin{equation}
S_{\cal A} = \frac{  \text{Area}({\cal E}_{\cal A} )  }{  4 \, \GN  }\,.
\label{eqn:RT_prescription}
\end{equation}
Should there be multiple extremal surfaces, we choose the one with minimal area (homologous to ${\cal A}$). 
The proper area of these surfaces diverges owing to the locality of the underlying QFT.  In our case we encounter potential divergences not only from the surface reaching out to the asymptotic boundary, but also from the presence of the sources driving the system. The physical result we are after is the finite  universal contribution $S^{\text{fin}}_{\cal A}$, which will measure the  entanglement created/destroyed as we drive the system away from thermal equilibrium.  Fortuitously, for our choice of scalar operator, there are no contributions due to the source, and hence we can simply regulate by background subtraction.\footnote{ Details of the divergent structure and the counter-terms necessary to compute the area functional in our set up can be found in Appendix \ref{sec:regent}.} As a result we will consider as our entanglement diagnostic, the following finite quantity 
\begin{equation}
\sreg(t)= \frac{4 \, \GN}{L_y} \, \big[  S_{\cal A}(t) - S_{\cal A}(t=0) \big]
\label{eqn:S_reg}
\end{equation}
where $L_y$ is the IR regulator in the non-compact translationally invariant direction. Since we drive the system away from thermal equilibrium, $S_{\cal A}(t=0)$ is the corresponding value of the entanglement entropy computed in the Schwarzschild-\AdS{4} geometry.
In what follows we will simply quote the results of our numerical simulations both for the behaviour of the extremal surfaces themselves and $\sreg(t)$.

\subsection{Extremal surfaces in the driven geometries}
\label{sec:extrsufaces}
The extent to which the extremal surfaces penetrate into the bulk can for the most part be determined from 
the location of the cap-off point which we parameterize as $(t^*, u^*=1/r^*,x=0)$.\footnote{ The coordinate $u =1/r$ is chosen such that the horizon remains at $u=1$ during the entire course of the evolution (the boundary is at $u=0$).} 
For very small regions we are reasonably close to the AdS boundary whence, the curves are approximately semi-circles 
$u^2 + x^2 \approx a^2$.  As we increase to larger strip widths the extremal surfaces start to probe the interesting regions of the driven geometry and thus allows us to see qualitative differences between the four phases.

Generically we see that the following statements hold irrespective of the phases we consider:
\begin{enumerate}
\item The radial depth and the temporal extent spanned by the surface evolves non-trivially as a function of $t_{\cal A}$. One consequence of working with ingoing coordinates \eqref{eq:bulkcy} is that the surfaces naturally dip back in time (see \cite{Hubeny:2013hz,Hubeny:2013dea}).  
\item The oscillatory driving of the system imprints itself in the profile of the extremal surfaces, with the scale of these oscillations set by the the driving parameters $\amp$ and $\per$. The periodic movement of the surface can be seen in pulsations of the turnaround point of the surface: $u^*$ and $t^*$ have oscillations of the same period superposed over some enveloping function. 
\item On average, the extremal surfaces reach further into the bulk with time; $u^*(t_{\cal A})$ is monotonically increasing for the range of parameters explored. To understand this note, we gauge fixed the bulk coordinate chart \eqref{eq:bulkcy} such that the horizon is at $u_+=1$. In these coordinates the proper size of the region ${\cal A}$ increases (due to $\gxx(t,r)$) which means that the surfaces want to get closer to the horizon to extremize the area functional. 
The rate at which this happens depends on both the amplitude and the frequency of the driving. 
We also note that surfaces dip less temporally, i.e., $t^* - t_{\cal A}$ is increasing.
\item We also note that the location of the extremal surface appears to be consistent with causality of entanglement entropy \cite{Headrick:2014cta}. While we have not explicitly checked that the surface lies in the casual shadow of the boundary region ${\cal A}$, one simple consistency check visible from our results for $t^*$ is that $t^*< t_{\cal A}-a$. We remind the reader that in \eqref{eq:bulkcy} lines of constant $t$ and $x$ are radially ingoing null geodesics. Causality at the very least requires that the cap-off point of the extremal  surface lies below the ingoing null geodesic from the domain of dependence. Since for the strip region the boundary domain of dependence is a diamond anchored at $(t_{\cal A}\pm a,0)$ and $(0,\pm a)$, we note that the ingoing light ray from the bottom tip of this diamond cannot signal to the cap-off point.  
\end{enumerate}

\begin{figure}
\begin{minipage}[c][11cm][t]{.65\textwidth}
  \vspace*{\fill}
  \centering
  \includegraphics[width=\textwidth]{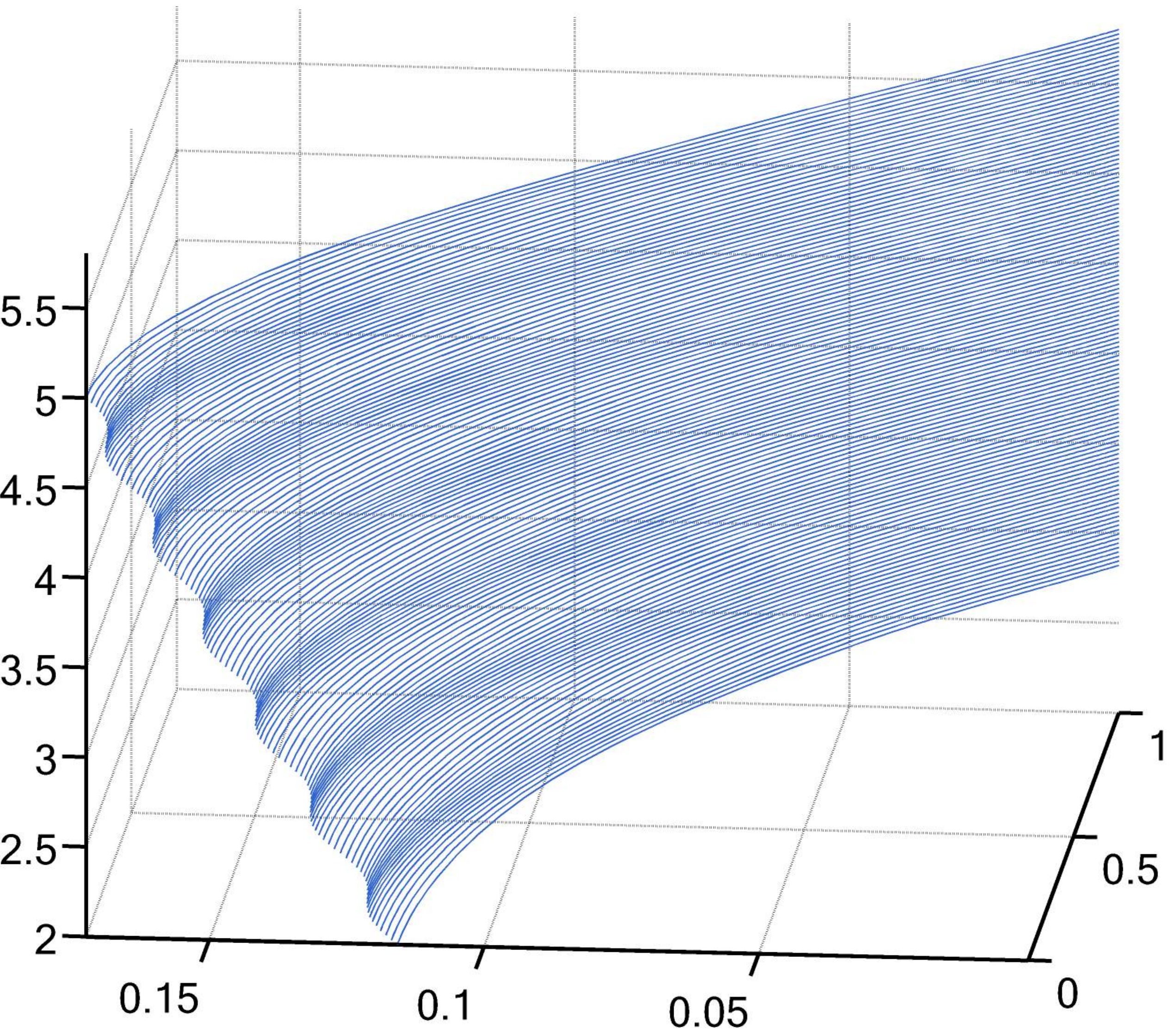}
        \put(-11,19){\makebox(0,0){\Large $\frac{x}{a}$}}
        \put(-140,-10){\makebox(0,0){\large $u$}}
        \put(-290,100){\makebox(0,0){\Large $\frac{t}{\per}$}}
\end{minipage}
\begin{minipage}[c][11cm][t]{.3\textwidth}
  \vspace*{\fill}
  \centering
  \includegraphics[width=\textwidth]{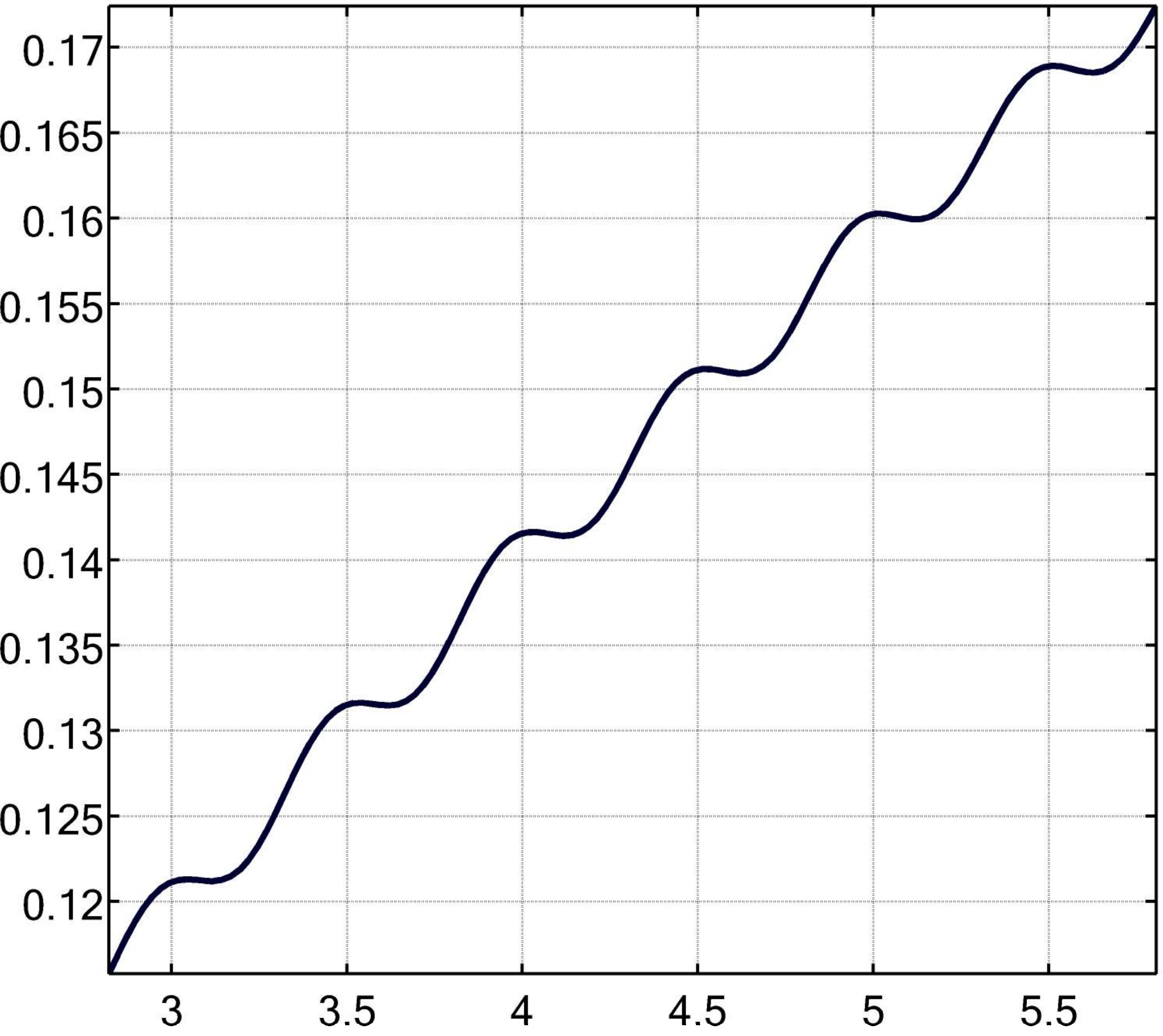}
        \put(-105,105){\makebox(0,0){\large $u^*$}}
\vspace{0.2cm}
  \includegraphics[width=\textwidth]{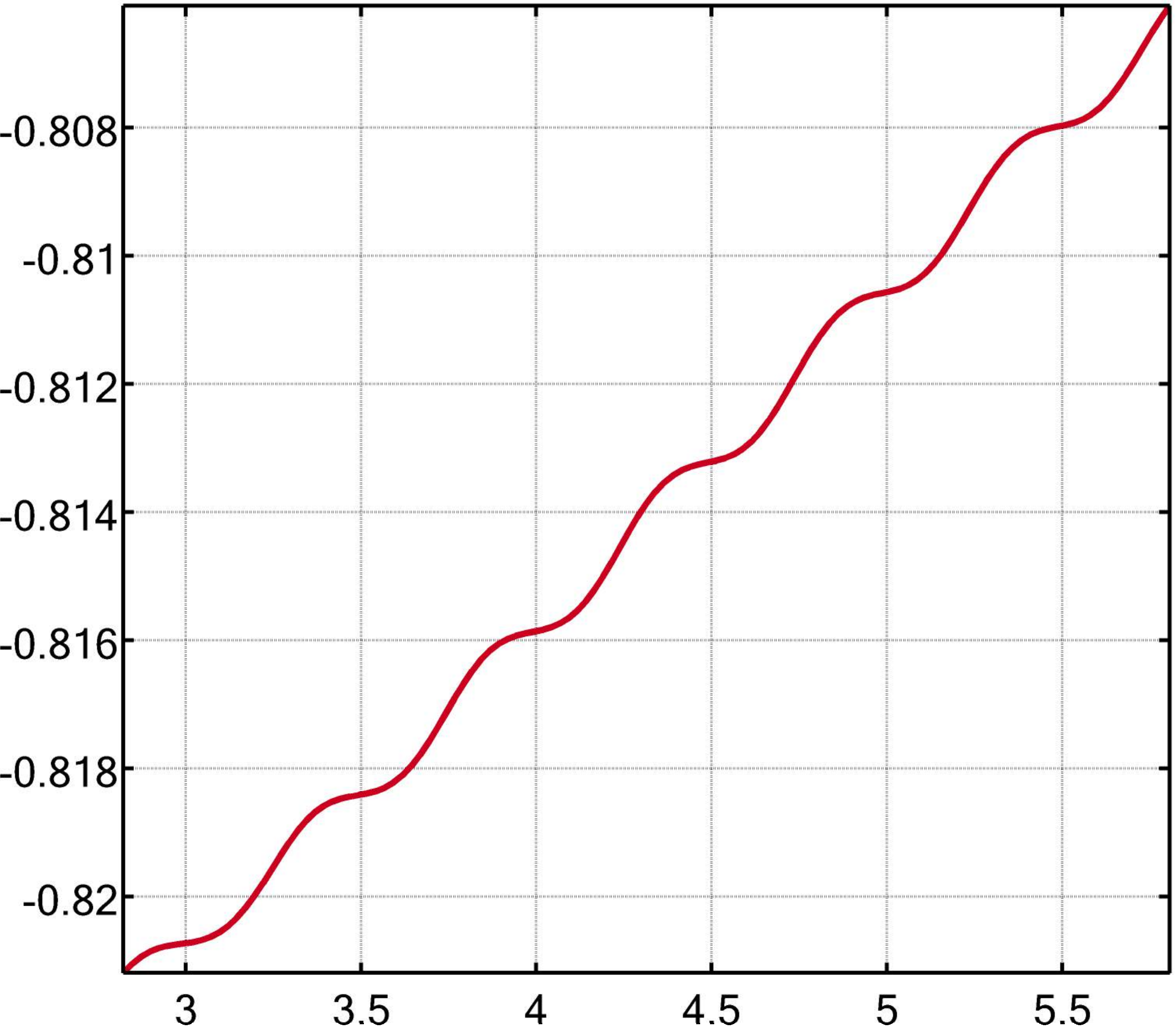}
        \put(-103,99){\makebox(0,0){\normalsize $\tilde{t}^*$}}
        \put(-56,-8.0){\makebox(0,0){\normalsize $t_{\cal A}/\per$}}
\end{minipage}
\caption{Evolution of the extremal surfaces for a strip of width $a=0.05$ with  driving parameters $\xi(\per=0.1,\amp=1)=0.1$ (phase I; dissipation-dominated). We pick a  UV cutoff $u_{\cal A}=10^{-3}$ and have defined $\tilde{t}^* \equiv (t^*-t_{\cal A})/\per$ to measure the cap-off $t^*$ point relative to the boundary.}
\label{fig:ES_evolution_per01_amp1}
\end{figure}

\begin{figure}
\begin{minipage}[c][11cm][t]{.65\textwidth}
  \vspace*{\fill}
  \centering
  \includegraphics[width=\textwidth]{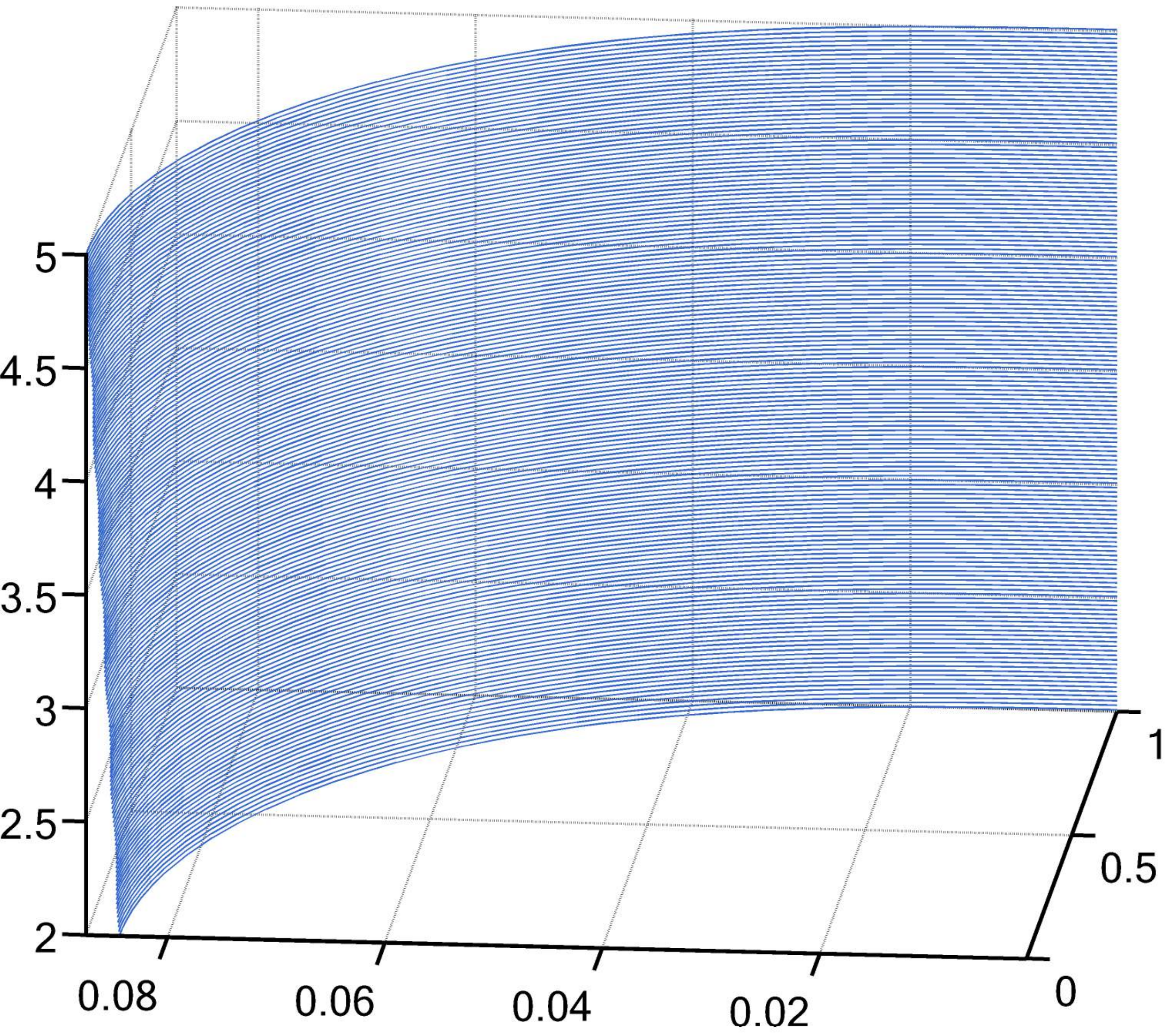}
        \put(-11,19){\makebox(0,0){\Large $\frac{x}{a}$}}
        \put(-140,-10){\makebox(0,0){\large $u$}}
        \put(-290,100){\makebox(0,0){\Large $\frac{t}{\per}$}}
\end{minipage}
\begin{minipage}[c][11cm][t]{.3\textwidth}
  \vspace*{\fill}
  \centering
  \includegraphics[width=\textwidth]{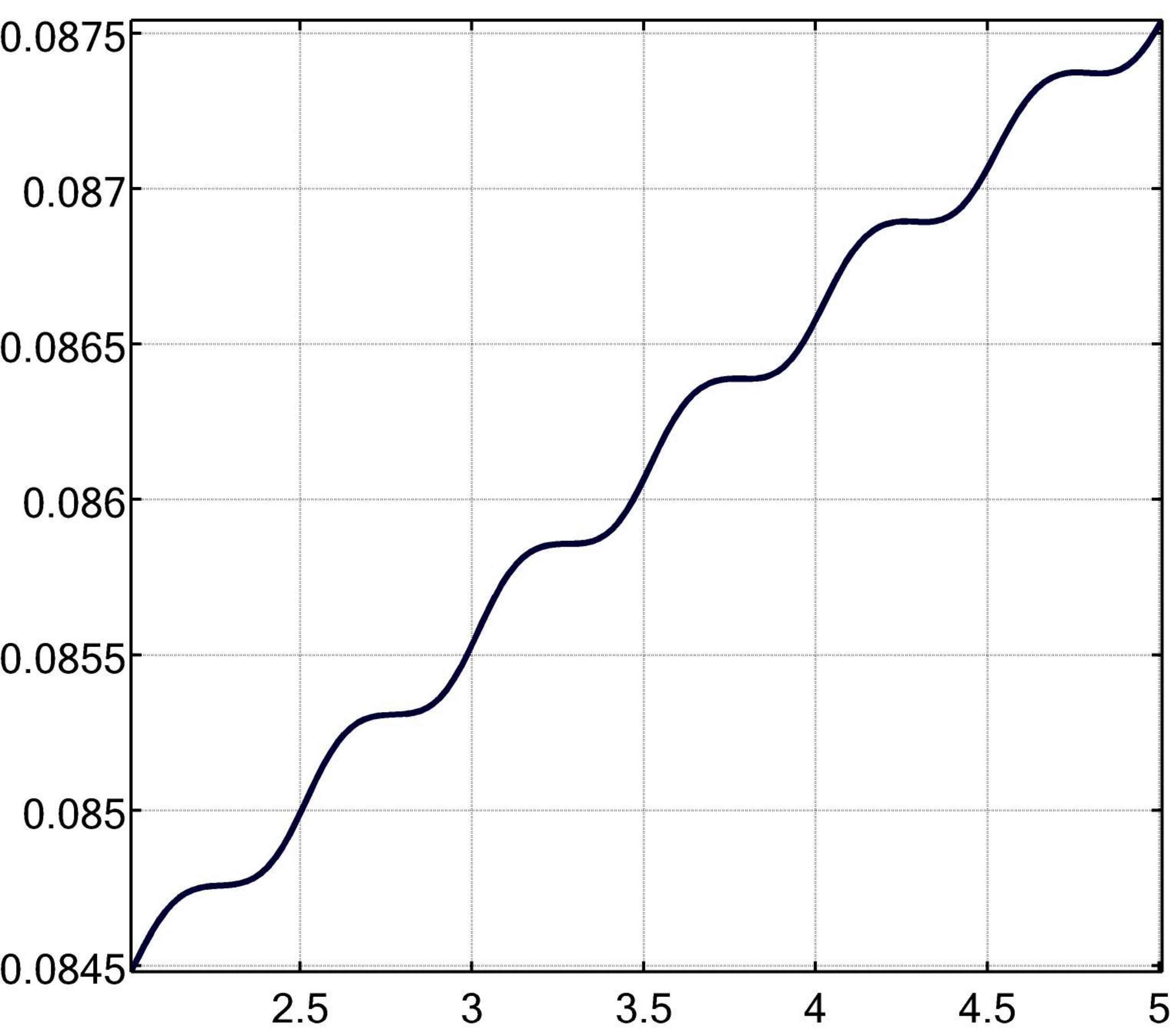}
        \put(-105,105){\makebox(0,0){\large $u^*$}}
\vspace{0.2cm}
  \includegraphics[width=\textwidth]{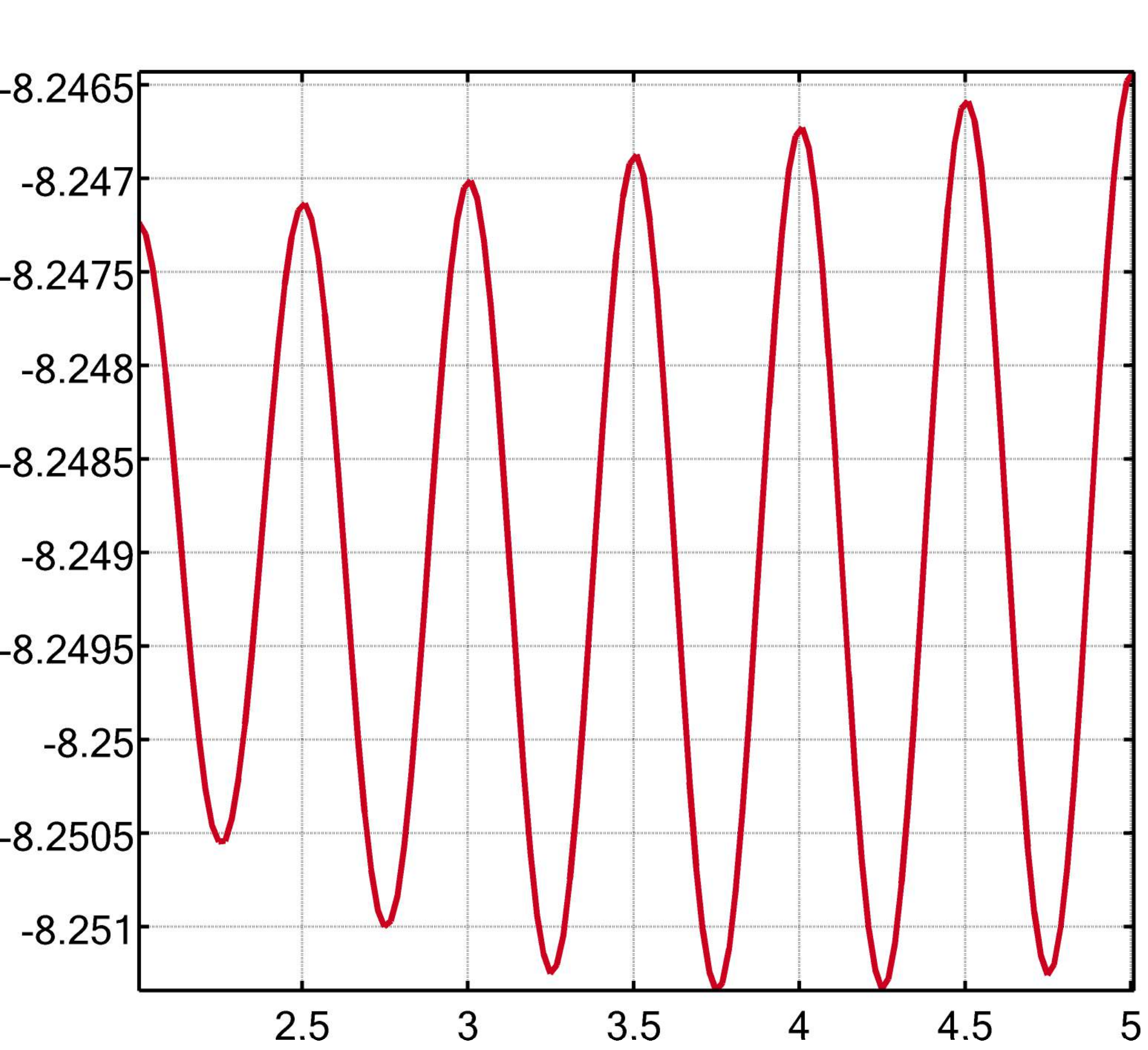}
        \put(-103,99){\makebox(0,0){\normalsize $\tilde{t}^*$}}
        \put(-116,114){\makebox(0,0){\tiny $\times 10^{-3}$}}
        \put(-56,-8.0){\makebox(0,0){\normalsize $t_{\cal A}/\per$}}
\end{minipage}
\caption{Evolution of the extremal surfaces for a strip of width $a=0.05$ with driving parameters $\xi(\per=10,\amp=1)=10$ (phase II; tilted).
Conventions described in Fig.~\ref{fig:ES_evolution_per01_amp1} apply.}
\label{fig:ES_evolution_per10_amp1}
\end{figure}

\begin{figure}
\begin{minipage}[c][11cm][t]{.65\textwidth}
  \vspace*{\fill}
  \centering
  \includegraphics[width=\textwidth]{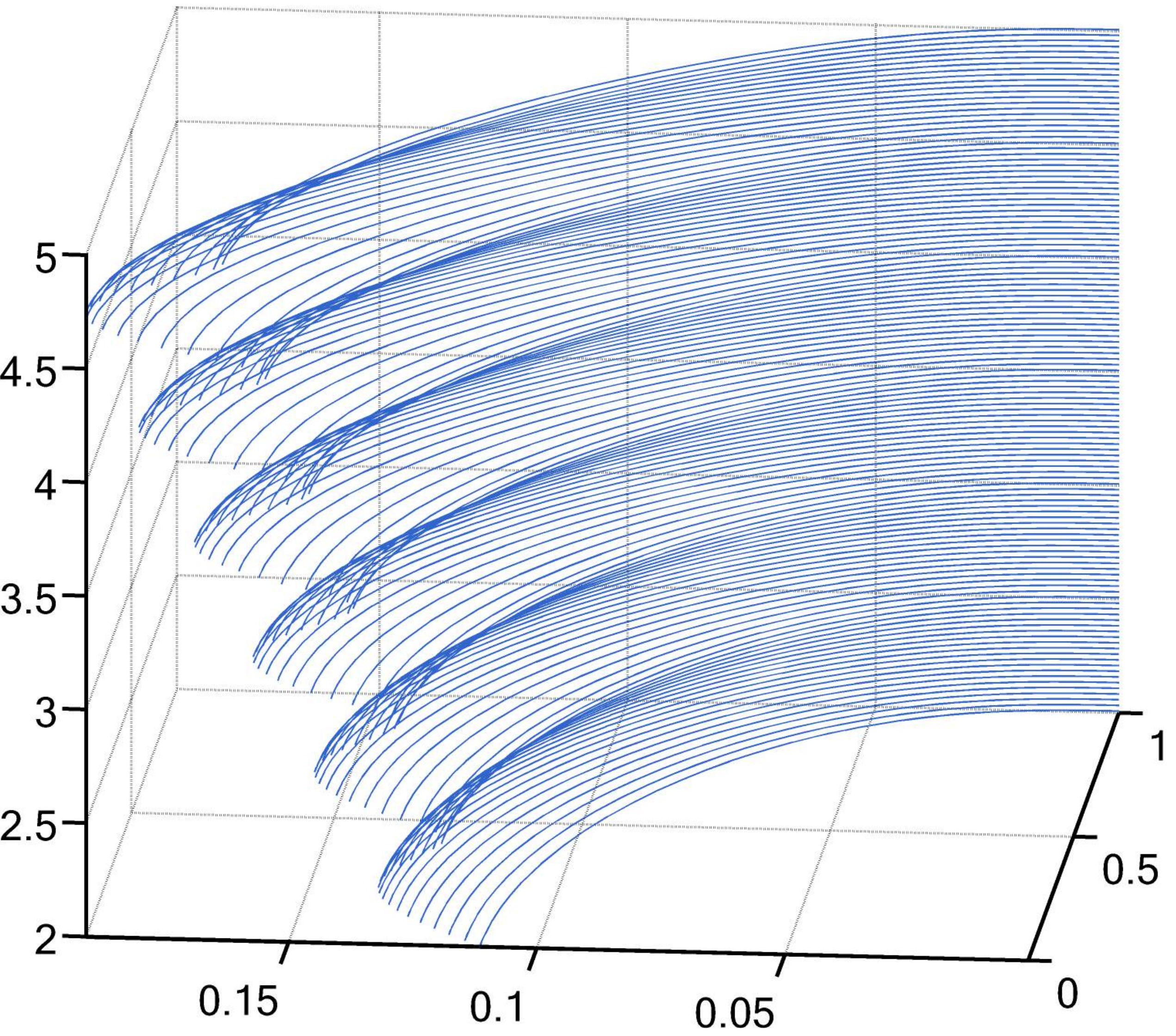}
        \put(-11,19){\makebox(0,0){\Large $\frac{x}{a}$}}
        \put(-140,-10){\makebox(0,0){\large $u$}}
        \put(-290,100){\makebox(0,0){\Large $\frac{t}{\per}$}}
\end{minipage}
\begin{minipage}[c][11cm][t]{.289\textwidth}
  \vspace*{\fill}
  \centering
  \includegraphics[width=\textwidth]{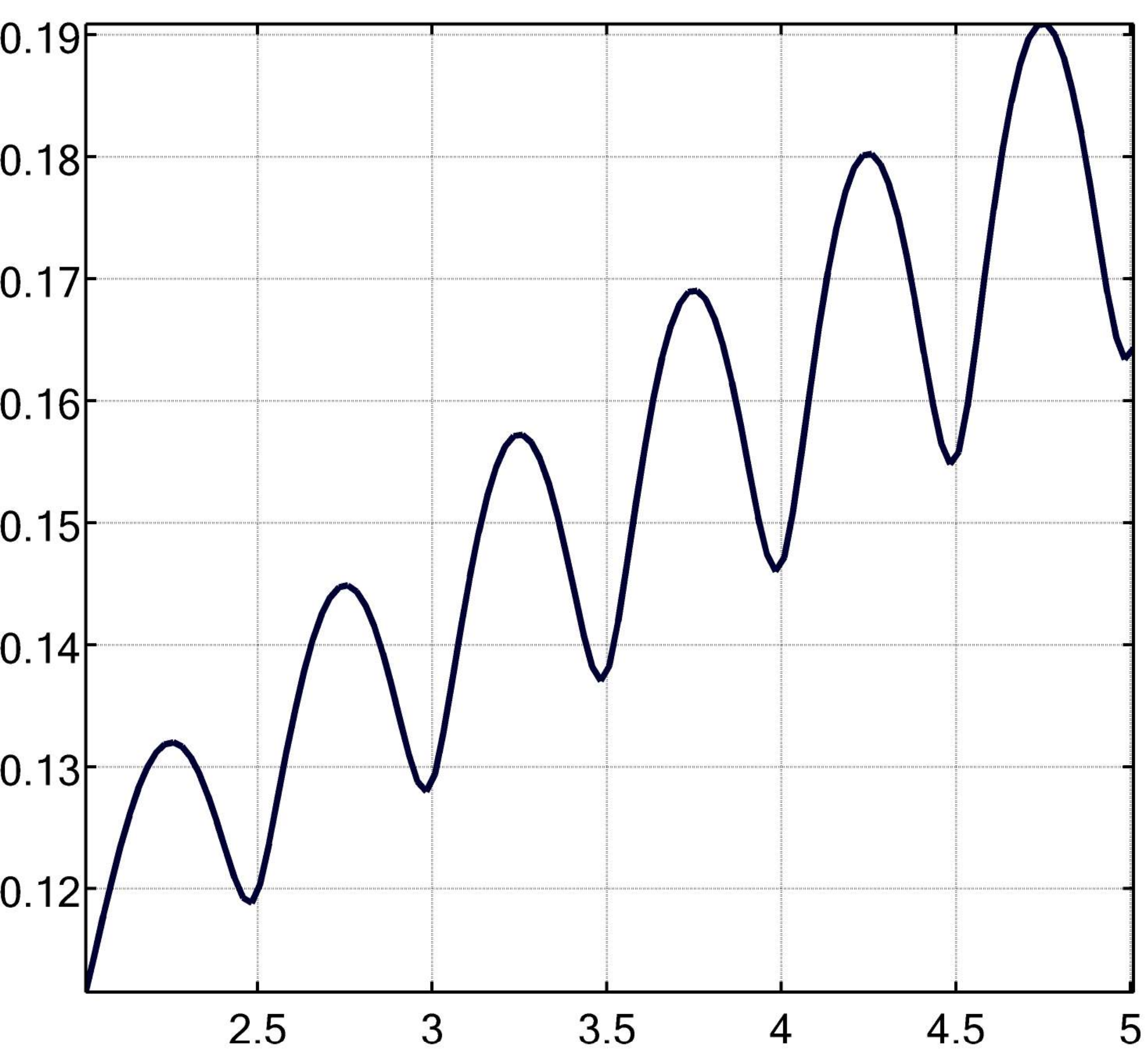}
        \put(-105,105){\makebox(0,0){\large $u^*$}}
\vspace{0.2cm}
  \includegraphics[width=\textwidth]{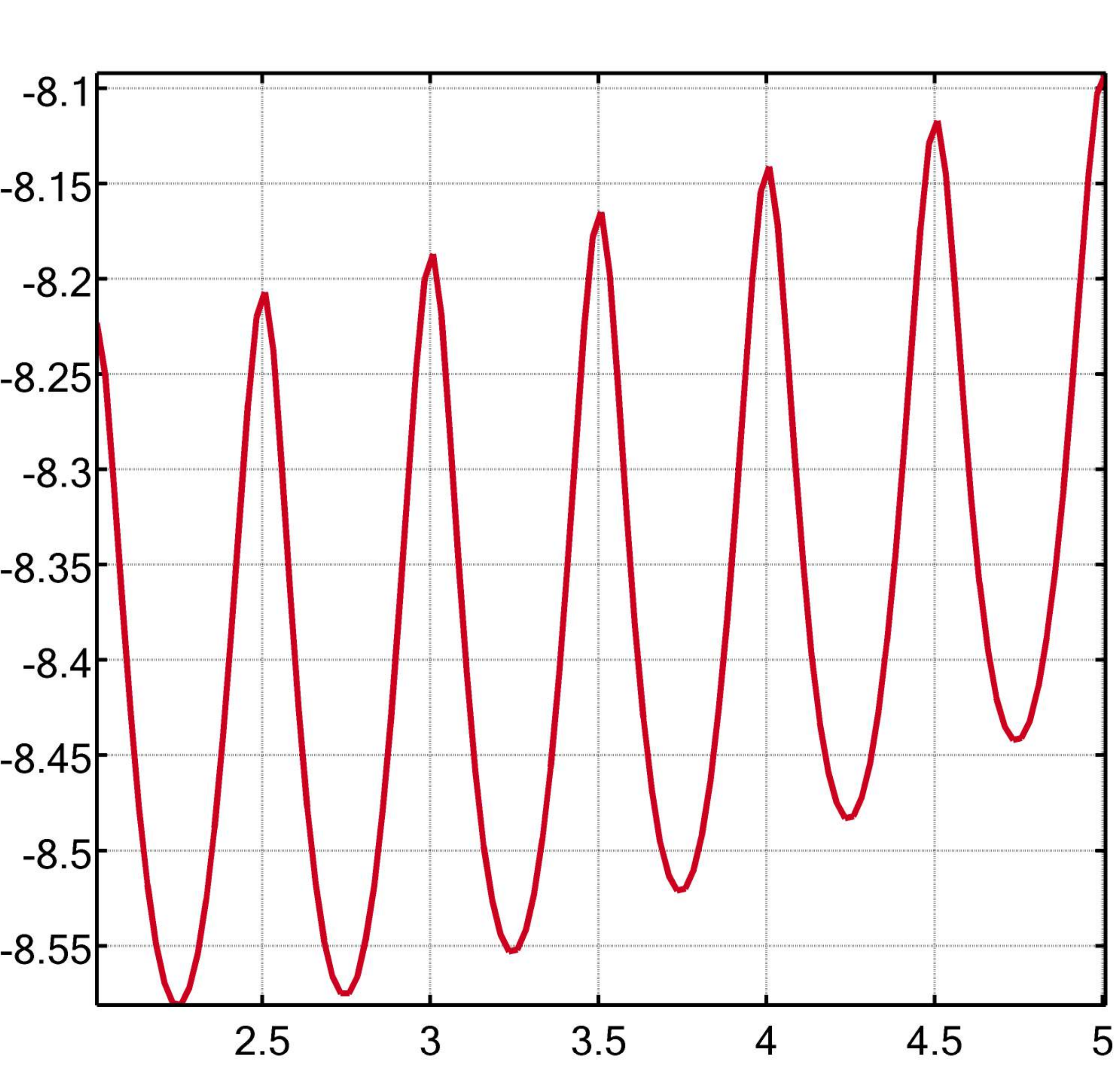}
        \put(-103,99){\makebox(0,0){\normalsize $\tilde{t}^*$}}
        \put(-110,114){\makebox(0,0){\tiny $\times 10^{-3}$}}
        \put(-56,-8.0){\makebox(0,0){\normalsize $t_{\cal A}/\per$}}
\end{minipage}
\caption{
Evolution of the extremal surfaces for a strip of width $a=0.05$ with driving parameters $\xi(\per=10,\amp=20)=200$ (phase III; unbounded amplification). 
Conventions described in Fig.~\ref{fig:ES_evolution_per01_amp1} apply.}
\label{fig:ES_evolution_per10_amp20}
\end{figure}

\begin{figure}
\begin{minipage}[c][11cm][t]{.65\textwidth}
  \vspace*{\fill}
  \centering
  \includegraphics[width=\textwidth]{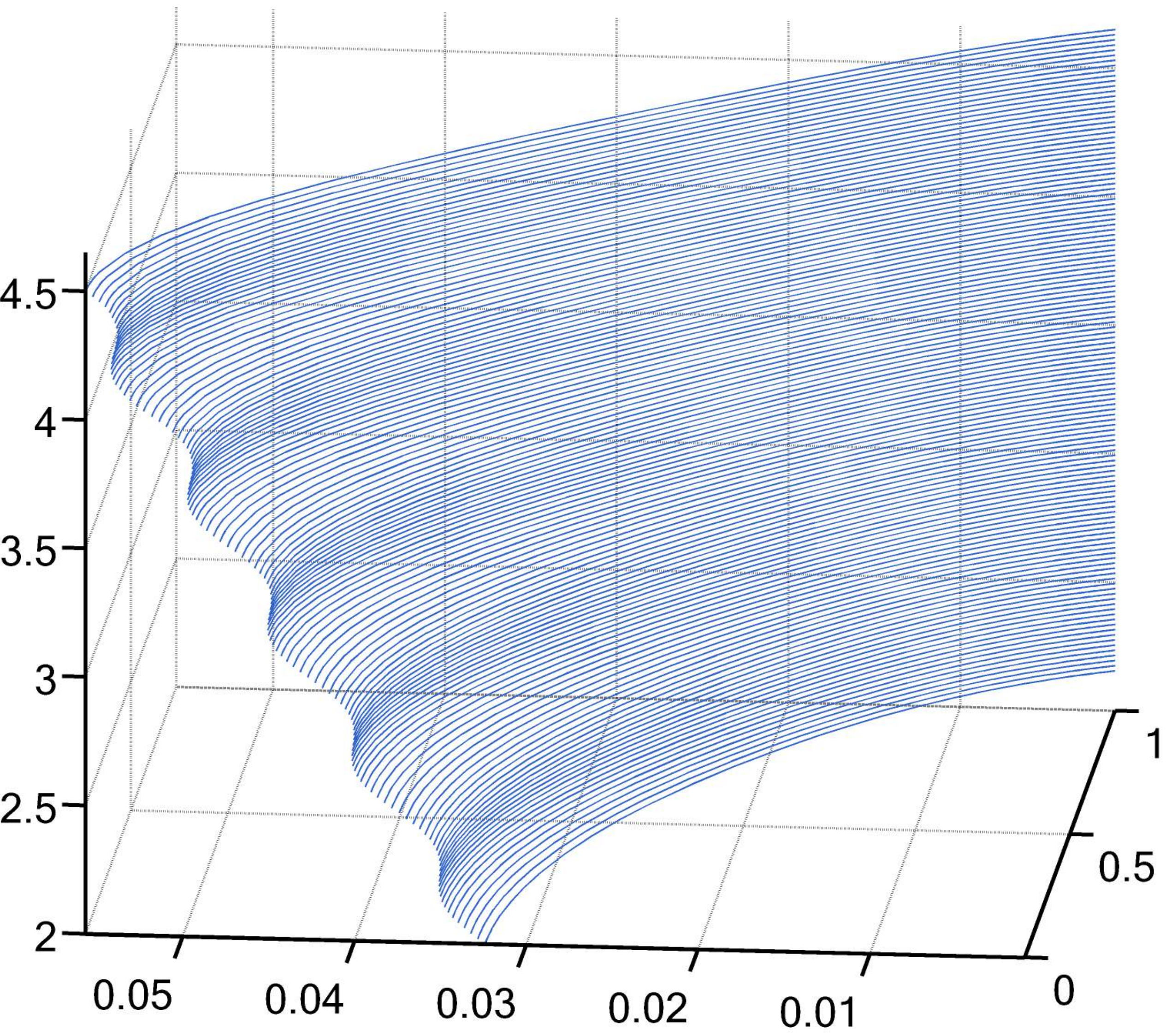}
        \put(-11,19){\makebox(0,0){\Large $\frac{x}{a}$}}
        \put(-140,-10){\makebox(0,0){\large $u$}}
        \put(-290,100){\makebox(0,0){\Large $\frac{t}{\per}$}}
\end{minipage}
\begin{minipage}[c][11cm][t]{.3\textwidth}
  \vspace*{\fill}
  \centering
  \includegraphics[width=\textwidth]{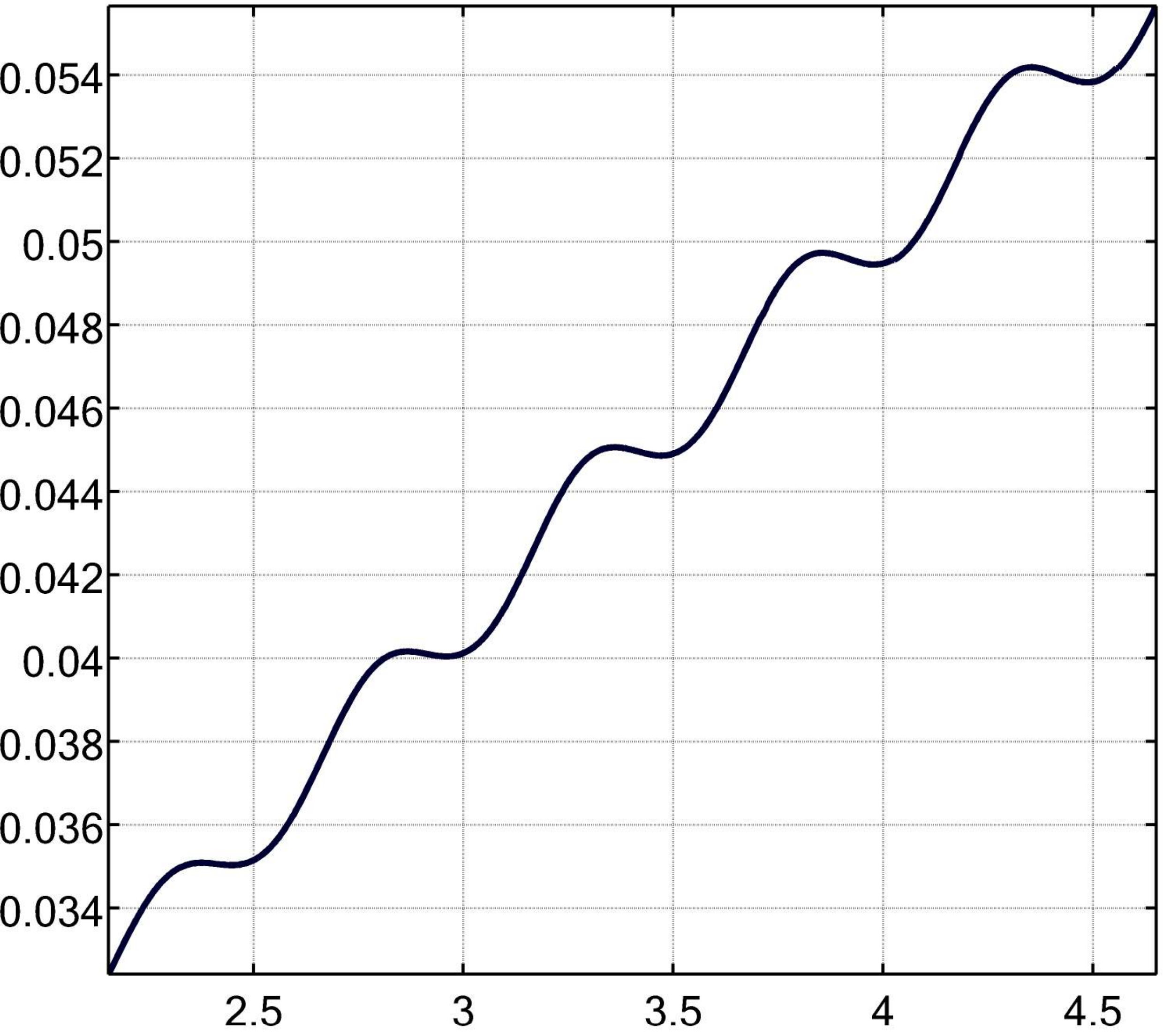}
        \put(-105,105){\makebox(0,0){\large $u^*$}}
\vspace{0.2cm}
  \includegraphics[width=\textwidth]{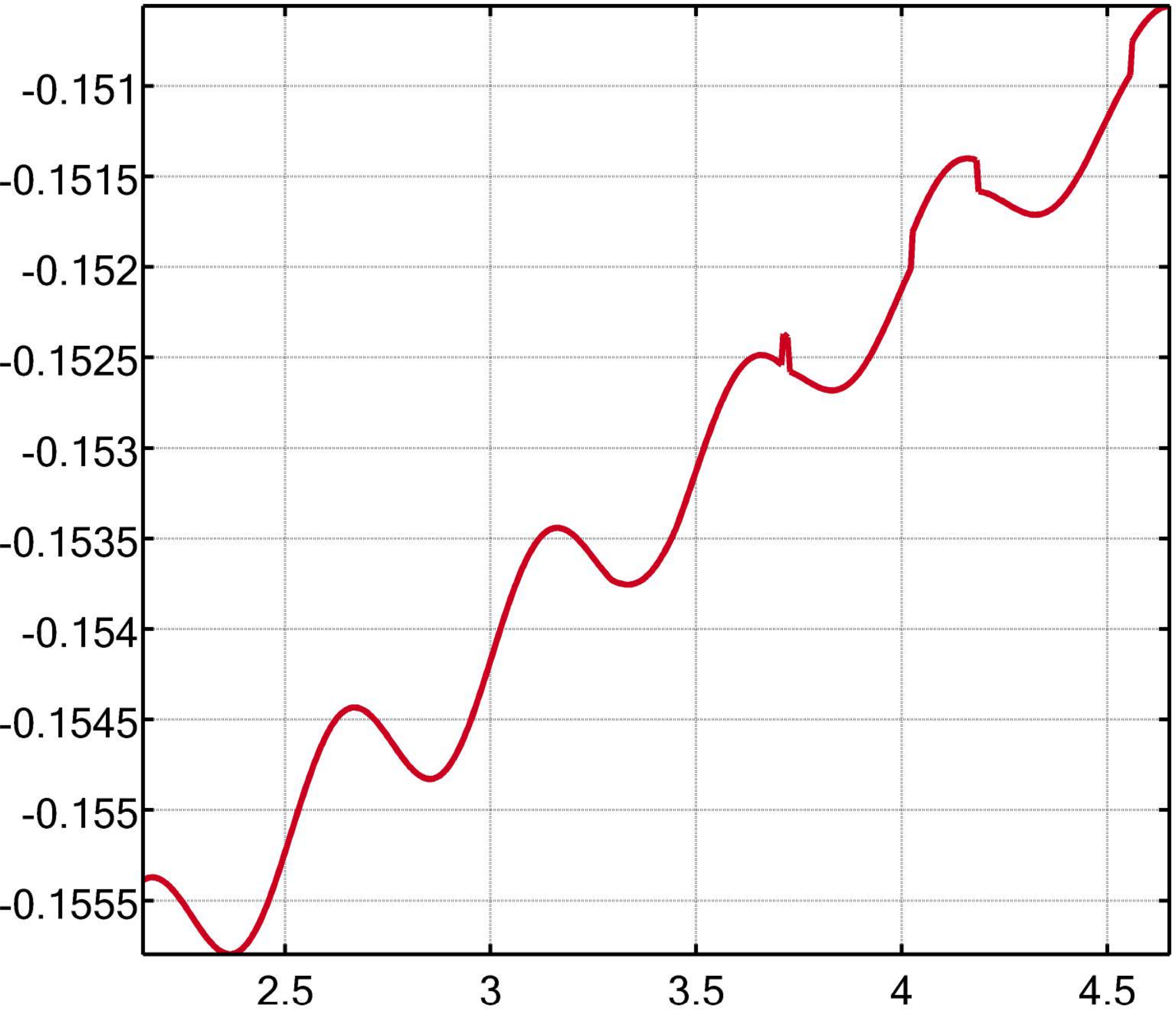}
        \put(-103,99){\makebox(0,0){\normalsize $\tilde{t}^*$}}
        \put(-56,-8.0){\makebox(0,0){\normalsize $t_{\cal A}/\per$}}
\end{minipage}
\caption{Evolution of the extremal surfaces for a strip of width $a=0.01$ with driving parameters $\xi(\per=0.1,\amp=20)=2$ (phase IIb; dynamical crossover). Conventions are as described in Fig.~\ref{fig:ES_evolution_per01_amp1}.}
\label{fig:ES_evolution_per01_amp20}
\end{figure}

In the following discussion we will illustrate the behaviour of the extremal surfaces more explicitly in each of our phases. We have been reasonably conservative in our analysis and have chosen to work only with surfaces that do not get too close to the horizon (in fact $u^* <0.2$). This is to avoid both numerical issues as well as to avoid complications from the existence of multiple extremal surfaces. We follow a single branch of solutions as described at the end of  Appendix \ref{sec:extrdet}. The primary results of the extremal surfaces are shown in the plots   Figs.~\ref{fig:ES_evolution_per01_amp1}, \ref{fig:ES_evolution_per10_amp1}, \ref{fig:ES_evolution_per10_amp20},   and \ref{fig:ES_evolution_per01_amp20}, where we show the evolution of the extremal surface as well   as $u^*(t_{\cal A})$ and $t^*(t_{\cal A})$.

%
%

\paragraph{Linear regime (small $\amp$):}

Although all phases display extremal surfaces that sink into the bulk with each driving cycle, the growth of $u^*$ in the linear regime of small amplitudes is most steady. We focus here on phases I (high frequency; dissipation-dominated) illustrated in Fig.~\ref{fig:ES_evolution_per01_amp1} 
and IIa (low frequency; tilted) illustrated in Fig.~\ref{fig:ES_evolution_per10_amp1},  which fall under this characterization.  As the frequency is lowered and we pass from the dissipation-dominated phase to the tilted phase, there is drastic reduction in the growth of $u^*$ per cycle. 

The evolution of $t^*$ in the two phases is also interesting; $t^*-t_{\cal A}$ is gradually increasing on average with time 
(recall that in the stationary geometry $t^*-t_{\cal A}$ would be constant). It turns out to be useful to look at  
a dimensionless parameter $\tilde{t}^* \equiv (t^*-t_{\cal A})/\per$ which measures the cap-off time relative to the boundary.  In this context, there is more time-lag in phase I i.e., $\tilde{t}^*_{\text{\tiny{I}}} \ll \tilde{t}^*_{\text{\tiny{IIa}}} \lesssim 0$,  which hints at the cause for why the surfaces do not penetrate as far deep in the bulk in phase IIa as opposed to phase I.\footnote{Note that in absolute terms however, $t^*$ in both regimes is comparable in magnitude.}
In addition we see strong oscillatory patterns in phase II in spite of having only a steady increase in $u^*$; such a feature is absent in phase I.

%
%
%

\paragraph{Non-linear regime (large $\amp$):}

We now turn to the phases III (low frequency; unbounded amplification) illustrated in 
Fig.~\ref{fig:ES_evolution_per10_amp20} and IIb (high frequency; wobbly) illustrated in
Fig.~\ref{fig:ES_evolution_per01_amp20} in the non-linear regime of high amplitude. Some of the features seen in the linear regime continue to pertain: we see more  pronounced oscillations in $\tilde{t}^*$ and a decreased tendency for the surfaces to lag behind in time at lower frequencies. 

In the unbounded amplification regime (phase III), we see significant bursts of growth of the extremal surfaces. The oscillatory driving is felt rather acutely by the surfaces and the evolution is considerably violent. On average however, $u^*$ appears to advance more serenely despite having large amplitude oscillations per cycle.

In the dynamical crossover wobbly regime (phase IIb), there is a considerable amount of instability. We chose here to work with smaller strip widths   $a=0.01$ (instead of $a=0.05$) to avoid complications of phase transitions between multiple competing extremal surfaces. 
The early part of the evolution is in line with what happens in the dissipation-dominated regime (phase I), but shortly after, there are discontinuities in the $\tilde{t}^*$ parameter with no noticeable effect in $u^*$. 
Around $t_{\cal A}/P \approx 4.0 - 4.2$ and $t_{\cal A}/P \approx 4.6$, we see an exchange of dominance in the extremal surface, which starts out at a higher value of $\tilde{t}^*$.

All in all, the extremal surfaces in the non-linear regime definitely has elements of intrigue owing to the large pulses of energy that affect the bulk geometry significantly.  Although we do not delve into extremal surfaces that are positioned deeper into the bulk, we notice in the course of our analysis that the surfaces tend towards the horizon as expected. More curiously, we also find that for larger regions we cannot find extremal surfaces that stay outside the apparent horizon. This is not surprising since we expect based on earlier results that there will be surfaces that penetrate the apparent horizon of the black hole (cf., \cite{AbajoArrastia:2010yt}). However, one of the disadvantages of our numerical scheme is that we are unable to explore this interesting regime due to the fact that the spacetime inside the apparent horizon has been excised. As explained in \cite{Chesler:2013lia}, this was to avoid complications with having caustics in the coordinate chart. Analysis of entanglement entropy however does require us to have the complete bulk geometry. 

\subsection{The evolution  of entanglement}
\label{sec:eegrow}
\begin{figure}
        \centering
        \begin{subfigure}[b]{0.42\textwidth}
                \includegraphics[width=\textwidth]{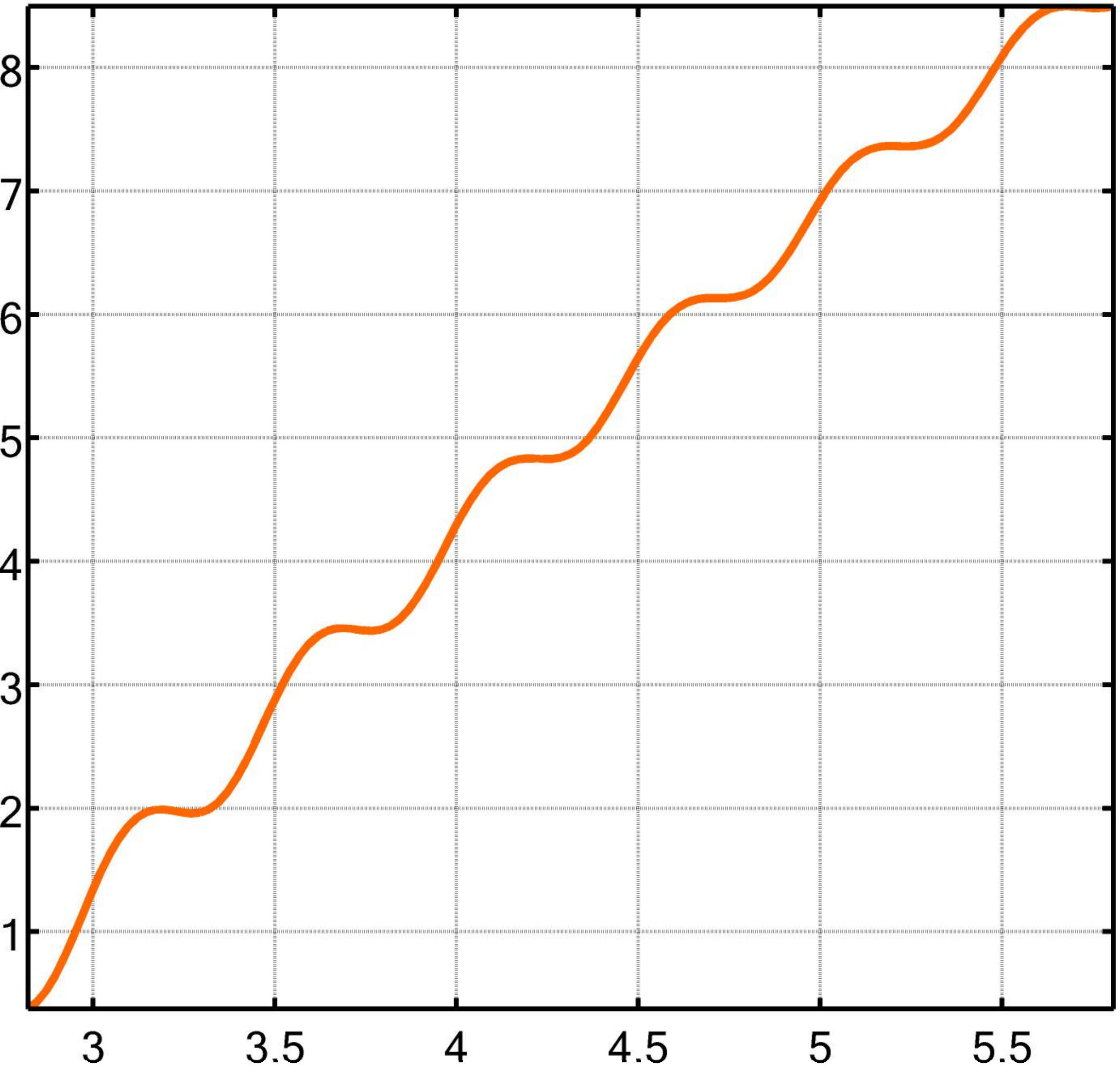}
                \put(-82,-7){\makebox(0,0){\normalsize $t_{\cal A}/\per$}}
                \put(-191,92){\makebox(0,0){\rotatebox{0}{$\sreg$}}}
                \caption{Phase I: $\xi(\amp=1,\per=0.1)=0.1$}
                \label{subfig:EE_reg_phase1}
        \end{subfigure}
        \qquad 
        \begin{subfigure}[b]{0.438\textwidth}
                \includegraphics[width=\textwidth]{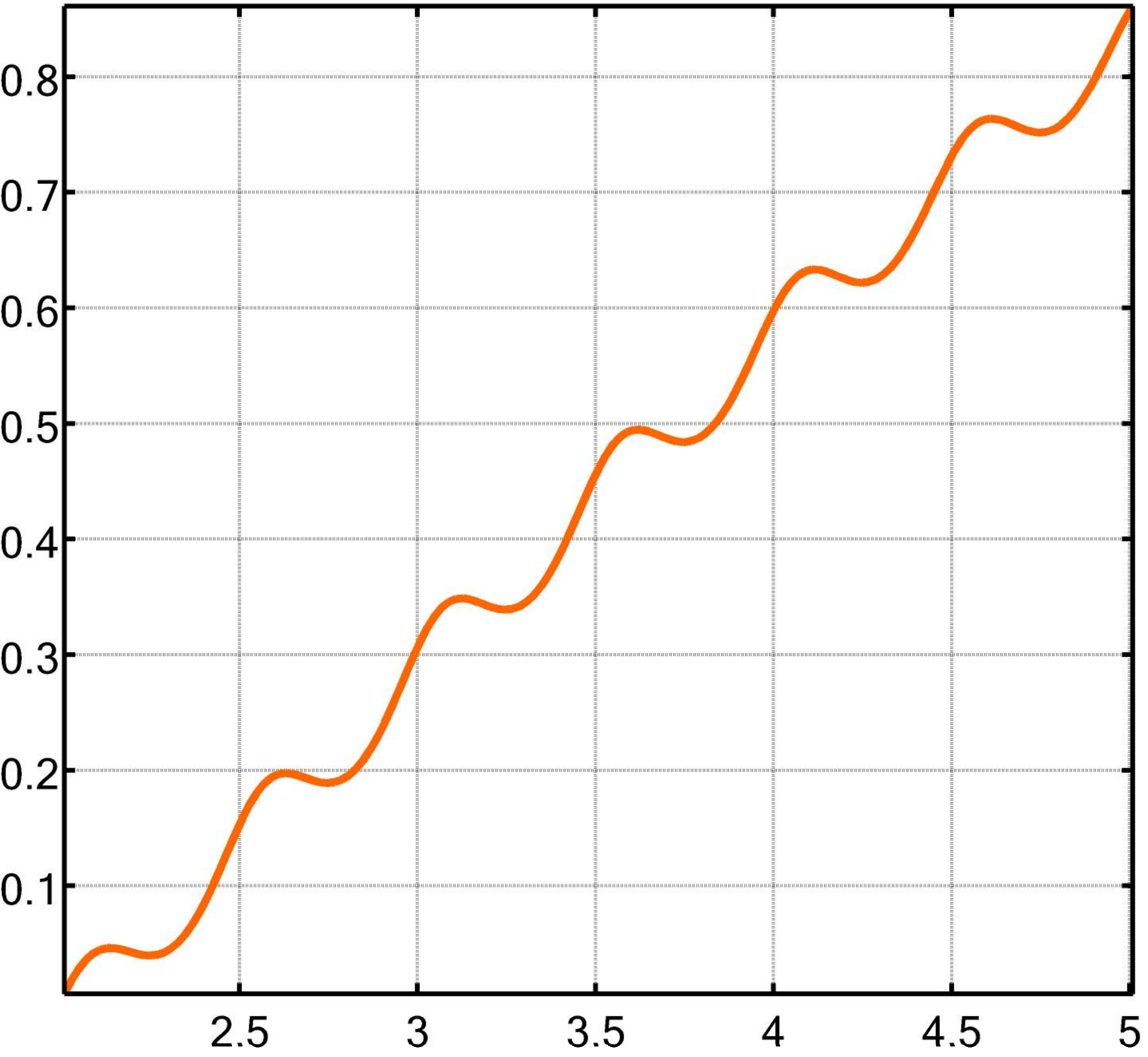}   
                \put(-82,-7){\makebox(0,0){\normalsize $t_{\cal A}/\per$}}
                \put(-198,92){\makebox(0,0){\rotatebox{0}{$\sreg$}}}
                \caption{Phase IIa: $\xi(\amp=1,\per=10)=10$}
                \label{subfig:EE_reg_phase2}
        \end{subfigure}

      \vspace{5mm}
        \begin{subfigure}[b]{0.42\textwidth}
                \includegraphics[width=\textwidth]{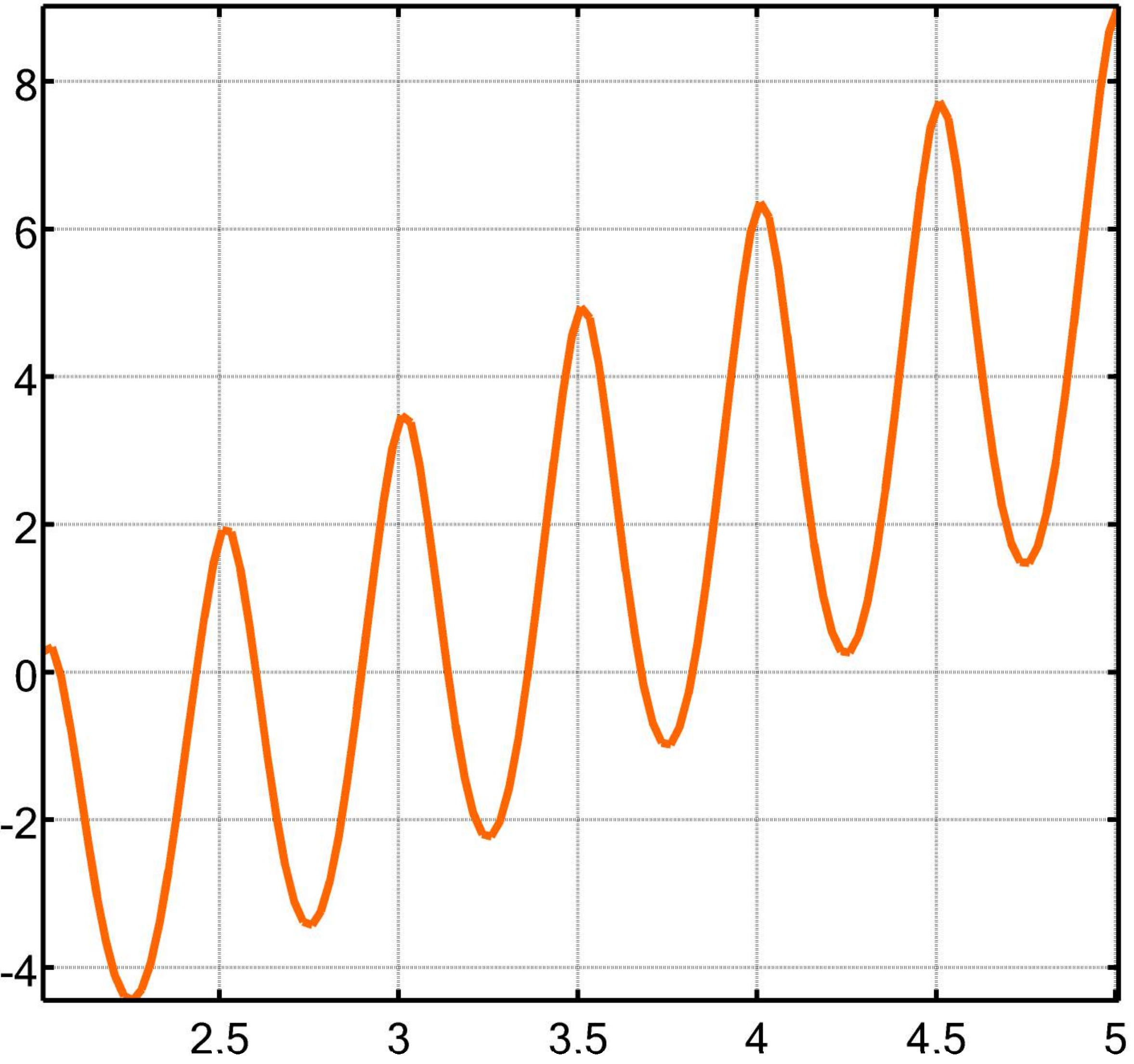}
                \put(-82,-7){\makebox(0,0){\normalsize $t_{\cal A}/\per$}}
                \put(-191,92){\makebox(0,0){\rotatebox{0}{$\sreg$}}}
                \caption{Phase III: $\xi(\amp=20,\per=10)=200$}
                \label{subfig:EE_reg_phase3}
        \end{subfigure}
        \qquad 
        \begin{subfigure}[b]{0.42\textwidth}
                \includegraphics[width=\textwidth]{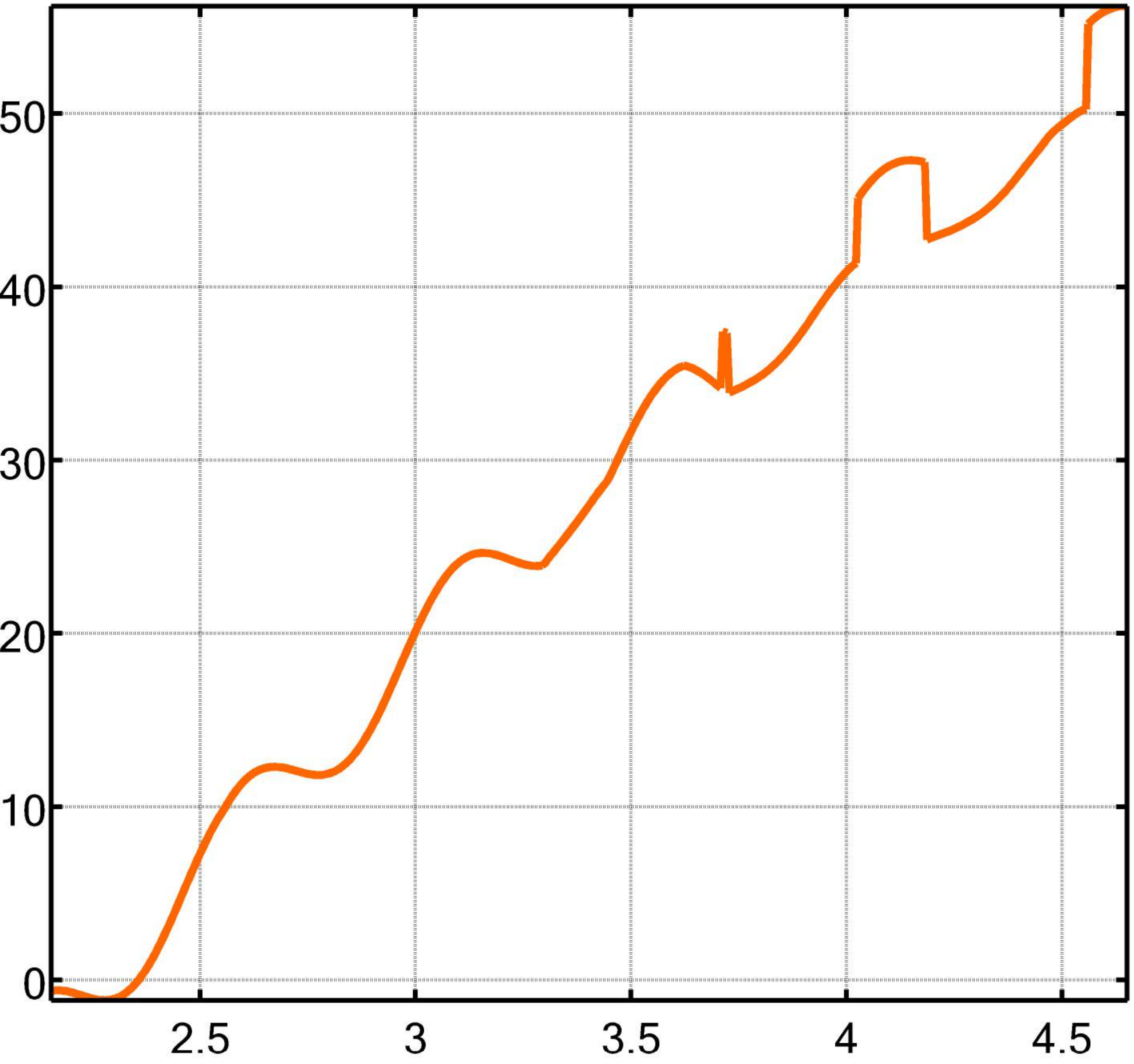}
                \put(-82,-7){\makebox(0,0){\normalsize $t_{\cal A}/\per$}}
                \put(-191,92){\makebox(0,0){\rotatebox{0}{$\sreg$}}}
                \caption{Phase IIb: $\xi(\amp=20,\per=0.1)=2$}
                \label{subfig:EE_reg_phase4}
        \end{subfigure}
        \caption{The evolution of the regularized entanglement entropy, $\sreg$ defined in Eq.~\eqref{eqn:S_reg}, for the four phases for a radial cutoff of $u_{\cal A} = 10^{-3}$. The strip widths are $a=0.05$ for panels (a), (b), (c), and $a=0.01$ for panel (d).}
        \label{fig:EE_unreg}
\end{figure}

We now turn to the evolution of the entanglement entropy; the results are presented in Fig~\ref{fig:EE_unreg} for the regulated quantity $\sreg$ as introduced in \eqref{eqn:S_reg}.

In the dissipation-dominated regime (phase I), the entanglement entropy gradually increases, though in each cycle of forcing there is a time period for which the growth is negligible.  We expect this feature is simply a consequence of the entanglement entropy tracking the thermal entropy. Even though we are not quite probing the full thermal contribution with the relatively small regions ${\cal A}$, it bears to reason that the  variation of the geometry is more or less equitable on all radial scales. This appears consistent with other probes of this phase. As we discussed in \S\ref{sec:dissdom} the weak driving allows the system to efficiently dissipate the energy induced by the source and the conductivity $\sigma(t)$ was purely imaginary. Basically the dominant effect here is the growth of the black hole horizon due to the driving and this in turn imprints itself into the growth of $\sreg$ seen in Fig.~\ref{subfig:EE_reg_phase1}.

\begin{figure}
        \centering
        \begin{subfigure}[b]{0.42\textwidth}
                \includegraphics[width=\textwidth]{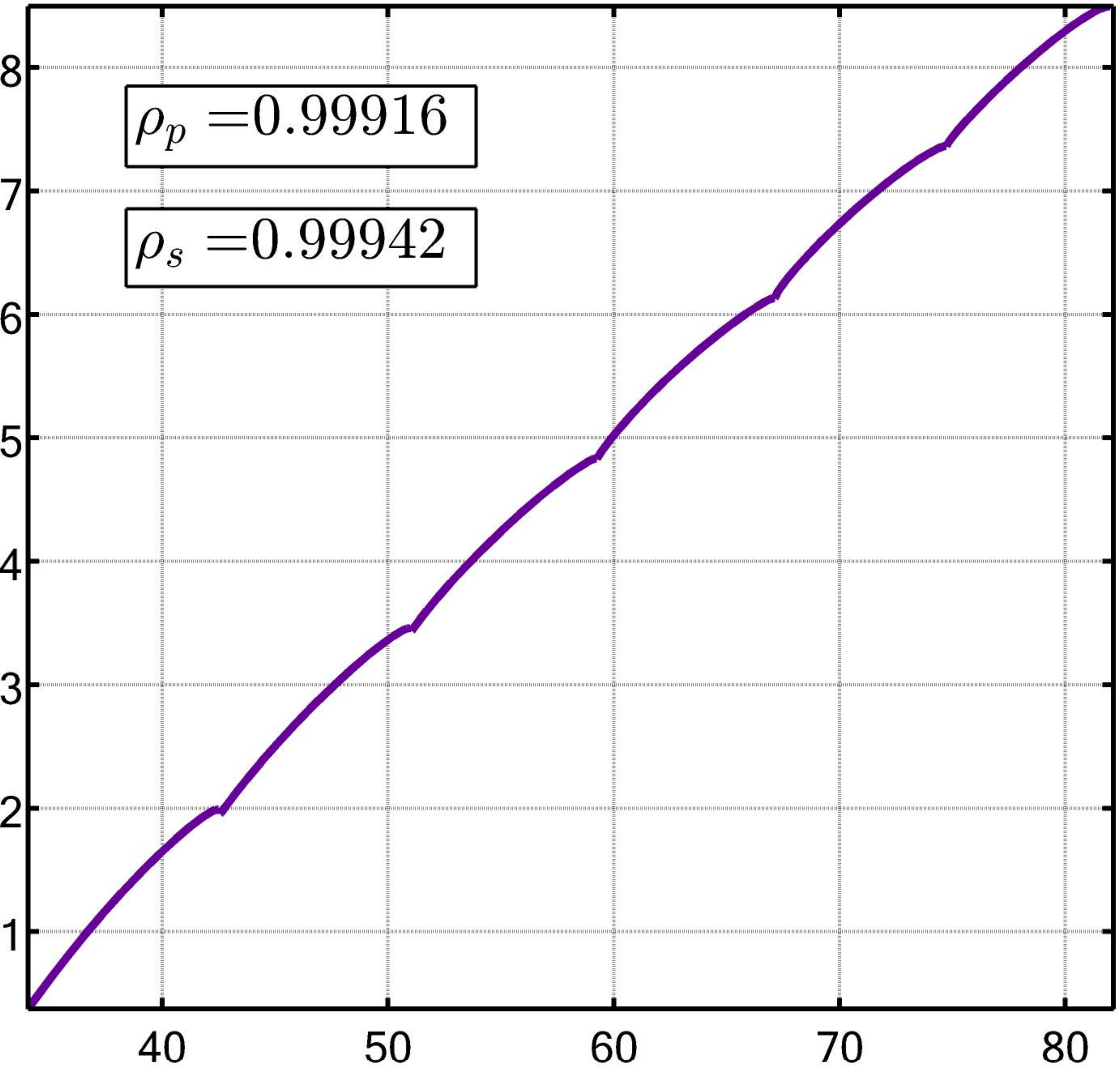}
                \put(-82,-7){\makebox(0,0){\normalsize $s/s_0$}}
                \put(-191,92){\makebox(0,0){\rotatebox{0}{$\sreg$}}}
                \caption{Phase I: $\xi(\amp=1,\per=0.1)=0.1$}
                \label{subfig:EE_regS_phase1}
        \end{subfigure}
        \qquad 
        \begin{subfigure}[b]{0.438\textwidth}
                \includegraphics[width=\textwidth]{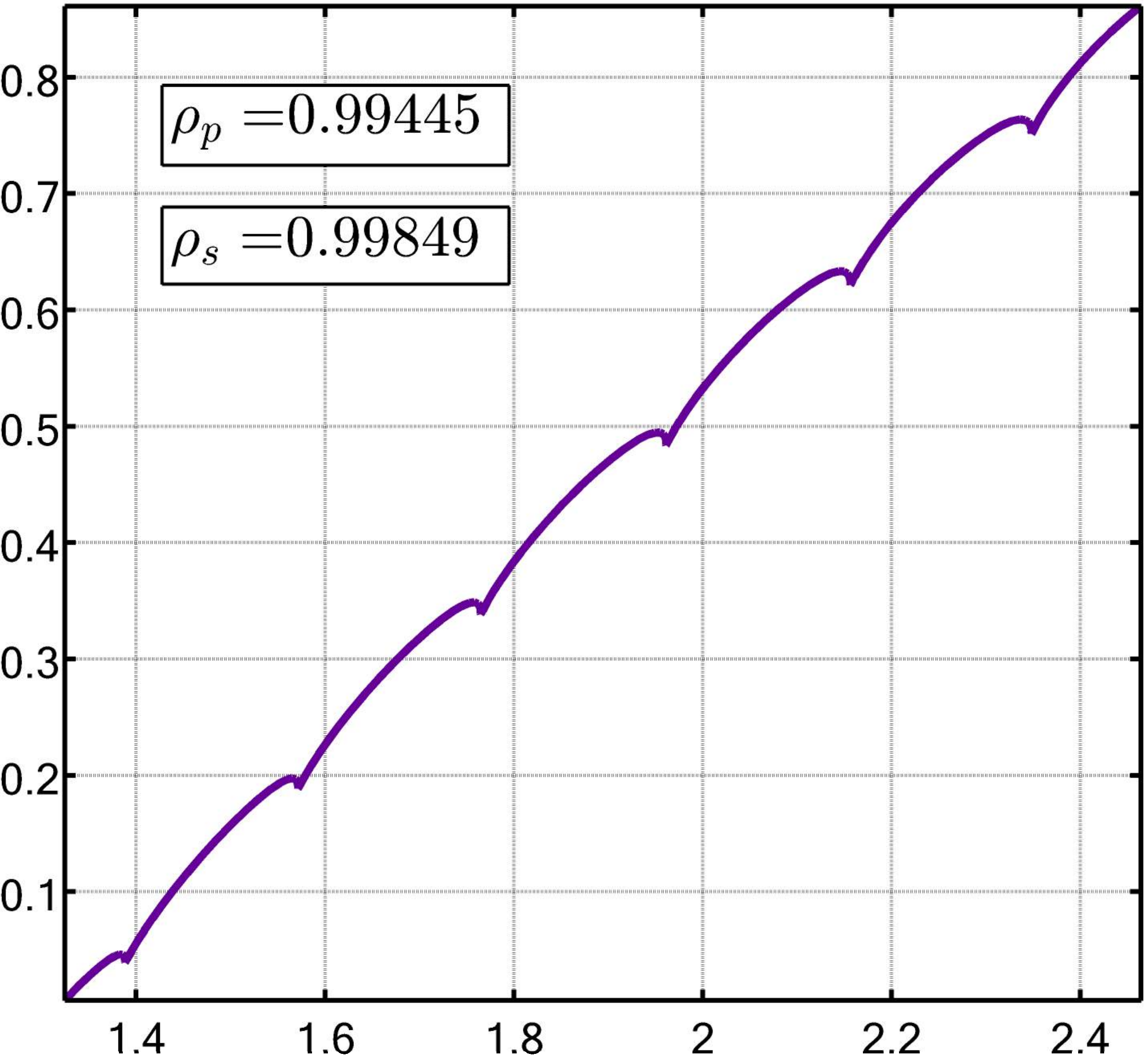}   
                \put(-82,-7){\makebox(0,0){\normalsize $s/s_0$}}
                \put(-198,92){\makebox(0,0){\rotatebox{0}{$\sreg$}}}
                \caption{Phase IIa: $\xi(\amp=1,\per=10)=10$}
                \label{subfig:EE_regS_phase2}
        \end{subfigure}

      \vspace{5mm}
        \begin{subfigure}[b]{0.42\textwidth}
                \includegraphics[width=\textwidth]{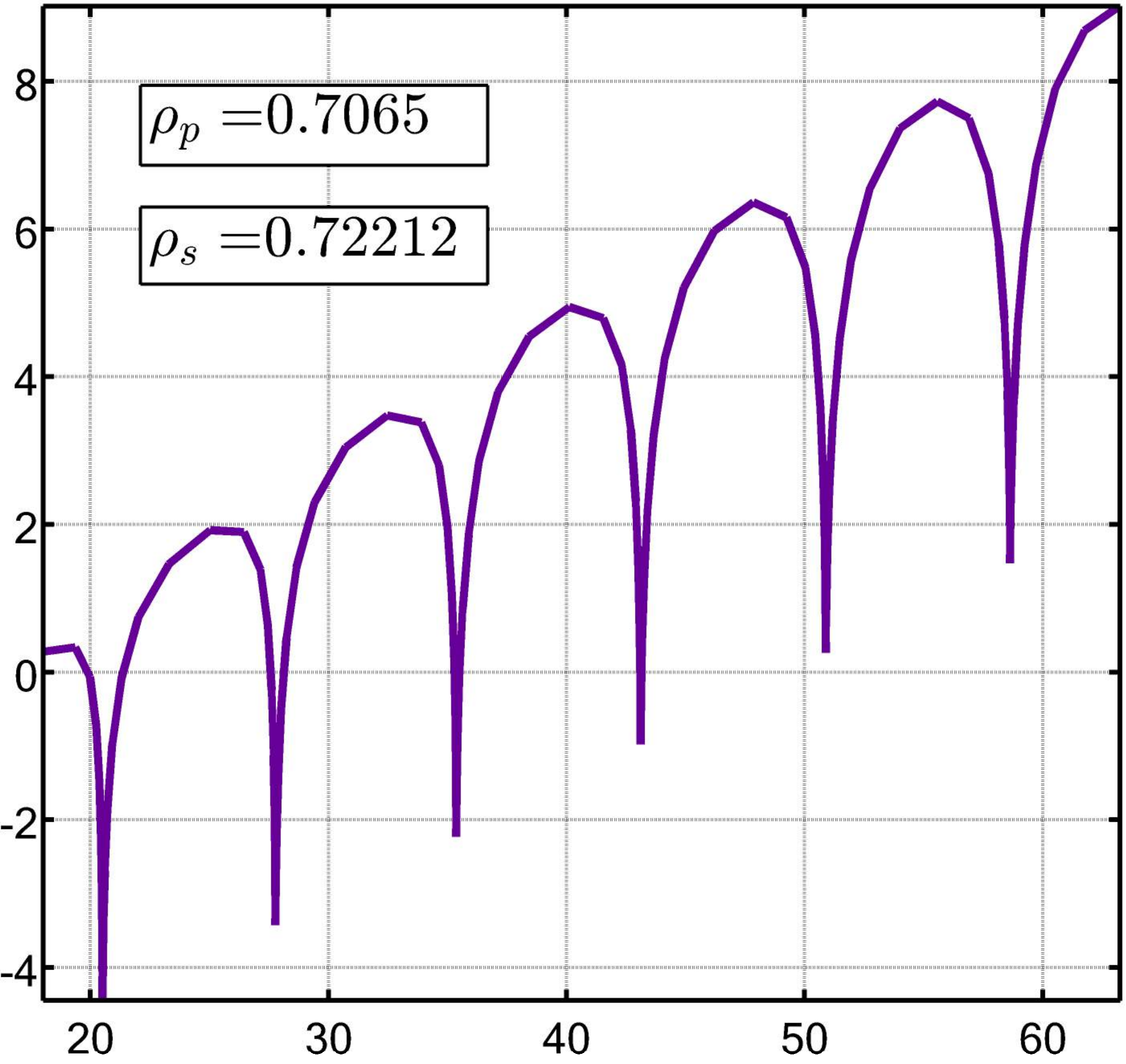}
                \put(-82,-7){\makebox(0,0){\normalsize $s/s_0$}}
                \put(-191,92){\makebox(0,0){\rotatebox{0}{$\sreg$}}}
                \caption{Phase III: $\xi(\amp=20,\per=10)=200$}
                \label{subfig:EE_regS_phase3}
        \end{subfigure}
        \qquad 
        \begin{subfigure}[b]{0.42\textwidth}
                \includegraphics[width=\textwidth]{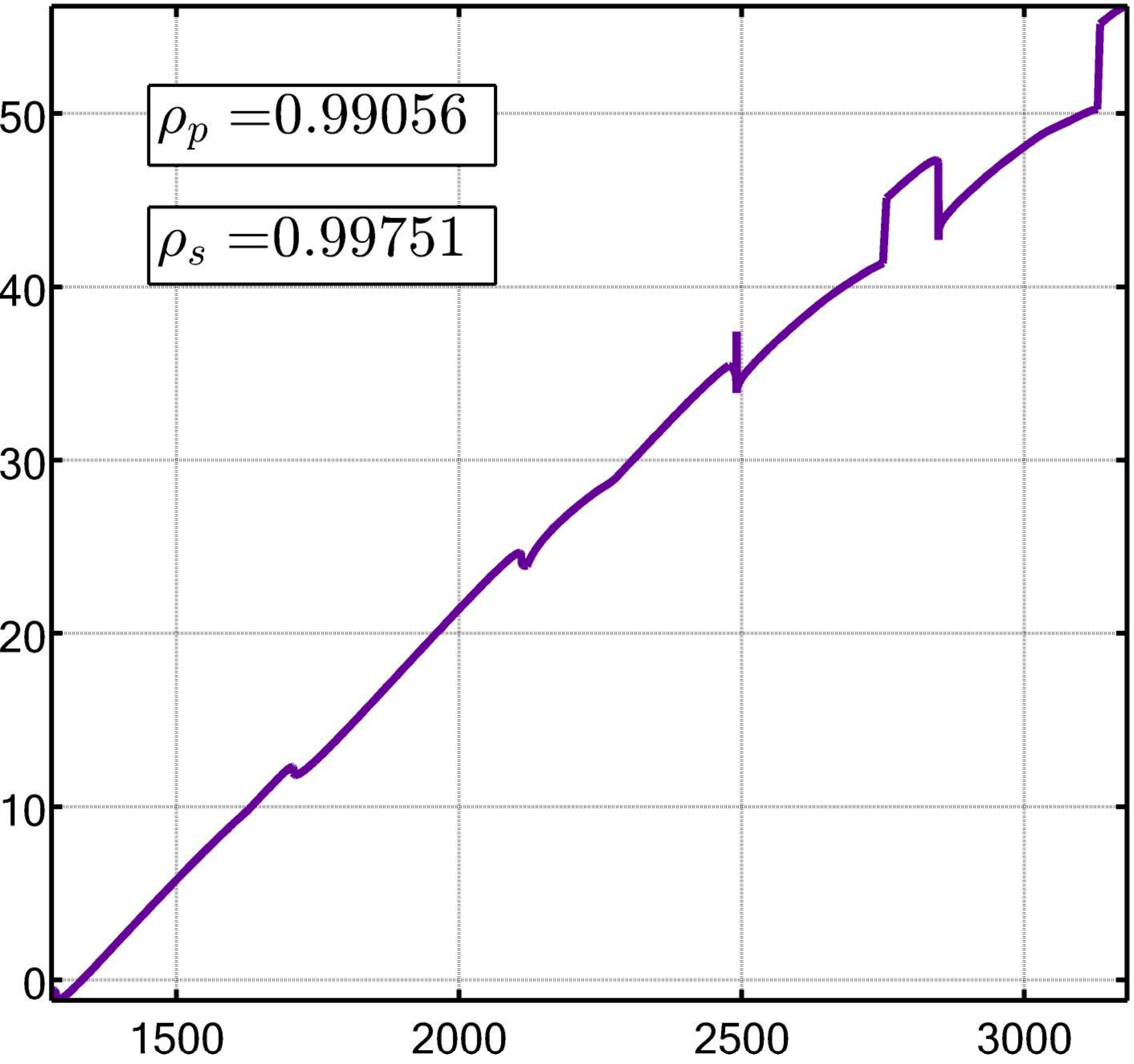}
                \put(-82,-7){\makebox(0,0){\normalsize $s/s_0$}}
                \put(-191,92){\makebox(0,0){\rotatebox{0}{$\sreg$}}}
                \caption{Phase IIb: $\xi(\amp=20,\per=0.1)=2$}
                \label{subfig:EE_regS_phase4}
        \end{subfigure}
        \caption{The evolution of the regularized entanglement entropy, $\sreg$ defined in Eq.~\eqref{eqn:S_reg}, against the normalized entropy of the black hole, $s/s_0 = s/s(t=0)$, for the four phases for a radial cutoff of $u_{\cal A} = 10^{-3}$. The strip widths are $a=0.05$ for panels (a), (b), (c), and $a=0.01$ for panel (d). We include the Spearman  and Pearson rank coefficients, $-1 \leq \rho_s \leq 1$ and $-1\leq \rho_p \leq 1$ respectively, for each plot to demonstrate the linearity of the correlation between the entanglement entropy and the thermal entropy (see text for explanation).}
        \label{fig:EE_reg_S}
\end{figure}

On the other hand when we reach phase IIa (tilted regime) by way of small amplitudes, we start to see definite oscillatory evolution of $\sreg$. In each oscillatory period we see a local reduction in $\sreg$. On the other hand the temporal radial depth attained by the extremal surface as measured by $u^*$ is almost similar to that in phase I by juxtaposing the behaviour in Fig.~\ref{fig:ES_evolution_per01_amp1} and Fig.~\ref{fig:ES_evolution_per10_amp1}.
 In phase IIa however, our extremal surfaces are closer to the boundary in contrast to phase I. We conjecture that the origin of the reduction in the $\sreg$ is associated with the sharp oscillations in $t^*$ or equivalently $\tilde{t}^*$. These imprint themselves into the actual value of the area despite the surface not getting too far into the bulk (which is possible since even the asymptotics of the geometry is sensitive to the driving, 
cf., \eqref{eq:metfg}). The onset of non-monotone growth of $\sreg$ in Fig.~\ref{subfig:EE_reg_phase2}  characterizes the departure from the linear regime to the non-linear domain in line with the behavior of the phase portrait which in turns modifies the conductivity (which picks up a real part $\sigmaIn > 0$ in phase IIa).

The temporal change of $\sreg$ is much more pronounced in the non-linear regime. 
In the unbounded amplification phase III (see Fig.~\ref{subfig:EE_reg_phase3}) and the dynamical crossover wobbly phase IIb  (see Fig.~\ref{subfig:EE_reg_phase4}), the $\sreg$ appears to track the time-coordinate of the cap-off point $\tilde{t}^*$ quite efficiently.  Indeed here we expect the non-linearities of the system to be the dominant effect. We know that the black hole grows quite rapidly in response to the energy injected into the system at the boundary from our discussion in \S\ref{sec:dc} and \S\ref{sec:unamp}. The behaviour in phase III is smooth with large amplitude oscillations, which qualitatively track quite well the behaviour of  
$\tilde{t}^*$. The dynamical crossover wobbly phase (phase IIb) exhibits a lot more drastic behaviour. We encounter for the first time a jumps in the family of extremal surface that minimize the area (satisfying the boundary conditions and the homology constraint). These jumps translate into continuous but non-differentiable kinks in $\sreg$ visible in Fig.~\ref{subfig:EE_reg_phase4}. We again note that the radial position of the cap-off point of the extremal surface behaves much more smoothly and the glitches appear in $\tilde{t}^*$. Furthermore, the growth of the entanglement itself is rather steep as we see about an order of magnitude difference in $\sreg$ between the low amplitude and high amplitude regimes.

It is interesting to contrast the change of entanglement entropy with the change in the thermal entropy to see how the two are correlated. As we have argued above, the fact that we have an ever increasing thermal entropy (the bulk black hole is constantly growing) implies that even for small sub-systems we will quickly see overwhelming thermal contribution. We display in 
Fig.~\ref{fig:EE_reg_S} the functional dependence of $\sreg$ on the (normalized) instantaneous thermal entropy $s(t)/s(t=0)$. \

It is immediately apparently by eyeballing the plots that there appears to be near-perfect correlation in three phases with 
Fig.~\ref{subfig:EE_regS_phase3} corresponding to phase III being the only outlier. To get a quantitative feeling for the correlation we have also indicated the Pearson correlation coefficient $\rho_p$ as well as the Spearman rank coefficient $\rho_s$. These are statistical markers for measuring correlations between two sets of data and are defined to take values in the interval $[-1,1]$. The values 
$\rho_s, \rho_p = 0, \pm 1$ signify zero, perfect positive and perfect negative correlation respectively. While the Spearman coefficient indicates that the observables in question are monotonically related, the Pearson coefficient provides an accurate measure of linear correlation. Indeed from the results quoted in Fig.~\ref{fig:EE_reg_S} we see that $\sreg(s)$ is a linear function to a very good approximation in phases I, IIa and IIb. It is curious that the linearity is respected even in the presence of the glitches in the growth of entanglement entropy (we do not see any drastic behaviour in the area of the apparent horizon). The unbounded amplification phase III clearly demonstrates the effects of non-linearities by decorrelating $\sreg$ and $s(t)$.

\section{Discussion}
\label{sec:discuss}

The non-equilibrium dynamics of  strongly coupled field theories is amenable to detailed quantitative exploration using the AdS/CFT correspondence. We have exploited this set-up to study the behaviour when a homogeneous thermal plasma is driven away from equilibrium by a periodically sourcing a relevant (composite) scalar operator.  The resulting dynamics exhibits a rather rich phase structure illustrated in Fig.~\ref{fig:PP_qualitative}. 

We identified four distinct phases, characterizing them in terms of the frequency and amplitude of the external driving force. Of these the dissipation dominated phase I is perhaps most intuitive for here the weakness of the driving, allows the system to to catch up with the driving. This is clearly visible in the various observables we studied; the complex conductivity of the response is purely real owing to the phase lag between the source and response and the evolution of entanglement is pretty quiescent. 

There is more structure when we ramp up either the period of driving, or the amplitude, for now the system departs quite rapidly away from equilibrium. The response therefore is more pronounced; we see more in phase response and greater temporal oscillations. In phases IIa to IIb there emerges a non-vanishing imaginary part to the conductivity, which in fact appears to capture the entire response for high values of the period and amplitude. We also notice that there are significant fluctuations in the energy density and the entanglement entropy and furthermore, the entropy density grows rather rapidly in this regime. Perhaps most intriguing is the  unbounded amplification of phase III, where we see sharp fluctuations and a highly non-linear response. We argue that this response appears to be not captured by polynomial self-interactions of the composite operator; the intricate dynamics of gravity in AdS appears to induce effective non-polynomial couplings in the effective action for the operator ${\cal O}$ we use to perturb the system away from equilibrium. We believe this fact is significant and should be taken into account when attempting to construct effective models distilling the effects of  gravitational interactions for strongly coupled systems .  

While our focus has been on computing the simplest set of observables, essentially one-point functions and entanglement entropy for small sub-systems, the power of holography is that we can do much more. In time independent equilibrium scenarios it is straightforward to use the holographic map to compute correlation functions (at least two point functions). In the genuine non-equilibrium scenarios as those we have focused on the technology for computing such observables, whilst present \cite{CaronHuot:2011dr,CaronHuot:2011dr} is still a bit cumbersome to work with (at least numerically). It would be interesting to develop these techniques further perhaps taking inspiration from the analytical models of \cite{Ebrahim:2010ra,Keranen:2014lna}. This would allow us with a direct probe of fluctuations in the plasma, which can be contrasted with the dissipation in the system, the latter being measured by the entropy production through the growth of the horizon.

Likewise our exploration of the behaviour of entanglement entropy has been restricted to analysis of small sub-systems for pragmatic reasons. While the sub-system under consideration was chosen to have fixed size, the fact that we are continuously driving the system leads to an ever increasing thermal contribution to the entanglement. Geometrically this is easy to understand since the horizon for our bulk solution is ever growing (as we have indicated that both the event and apparent horizons are required to be monotonic in our set-up) and 
reaches out towards the boundary in the course of the evolution. As  a result, the local thermal scale can overwhelm the relative smallness of the sub-region we choose. To have precise mapping of the entanglement structure we need to be able to ascertain the true minimum of the area functional in such scenarios bearing in mind that the extremal surface can (and often does) penetrate various horizons. A significant obstacle in ascertaining this is the fact that the characteristic method for solving Einstein's equations developed in \cite{Chesler:2013lia} excises the region of the spacetime behind the apparent horizon. While this is a technical obstacle, overcoming it would not only enable us to probe the interior of a highly non-equilibrium black hole using holographic entanglement, but it could also allow us to explore other interesting scenarios such as the effect of perturbing the ground state of the system by external sources.

\acknowledgments 

We would like to thank Veronika Hubeny and Henry Maxfield for useful discussions.
M.~Rangamani and M.~Rozali would like to acknowledge the hospitality of Yukawa Institute for Theoretical Physics, Kyoto during the course of the project. In addition M.~Rangamani would also like to acknowledge the hospitality of IAS, Princeton, University of Amsterdam and Aspen Center for Physics.

 M.~Rangamani acknowledges support from the Ambrose Monell foundation, by the 
 National Science Foundation under Grant 1066293,   by the FQXi  under grant ``Measures of Holographic Information" (FQXi-RFP3-1334),  by the STFC Consolidated Grant ST/L000407/1, and  the European Research Council under the European Union's Seventh Framework Programme (FP7/2007-2013), ERC Consolidator Grant Agreement ERC-2013-CoG-615443: SPiN (Symmetry Principles in Nature). M.~Rozali and A.~Wong are supported by NSERC.

\appendix
\section{Holographic Renormalization}
\label{sec:holren}

We collect here some salient results for the computation of physical field theory quantities using standard holographic techniques.

\subsection{Scalar deformations}
\label{sec:}

The bulk action \eqref{eq:bulkS} should be supplemented by boundary counter-terms to ensure that (a) the bulk equations of motion follow from a consistent variational principle and 
(b) the on-shell action evaluated on the solutions is finite.

In standard Poincar\'e-\AdS{d+1}
\begin{equation}
ds^2 = r^2\, \eta_{\mu\nu}\, dx^\mu\,dx^\nu + \frac{dr^2}{r^2} \equiv \frac{\eta_{\mu\nu}\, dx^\mu \, dx^\nu + dz^2}{z^2}
\label{}
\end{equation}	
the scalar field behaves asymptotically as 
\begin{align}
\phi(r,x) &\to \frac{1}{r^{d-\Delta}} \, \phi_0 + \frac{1}{r^\Delta}\, \phi_1
\nonumber \\
\phi(z,x) &\to z^{d-\Delta} \, \phi_0 + z^\Delta \, \phi_1 
\label{}
\end{align}	
We will work with standard quantization (Dirichlet boundary conditions) for the scalar field, which involves treating the mode that fall-off as $r^{\Delta-d}$ as the source for the scalar field.

In the presence of the source we let the metric to take the FG form,
\begin{equation}
ds^2 = \frac{dz^2}{z^2} + \frac{g_{\mu\nu}(x,z) \,dx^\mu\, dx^\nu}{z^2}
\label{}
\end{equation}	
where $g_{\mu\nu}(z,x) = \gamma_{\mu\nu} + {\cal O}(z)$.
If necessary we will denote by $\gamma_{\epsilon}$ the induced metric on the surface $z = z_\epsilon$ which differs from the boundary metric by a conformal transformation by $z_\epsilon^2$. We will ignore this issue for most part and write the counter-terms in terms of $\gamma_{\mu\nu}$ below for simplicity.

With these conventions we find  the following boundary counter-terms:
\begin{align}
 S_{bdy} &= \frac{1}{16\pi\, G_N}\;  \int d^d x \; \sqrt{-\gamma} \, \left( 2 K - 2\, (d-1) -\frac{1}{d-2}\, ^\gamma R  
 \right.
 \nonumber \\
 & \left. \qquad \qquad \qquad 
 -\frac{1}{2}\, \Delta_- \, \phi^2 + \frac{1}{2\,(2\,\Delta-d-2)}\;\left[ (\partial \phi)^2 + c_1 \; ^\gamma R\; \phi^2 \right]\right)
\label{}
\end{align}	
We are using conventional AdS/CFT definitions:
\begin{equation}
\Delta_\pm = \frac{d}{2} \pm \sqrt{\frac{d^2}{4}  + m^2} = \frac{d}{2} \pm \nu
 \label{}
\end{equation}	

Our interest concerns conformally coupled scalar field which has a mass in AdS units given by 
\begin{equation}
m_c^2 = -\frac{d^2-1}{4} \qquad \Longrightarrow \qquad \Delta_\pm = \frac{d\pm1}{2}
\label{}
\end{equation}	

To compute the boundary energy momentum tensor we vary
\begin{equation}
T^{\mu\nu} = \frac{2}{\sqrt{-\gamma}} \, \frac{(\delta S_{bulk} + S_{bdy})}{\delta\gamma_{\mu\nu}}
\label{}
\end{equation}	
where we should take care to include the appropriate radial dependence in the definition of $\gamma_{\mu\nu}$.

Lets split the contribution from the graviton and the scalar and write
\begin{equation}
T^{\mu\nu} = T^{\mu\nu}_g + T^{\mu\nu}_\phi
\label{}
\end{equation}	
where the split is determined by the requirement that $T^{\mu\nu}_\phi \propto \phi$.  Then the two pieces can be computed efficiently as follows:

\begin{align}
T^{\mu\nu}_g =  \frac{1}{16\pi \,G_N}\; \frac{2}{\sqrt{-\gamma}} \;\frac{\delta}{\delta \gamma_{\mu\nu}} 
& \left[
\int d^{d+1} x\; \sqrt{-g} \, (R+d(d-1))  
\right. \nonumber \\
& \left.
+\int d^dx\;\sqrt{-\gamma} \left(2 \, K - 2\, (d-1) - \frac{1}{d-2} \; {}^\gamma R \right)
\right]
\label{tmng1}
\end{align}	
which one can show evaluates to a nice covariant expression:
\begin{equation}
T^{\mu\nu}_g = \frac{2}{16\pi \,G_N} \; \left(K^{\mu\nu} - K\, \gamma^{\mu\nu}  + (d-1)\, \gamma^{\mu\nu} -\frac{1}{d-2}\, \left( {}^\gamma R^{\mu\nu} - \frac{1}{2} {}^\gamma R \, \gamma^{\mu\nu}\right) \right)
\label{tmng}
\end{equation}	
where $z_\epsilon$ is the location of the cut-off surface. 

The scalar contribution can be evaluated by using the fact that we are interested in the boundary variations to obtain:
\begin{align}
T^{\mu\nu}_{\phi} =  \frac{1}{16\pi \,G_N}\; \frac{2}{\sqrt{-\gamma}} \;\frac{\delta}{\delta \gamma_{\mu\nu}} 
 \left[
\int d^dx \sqrt{-\gamma}\; \left(\frac{1}{2\, z_\epsilon^{d-1}} \phi\, \partial_z \phi - \frac{1}{2\,z_\epsilon^d} \,\Delta_- \phi^2  + \cdots \right)
\right]
\label{tmnphi1}
\end{align}	
where $\cdots$ indicate the contribution from the higher order counter-terms and we have put back the powers of $z_\epsilon$  now. The details now depend on the asymptotic expansion of $\phi$. For general $\Delta$ we have to worry about the fact that the Taylor series solution in the neighbourhood of $z \simeq 0$ looks like
\begin{equation}
\phi(z,x) = \phi_0 \, z^{\Delta_-} + a_1(\phi_0) \, z^{\Delta_- -2} + \cdots + \phi_1 \, z^{\Delta_+} + \cdots
\end{equation}	
and we need to know the various intermediate pieces to complete the analysis.  The case we are interested in is rather special, where there are no powers of $z$ in the Taylor expansion between the source and the vev, so let us simply record the result for this case for now leaving a more general analysis for later.

Before proceeding though, let us note that we can express \eqref{tmnphi1} covariantly as follows ($ r=z^{-1}$):
\begin{align}
T^{\mu\nu}_{\phi} =  \frac{1}{16\pi \,G_N}\; \frac{2}{\sqrt{-\gamma}} \;\frac{\delta}{\delta \gamma_{\mu\nu}} 
 \left[
\int d^dx \sqrt{-\gamma_\epsilon}\; \left(\frac{1}{2\, r_\epsilon} \; \phi \, n^A \nabla_A\phi- \frac{1}{2} \,\Delta_- \phi^2  + \cdots \right)
\right]
\label{tmnphi2}
\end{align}	
where $n^A$ is the unit normal perpendicular to the cut-off surface.

\subsection{Specializing to $\Delta_- - \Delta_+ < 2$ ($\Delta_+ >d-2$)}
\label{sec:}
In this case the asymptotic expansion belongs to the special kind where 
\begin{equation}
\phi(z,x) = \phi_0 \, z^{\Delta_-} + \phi_1 \, z^{\Delta_+} + \cdots
\end{equation}	
where we are allowed to use the fact that $\phi_0\, z^{\Delta_-}$ is the beginning of an independent Taylor series where the powers of $z$ change by $2$ units (use the fact that the Lagrangian has $\phi \to -\phi$ symmetry). 
This corresponds to the  case we are interested where $\Delta_+ = 2$, $\Delta_- =1$ in $d= 3$.

In this circumstance we can simply use the terms written explicitly in \eqref{tmnphi1} to obtain
\begin{align}
T^{\mu\nu}_{\phi} =  \frac{1}{16\pi \,G_N}\; \frac{1}{2} (2\,\Delta_+ -d) \, \phi_0\, \phi_1\, \gamma^{\mu\nu}
\end{align}	

 Then we find
 \begin{equation}
T_{\mu\nu} = \frac{1}{16\pi\,G_N} \, \left(K_{\mu\nu} - K\, \gamma_{\mu\nu}  + (d-1)\, \gamma_{\mu\nu} + \frac{1}{2}\, (2\Delta_+ -d)\, \phi_0 \, \phi_1 \, \gamma_{\mu\nu}\right)
  \end{equation}	
 %
 
\subsection{$m^2 =-2$ in $d=3$}
\label{sec:}
 
Now, we can get the final answer for the case of interest either by working with the Fefferman-Graham expansion in which case we need to know that 
 \begin{align}
g_{\mu\nu}(z,x) dx^\mu\, dx^\nu&= - \left(1 -\frac{1}{4}\, \phi_0^2\, z^2+ \frac{4}{3}\, a_3 \, z^3 + \cdots \right) dt^2 
\nonumber \\
& \qquad + \left(1- \frac{1}{4}\, \phi_0^2 \, z^2- \frac{2}{3}\, (a_3 + \phi_0\, \phi_1)  \, z^3 + \cdots \right) \, (dx^2 + dy^2)
 \label{eq:metfg}
 \end{align}	
 The metric fall-offs allow us to compute the pieces in the boundary stress tensor directly since the $z^3$ term above is the correct answer. 

 Using this or directly computing from the CY-ansatz \eqref{tmnphi2} we claim to obtain (rescaled the result by a factor of $3/2$).
 \begin{equation}
T^\mu_\nu = \text{diag} \bigg\{2\, a_3 + \, \phi_0\, \phi_1 ,  -a_3, - a_3\bigg\}
\end{equation}	
 We can check that this satisfies the Ward identities:
 \begin{equation}
T^\mu_\mu = \phi_0\, \phi_1 = {\cal J}\, {\cal O}_2 \,, \qquad \nabla_\mu T^{\mu0} = -2\, \dot{a_3} - \phi_1\, \dot{\phi}_0 - \phi_0\,\dot{\phi}_1 = -\phi_1\, \dot{\phi}_0 = {\cal O}\, \nabla^\nu\, {\cal J}
\end{equation}	
 where we used the boundary conservation law derived from the solution
 $\dot{a}_3 = -\frac{1}{2}\, \phi_0\, \dot{\phi}_1$.

\section{Extremal surfaces and entanglement for strips}
\label{sec:eeapp}

In this appendix we describe our methodology for finding extremal surfaces relevant for the computation of entanglement entropy. 
For simplicity we will focus on regions which exploit the symmetry of our set-up and consider ${\cal A}$ to be a strip 
extended along one of the translationally invariant directions, say $y$ without any loss of generality, as in 
Eq.~\eqref{eqn:spatial_strip_defn}. 
We need a bulk codimension-2 surface that ends on the boundary of this region i.e., at 
$x=\pm a$ (at the chosen instant of boundary time $t_{\cal A} $).  We describe our strategy for finding this surface and computing its  (regulated) area  below.

\subsection{Determining extremal surfaces}
\label{sec:extrdet}

To find the extremal surface, we start by gauge fixing the reparameterization invariance on the surface. We take $y$ to be one of the coordinates. Dimensionally reducing in this direction, we construct an effective action for a curve in the remaining directions and pick a proper-length parameter $\lambda$ as the second coordinate. Thus, the extremal surface ${\cal E}_{\cal A}$ is embedded in the bulk as 
\begin{equation}
X^{\mu} = \left( t(\lambda),r(\lambda),x(\lambda) ,y\right) .
\label{eqn:embedding_coordinates}
\end{equation}
We choose the proper-length parameter to ensure that $\sqrt{ \text{det} \gamma_{ab} } = 1$, which implies that  the unregulated area of the extremal surface is given as 
\begin{equation}
\text{Area}({\cal E}_{\cal A}) = L_y\,  \int_{{\cal E}_{\cal A}} d\lambda \sqrt{ \text{det} \gamma_{ab} } = \lambda_{{\cal E}_{\cal A}} \, L_y \,,
\label{eqn:area_proper_length_param}
\end{equation}
in terms of parameter distance $\lambda_{{\cal E}_{\cal A}}$ spanned by the curve and the IR regulator $L_y$. 

In practical terms we work with the effective Lagrangian 
\begin{equation}
\mathcal{L}= \gxx^{2}\, \left[ 2 \, t'\,  e^{2 \br} \left( r' - t'\, \gtt \right)+\gxx^2 \, x'^2\right]
\label{eqn:detgamma}
\end{equation}
where the metric functions $\rho$, $\gtt$, $\br$ are obtained by interpolation of our numerical solutions.  This is a geodesic problem, with some non-minimal coupling from the dimensional reduction along the translationally invariant direction of the strip.  Instead if using the geodesic equations,  we found it convenient to pass to a set of  six \emph{first-order} Hamilton-like equations 
by introducing $P_{t} = t'$, $P_{x} = x'$, and $P_{+} = r' - \gxx \,t'$ which are related to the conjugate momenta. 
The equations we solve are the above three and 
\begin{equation}
\begin{split}
P'_{x} &= -\frac{4 P_x}{\gxx} \left( (P_t \gtt+P_{+} )\partial_r \gxx + P_t \partial_t \gxx \right) =0 \\
P'_{t} &= 2 P_x^2 \gxx e^{-2 \chi} \partial_r \gxx-P_t^2 \left(\partial_r \gtt+2 \gtt \partial_r \chi +2 \partial_t \chi \right)-\frac{2 P_t^2}{\gxx} \left(\gtt \partial_r \gxx + \partial_t \gxx \right) =0 \\
P'_{+} &= 2 P_x^2 \gxx e^{-2 \chi} \left(\gtt \partial_r \gxx + \partial_t \gxx \right) +  P_t P_+ \partial_r \gtt- \frac{2 P_+^2}{\gxx} \left(\partial_r \gxx+ \gxx \partial_r \chi \right) =0
\end{split}
\label{eqn:P_eom}
\end{equation}
We start from $x=0$ in the bulk at some smooth cap-off point $(x=0,t^*,r^*)$ where $t'=r'=0$.\footnote{ This cap-off point is not necessarily the deepest point in the bulk; for the examples shown in this paper it however does turn out to coincide.} and propagate out to the boundary. We evolve until a  with a fixed UV cut-off  at $r_{\eebdy}$ and regulate the final answer for the entanglement entropy by background subtraction (see below).

In the main text we illustrate the temporal dependence of the extremal surfaces and $S_{\cal A}^\text{reg}$ for each of the four phases (I-IV) of Fig.~\ref{fig:PP_qualitative} for fixed strip width $a$. Since we numerically control the data of the cap-off point we work iteratively: we start by fixing a suitable strip width $a$ by tuning $r^*$ and $t^*$, then we evolve the extremal surfaces by increasing $t^*$ and re-adjusting $r^*$ such that the strip width remains as $a$.  We note that we assume that there are no discontinuities or multi-valuedness in the map from $(r^*,t^*) \to (a,r_{\eebdy})$, which we believe makes sense for small strip widths.\footnote{ Such behaviour was noticed in extremal surface computation in global Vaidya-AdS by \cite{Hubeny:2013dea}.} Finally, to work in a compact domain we choose $u = 1/r \in [0,1]$ which we will use to explain the properties of the extremal surfaces.

\subsection{Regulated entanglement entropies}
\label{sec:regent}

Since the extremal surfaces reach out all the way to the boundary, the proper area is divergent with the coefficient of the leading divergent term fixed by the area of the entangling surface $\partial {\cal A}$. 
For a state of the CFT with vanishing sources for operators it is well known \cite{Ryu:2006ef} that the entanglement entropy behaves as
\begin{equation}
S_{\cal A}= \frac{ \text{Area}(\partial {\cal A}) }{u_{\cal A}} + S_{\cal A}^\text{fin} + {\cal O}(u_{\cal A}) \,.
\label{eq:sdiv}
\end{equation}
where $S_{\cal A}^\text{fin}$ is finite in the limit $u_{\cal A} \to 0$. In normalizable states of the field theory $S_{\cal A}^\text{fin}$ is the universal contribution to entanglement which should be independent of the cutoff value $u_{\cal A}$.\footnote{ For the vacuum state of a CFT$_3$ with ${\cal A}$ being a circular disc $S_{\cal A}^\text{fin}$ would give the F-function \cite{Myers:2010xs,Jafferis:2011zi} (the latter defined as the logarithm of the partition function of the theory a three-sphere). In fact, this can be used to define a UV finite quantity without recourse to background subtraction: following \cite{Liu:2012eea,Casini:2012ei} we can just as well consider 
$\left(R\,\frac{d}{dR} -1\right) S_{\cal A}$, with $R$ being the disc radius, as the measure of entanglement growth.}  One natural way for us to extract this quantity is to measure the entanglement relative to the $t=0$ thermal Schwarzschild state
$\Delta S_{\cal A}(t) = S_{\cal A}(t) - S_{\cal A}(t=0)$, which can be extracted simply by vacuum subtraction.

Usually, when we turn on sources for relevant operators, these can contribute additional divergences to the entanglement entropy \cite{Hung:2011ta}. In general in the presence of additional relevant scales one naively expects there to be logarithmically divergent terms polluting \eqref{eq:sdiv} and rendering vacuum subtraction meaningless. Fortuitously, this does no happen for the problem at hand. This can be extracted by examining the detailed discussion of \cite{Hung:2011ta}, which we paraphrase below.

There is however a quick argument for the absence of logarithmic terms which we now describe. For scalar operators in CFT$_d$ with operator dimension  $\frac{d}{2}< \Delta < \frac{d}{2}+1$, as we have considered, it is well known in AdS/CFT that the corresponding bulk field has mass in the window where both asymptotic fall-offs are normalizable, i.e., $m^2 \in \left(m_{BF}^2, m_{BF}^2 +1\right)$ with the Breitenlohner-Freedman bound mass $m_{BF}^2 = -\frac{d^2}{4}$ as usual.\footnote{ Implicit in this statement is the fact that we are quantizing the scalar field with standard (Dirichlet) boundary conditions, so that the dimension of the dual operator is $\Delta = \Delta_+$.} In this window note that 
$\Delta_- - \Delta_+ <2$  and we have the Legendre transformed theory with an operator of 
dimension $\Delta_-$ by switching to alternate quantization \cite{Klebanov:1999tb}.

Turning on a source for the faster-fall off mode $\Delta_+$ is equivalent, insofar as  the leading back-reaction on the metric, to considering instead a state in the Legendre transformed theory where the alternate quantized operator with dimension $\Delta_-$ acquires a vacuum expectation value. However, since the divergence structure of entanglement is the same in all states of the field theory, and the conformal vacua of the two theories (standard and alternate quantization)  coincide, it follows that the divergence structure of $S_{\cal A}$ should be unchanged from \eqref{eq:sdiv}, even with ${\cal J}(x) \neq 0$. Our story is of course  a special case with $\Delta_+ = 2, \Delta_- = 1$ in $d=3$. This observation is consistent with the results of 
\cite{Hung:2011ta} and the counter-terms used in \cite{Auzzi:2013pca}. 

To explicitly analyze the structure of the divergences in the entanglement entropy, let us consider  the metric given in 
\eqref{eq:metfg}.  Since the details of the divergences are blind to the boundary spatio-temporal behaviour of the sources we will examine the somewhat simplified setting where $\phi_0 = \text{const}$ to glean the relevant information.

With the time-translational symmetry restored by this choice, the Lagrangian for the extremal surface  (which now is minimal) is simpler:
\begin{equation}
 {\cal L} = \frac{\sqrt{g_{ii}(z)}}{z^2}\, \sqrt{g_{ii} (z)+ z'(x)^2} \, 
\end{equation}	
where $g_{ii}(z)$ is the spatial component of the metric in \eqref{eq:metfg}.  This system has a conserved Hamiltonian, which we can exploit to write down an expression for the area directly. 
Introducing, $z_*$ which captures depth to which the minimal surface penetrates into the bulk, we have for the on-shell value of the area
\begin{equation}
\text{Area }({\cal E}_{\cal A}) \propto \int_\epsilon^{z_*} \, dz\; \frac{\sqrt{g_{ii}}}{z^2} 
\left(1- \frac{g_{ii}(z_*)^2\, z^4}{ g_{ii}(z)^2\, z_*^4} \right)^{-\frac{1}{2}} 
\end{equation}	
Using the explicit form of $g_{ii}$, the second term is at least $z^4$ near the boundary so we can forget about it. The first term is all that matters, so lets look at
\begin{equation}
\frac{\sqrt{g_{ii}}}{z^2} = \frac{1}{z^2}- \frac{1}{4}\, \phi_0^2 - \frac{2}{3}\, (a_3 + \phi_0\, \phi_1)  
\, z + \cdots
\end{equation}	
which has the $z^{-1}$ divergence expected upon integration, but no further contribution of relevance in $z \to 0$ limit. 
From the $\phi_0^2$ term we get a contribution to the finite part of the  entanglement, and this is indeed the physically relevant answer. It should be clear from this discussion is not specific to the  choice $m^2 = -2$ in $d=3$, but should hold for $\frac{d}{2}< \Delta_+ < \frac{d}{2}+1$ as we argued abstractly above.



\providecommand{\href}[2]{#2}\begingroup\raggedright\endgroup

\end{document}